\documentclass[a4paper]{aa} 
\pdfoutput=1
\usepackage{geometry}
\usepackage{graphicx}
\usepackage[varg]{txfonts}
\usepackage{graphicx}
\usepackage{psfrag}
\usepackage{hyperref}
\usepackage{gensymb}
\usepackage{xcolor}
\usepackage{tabularx}
\usepackage{lipsum}
\usepackage[normalem]{ulem}
\usepackage{amsmath}
\usepackage[labelfont=bf, labelsep=period]{caption}
\usepackage{subcaption}
\usepackage{mathtools}
\usepackage{balance}
\usepackage{adjustbox}
\usepackage{float} 
\usepackage{placeins}
\usepackage{caption}
\usepackage{stfloats}
\usepackage{pifont}
\usepackage{xcolor}
\usepackage{cancel}

\newcommand{\cmark}{\textcolor{green!80!black}{\ding{51}}}
\newcommand{\xmark}{\textcolor{red}{\ding{55}}}

\usepackage{siunitx} 

\bibpunct{(}{)}{;}{a}{}{,} 

\begin{document} 

\title{Analysis of NILC performance on B-modes data from sub-orbital experiments}
\subtitle{}
\author{Alessandro Carones\inst{1,2} 
\and Marina Migliaccio\inst{1,2}
\and Domenico Marinucci\inst{3}
\and Nicola Vittorio\inst{1,2}
}
\institute{Dipartimento di Fisica, Universit\`a di Roma ``Tor~Vergata'', via della Ricerca Scientifica 1, I-00133, Roma, Italy \label{inst1}
 \and Sezione INFN Roma~2, via della Ricerca Scientifica 1, I-00133, Roma, Italy \label{inst2}
\and Dipartimento di Matematica, Universita’ di Roma Tor Vergata, via della Ricerca Scientifica 1, I-00133, Roma, Italy \label{inst3}}
\offprints{alessandro.carones@roma2.infn.it}

\abstract
{The observation of primordial $B$ modes in Cosmic Microwave Background (CMB) polarisation data represents the main scientific goal of most of the future CMB experiments. This signal is predicted to be much lower than polarised Galactic emission (foregrounds) in any region of the sky, pointing to the need for effective components separation methods.}
{Among all the techniques, the blind Needlet-ILC (NILC) is of great relevance given our current limited knowledge of the $B$-mode foregrounds. In this work, we explore the possibility of employing NILC for the analysis of $B$ modes reconstructed from partial-sky data, specifically addressing the complications that such an application yields: E-B leakage, needlet filtering, and beam convolution.}
{We consider two complementary simulated data-sets of future experiments: the balloon-borne SWIPE telescope of the Large Scale Polarisation Explorer, which targets the observation of both reionisation and recombination peaks of the primordial CMB $B$-mode angular power spectrum, and the ground-based Small Aperture Telescope of Simons Observatory, which, instead, is designed to observe only the recombination bump at $\ell \sim 80$. We assess the performance of two alternative techniques to correct for the CMB E-B leakage: the \emph{recycling technique} and the \emph{ZB method}.}
{We find that both techniques reduce the E-B leakage residuals at a negligible level given the sensitivity of the considered experiments, except for the recycling method in the SWIPE footprint at $\ell < 20$. Thus, we implement two extensions of the pipeline, the iterative B-decomposition and the diffusive inpainting, which enable us to recover the input CMB $B$-mode power for $\ell \geq 5$. For the considered experiments, we demonstrate that needlet filtering and beam convolution do not affect the CMB $B$-mode reconstruction. Finally, with an appropriate masking strategy, we find that NILC foregrounds subtraction allows to achieve sensitivities for the tensor-to-scalar ratio in agreement with the targets of the considered CMB experiments.}
{}

\authorrunning{}

\titlerunning{NILC method for B-modes data of sub-orbital experiments}

\keywords{Cosmology: Cosmic background radiation, observations - Methods: data analysis}

\maketitle

\section{Introduction}
Observations of the temperature and polarisation anisotropies of the Cosmic Microwave Background (CMB) have led to the establishment of a cosmological concordance scenario, the $\Lambda$CDM model, whose parameters are now tightly constrained (see, e.g., Boomerang:~\citealt{2006ApJ...647..799M}; WMAP:~\citealt{2013ApJS..208...19H}; Planck: \citealt{2020A&A...641A...6P}). 
CMB polarisation data are usually decomposed into the so-called $E$ and $B$ modes, of even and odd parity, respectively, in the sky \citep{1997PhRvL..78.2058K, 1997PhRvD..55.1830Z}. \\
$E$ modes, which are mostly generated on the last scattering surface by scalar perturbations, have been clearly detected by a number of experiments and contribute to confirm the $\Lambda$CDM cosmological scenario (see, e.g., {DASI:~\citealt{2002Natur.420..772K}}; {WMAP:~\citealt{2003ApJS..148..175S}; Boomerang:~\citealt{2006ApJ...647..813M}; Planck:~\citealt{2020A&A...641A...6P}). \\
$B$ modes, instead, are thought to be associated to at least two independent mechanisms. On small angular scales (typically a few arcminutes), they are mainly sourced by primordial $E$ modes distorted into $B$ modes via weak gravitational lensing by the intervening large-scale structures along the line of sight from the last scattering surface to us. This signal is commonly known as \emph{lensing $B$ modes} \citep{1998PhRvD..58b3003Z} and has been observed by several experiments, although the detections are still with a modest signal-to-noise ratio or in very small patches of sky \citep{2013PhRvL.111n1301H, 2014PhRvL.113b1301A, 2015Planck_GL, 2016ApJ...833..228B, 2015ApJ...807..151K, 2017ApJ...848..121P}. \\ 
On large scales, $B$ modes are expected to originate from tensor perturbations (primordial gravitational waves) generated in the very early Universe by a phase of cosmic inflation \citep{1997PhRvL..78.2058K}. Their magnitude is set by the tensor-to-scalar ratio $r$, which is defined as the amplitude of primordial gravitational waves over the one of initial density perturbations.
Such a primordial $B$-mode signal still escapes detection so far with an upper limit of $r \leq 0.032$ at $95\,\%$ CL \citep{2022PhRvD.105h3524T}. The detection of the primordial gravitational waves background would represent a powerful test of cosmic inflation and a means to discriminate among numerous theoretical models. \\
The main features in the BB tensor angular power spectrum are the reionisation bump at very large angular scales ($\ell \lesssim 10$), 
which is associated to the integral of the linear polarisation generated by quadrupolar anisotropies from the last scattering surface as seen by each free electron after reionisation,
and the recombination bump on smaller scales ($\ell \sim 80$), which represents the imprint of the primordial gravitational waves on the physics of the last scattering surface. \\
Many experiments have been designed or proposed to observe $B$-mode polarisation, either from the ground:  
POLARBEAR \citep{2010SPIE.7741E..1EA}, QUBIC \citep{2011APh....34..705Q}, BICEP \citep{2016JLTP..184..765W}, Keck-Array \citep{2003SPIE.4843..284K}, LSPE-STRIP \citep{2012SPIE.8446E..7CB}, ACT \citep{2020JCAP...12..047A}, SPT \citep{2020PhRvD.101l2003S}, Simons Observatory \citep{2019JCAP...02..056A}; 
from balloons: 
SPIDER \citep{2010SPIE.7741E..1NF}, LSPE-SWIPE \citep{2012SPIE.8452E..3FD}; 
or from space: 
LiteBIRD \citep{2014JLTP..176..733M}, PICO \citep{2019arXiv190210541H}. \\
The development of effective component separation methods has become one of the crucial aspects of data analysis for all these experiments. The reason being that the quest for primordial $B$ modes is made much more difficult by the presence of instrumental noise and polarised foregrounds, especially Galactic thermal dust and synchrotron emission. These methods are usually divided into blind, parametric and template-fitting techniques. Blind methods, such as ILC (\citealt{2003ApJS..148...97B}, \citealt{2003PhRvD..68l3523T}), NILC \citep{2012MNRAS.419.1163B}, FastICA \citep{2002MNRAS.334...53M}, SMICA \citep{2003MNRAS.346.1089D}, usually linearly combine multi-frequency maps in such a way as to minimise some meaningful statistical quantity of the final map. Parametric methods, such as Commander \citep{2008ApJ...676...10E}, FGBuster \citep{2009MNRAS.392..216S}, FastMEM \citep{2002MNRAS.336...97S}, instead, explicitly model the frequency properties of the foregrounds by means of a set of parameters that are fitted to the data. Such methods provide an easy way to characterise and propagate foregrounds residual errors, but their effectiveness depends on how reliable the adopted model is. Finally, template-fitting algorithms, such as SEVEM \citep{2003MNRAS.345.1101M}, try to construct internal foregrounds templates either in pixel or harmonic space using multi-frequency observations. \\
There is no clear evidence (especially in polarisation) on which approach is more effective in reconstructing the CMB signal. 
Thus, applying several different methods on the same data-set allows performing comparisons and evaluating the robustness of the results. \\
In this work, we focus our attention on the so-called Internal Linear Combination (ILC) algorithms. They are of great relevance in the context of CMB data-analysis because they perform noise and foregrounds subtraction with minimal prior information on their properties and, hence, they are not prone to systematic errors due to mis-modelling of the spectral energy distribution of the Galactic emission components. ILC methods will then play a key role in the analysis of future CMB experiments, given our still limited knowledge of the properties of the polarised foregrounds. One of the drawbacks of such approaches is that the estimation of foregrounds residuals usually relies on Monte Carlo simulations or other bootstrapping techniques. \\
Throughout the years, several extensions have been proposed; the Needlet ILC (NILC), which performs variance minimisation in needlet space, has proven to be one of the most promising blind techniques. NILC has been extensively applied to full-sky satellite data, but in the near future new observations of the CMB polarisation will be obtained mainly from ground-based and balloon-borne experiments, which will only observe a portion of the sky. Therefore, in this work, we explore the possibility of employing NILC for the analysis of partial-sky CMB $B$-mode data, specifically addressing the further challenges that such an extension yields: E-B leakage, needlet filtering, and beam convolution.
The importance of accurately characterising the impact of these operations has also recently been pointed out by \citet{2022JCAP...07..044Z}. \\
In our analysis, we consider two complementary experiments:
	\begin{itemize}
	    \item the LSPE-SWIPE, a balloon-borne telescope that avoids most of the atmospheric contamination, whose scientific goal is the observation of both the reionisation and recombination peaks
	    \item The Simons Observatory, which is a ground-based experiment targeting the detection of the recombination bump.
	\end{itemize}
The Short Wavelength Instrument for the Polarisation Explorer (SWIPE, \citealt{2012SPIE.8452E..3FD})
is one of the two instruments of the
LSPE (Large Scale Polarisation Explorer) experiment \citep{2021JCAP...08..008A}. 
It is a balloon-based array of bolometric polarimeters, that will survey the sky in three frequency bands centred at $145,\ 210$ and $240$ GHz, expecting to constrain the tensor-to-scalar ratio down to $r=0.015$ (at $95\%$ CL).
The SWIPE $145$ GHz band will be the main channel for CMB science, while observations at $210$ and $240$ GHz will monitor thermal dust contamination. SWIPE is scheduled to be launched from the Svalbard islands and will observe a large sky fraction (around $35\%$) with an angular resolution of Full Width at Half Maximum (FWHM) $85^{\prime}$ for around 15 days, exploiting the optimal observational conditions offered by the Arctic night (at latitude around $78^\circ$N). \\
The Simons Observatory (SO) consists of a Large Aperture Telescope (LAT) with a $6$-meter primary mirror and three $0.5$-meter refracting Small Aperture Telescopes (SATs) \citep{2019JCAP...02..056A}. SO will be located in the Atacama Desert at an altitude of $5.200$ metres in Chile’s Parque Astronomico and will collect data in the frequency range between $25$ and $280$ GHz. It is designed to measure a signal at the $r = 0.01$ level at a few $\sigma$ significance or to exclude it at similar significance, using the $B$-mode amplitude around the recombination bump. The reionisation peak, instead, cannot be observed by SO due to atmospheric contamination that will allow to sample only cosmological modes with $\ell > 30$. \\
The paper is organised as follows. In Sect. \ref{sec:maps}, we present the considered simulated multi-frequency data-sets; in Sect. \ref{sec:extensions} we describe the ILC methods considered in this work and how to extend the application of the NILC algorithm to partial-sky $B$-mode data; in Sect. \ref{sec:results_ideal} you can find the procedures adopted to assess the performance of ILC methods; in Sect. \ref{sec:results} the obtained results from the application of NILC to LSPE-SWIPE and SO simulated data; finally, we report our conclusions in Sect. \ref{sec:conclusion}.
\section{Simulated data-sets}
\label{sec:maps}
The simulated data-sets employed in this analysis include the two aforementioned CMB experiments: LSPE-SWIPE and the Small Aperture Telescope of Simons Observatory (SO SAT). We do not consider the Large Aperture Telescope of Simons Observatory, because it will not be devoted to primordial $B$-mode science. In the case of LSPE-SWIPE, realistic simulated Planck maps have been included in the component separation pipeline to have a broader frequency coverage to better trace the Spectral Energy Distributions (SEDs) of $B$-mode foregrounds. \\ 
The main properties of each data-set are listed in Table \ref{tab:set-up} (see \citealt{2021JCAP...08..008A}, \citealt{2020A&A...641A...1P} and \citealt{2019JCAP...02..056A}).
\begin{table*}[htbp!]
\centering
\setlength{\tabcolsep}{3pt}
\renewcommand{\arraystretch}{1.2}
\begin{tabular}{|c|c|c|c|}
\hline
 & \textbf{SWIPE} & \textbf{SO SAT} & \textbf{Planck} \\
\hline\hline
\text{Frequency (GHz)} & $145,\ 210,\ 240$ & $27,\ 39,\ 93,\ 145,\ 225,\ 280$ & $30,\  44,\  70, \ 100,\  143,\  217,\  353$  \\
\hline
\text{Beam FWHM (arcmin)}  & $85$ & $91,\ 63,\ 30,\ 17,\ 11,\ 9$ & $32,\ 28,\ 13,\ 10,\ 7,\ 5,\ 5$ \\
\hline
\text{Sky coverage} ($\%$) & $37$  & $34$ & $100$ \\
\hline
\text{Sensitivity} $\sigma_{Q,U}\, (\mu K_{CMB}\cdot$arcmin)  & $10,\ 17,\ 34$ & $35,\ 21,\ 2.6,\ 3.3,\ 6.3,\ 16$ & $210,  240,  300,  118,  70, 105, 439$ \\
\hline
\end{tabular}
\medskip
\caption{Instrumental properties of the different CMB experiments considered in this work.}
\label{tab:set-up}
\end{table*}
All maps are generated by adopting the HEALPix pixelisation scheme \citep{2005ApJ...622..759G} with $N_{side}=128$, which corresponds to a pixel resolution of $27.5^{\prime}$. \\
The Q and U maps at the different frequencies are obtained from the co-addition of three separate components: CMB, Galactic emission (synchrotron and thermal dust) and Gaussian white and isotropic noise. We do not include polarised extra-Galactic sources, since they are expected to be
dominant on angular scales smaller than those of main interest for inflationary $B$-mode science ($\ell > 100$) \citep{2018ApJ...858...85P}.\\
The CMB maps are generated with the HEALPix Python package\footnote{\url{https://github.com/healpy/healpy}} as random Gaussian realisations from the angular power spectra of the best-fit Planck-$2018$ parameters \citep{2020A&A...641A...6P} with tensor-to-scalar ratio $r=0$, unless otherwise specified. \\
Instrumental noise is simulated as white and isotropic Gaussian realisations with standard deviations in each pixel associated to the polarisation sensitivities listed in Table \ref{tab:set-up}. \\
Galactic emission is generated using the \textit{PySM} Python package\footnote{\url{https://github.com/healpy/pysm}} \citep{2017MNRAS.469.2821T}. Synchrotron polarised emission is modelled with a simple power law \citep{Rybicki}:
\begin{equation}
	X_{s}(\hat{n},\nu) = X_{s}(\hat{n},\nu_{0}) \cdot  \Bigg(\frac{\nu}{\nu_{0}} \Bigg)^{\beta_{s}(\hat{n})}
	\label{eq:sync}
\end{equation}
with $X=\{Q,U \}$, $\hat{n}$ the position in the sky and $\nu$ the frequency. $X_{s}(\hat{n},\nu_{0})$ represents the synchrotron template at a reference frequency $\nu_{0}$. Thermal dust emission is modelled with a modified black-body (MBB):
\begin{equation}
	X_{d}(\hat{n},\nu) = X_{d}(\hat{n},\nu_{0}) \cdot  \Bigg(\frac{\nu}{\nu_{0}} \Bigg)^{\beta_{d}(\hat{n})} \cdot \frac{B_{\nu}(T_{d}(\hat{n}))}{B_{\nu_{0}}(T_{d}(\hat{n}))},
	\label{eq:dust}
\end{equation}
where $B_{\nu}(T)$ is the black-body spectrum. \\
We consider two different Galactic models, commonly adopted within the forecast analyses of CMB polarisation experiments:
\begin{itemize}
    \item \texttt{d0s0}, where $\beta_{s}=-3,\ \beta_{d}=1.54$ and $T_{d}=20$ K are assumed to be constant across the sky;
    \item \texttt{d1s1}, where the spectral indices of synchrotron and thermal dust emission depend on the position in the sky.
\end{itemize}
In both cases, the dust template is the estimated dust emission at $353$ GHz in polarisation from the Planck-2015 analysis \citep{2016A&A...594A..10P}, smoothed with a Gaussian kernel of FWHM=$2.6\degree$ and with small scales added by the procedure described in \citet{2017MNRAS.469.2821T}. The synchrotron template is the WMAP $9$-year $23$-GHz Q/U map \citep{2013ApJS..208...20B}, smoothed with a Gaussian kernel of FWHM=$5\degree$ and small scales added. \\
In the \texttt{d1s1} model, dust temperature and spectral index maps are obtained from the Planck data using the Commander pipeline \citep{2016A&A...594A..10P}. The synchrotron spectral index map is derived using a combination of Haslam $408$ -MHz observations and WMAP $23$-GHz 7-year data \citep{2008A&A...490.1093M}. The \texttt{d0s0} model is a simplified representation of Galactic polarised emission, because we know that the spectral properties of the foregrounds vary in the sky. However, it still represents an important starting point for assessing the performance of blind methods on polarisation $B$-mode data-sets. On the other hand, the \texttt{d1s1} is surely closer to what the polarised Galactic emission is expected to be. \\ 
The maps of all components are smoothed with the beam of the channel of the considered data-set with the lowest angular resolution and then added together. \\
The sky patches observed by LSPE-SWIPE and SO SAT are shown in Fig. \ref{fig:patches}. The SO SAT footprint has a sky coverage of $f_{sky}=34\%$. However, due to the highly inhomogeneous scanning of the telescope, some regions of the sky will be much better observed than others, leading to an effective sky fraction of $\sim 10\%$ (see \citealt{2019JCAP...02..056A}). Therefore, to perform a realistic foregrounds subtraction, the ILC methods are applied to the simulated SO SAT data-set within the reduced patch shown in Fig. \ref{fig:patch_so} and obtained considering only the $10\%$ of the pixels with the highest values in the hit counts map provided by the SO Collaboration. Consistently, the sensitivity values reported in Table \ref{tab:set-up} refer to a homogeneous hit counts map with a sky fraction of $f_{sky}=10\%$ \citep{2019JCAP...02..056A}.
\begin{figure*}
	\centering
	\includegraphics[width=0.48\linewidth]{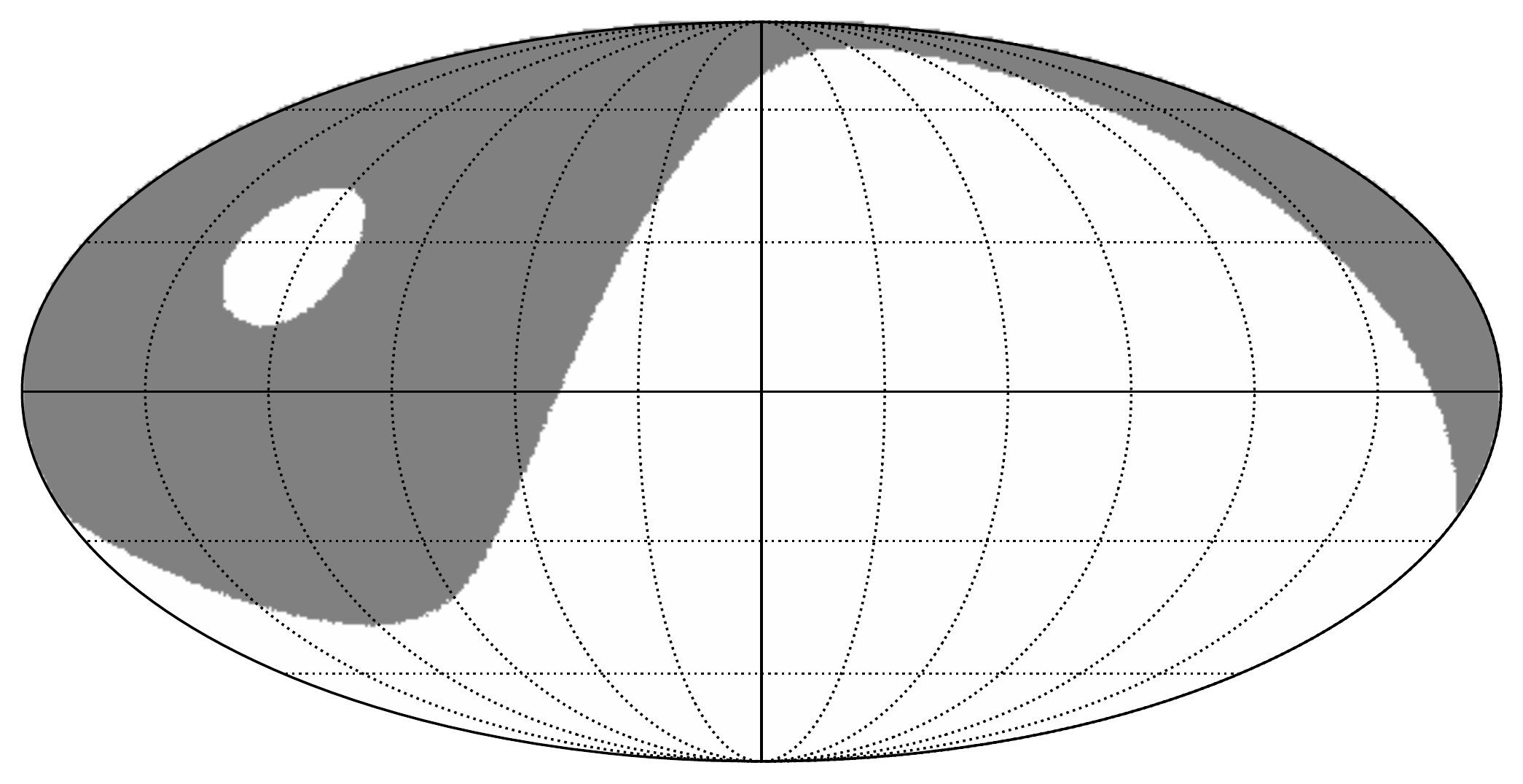}
	\includegraphics[width=0.48\linewidth]{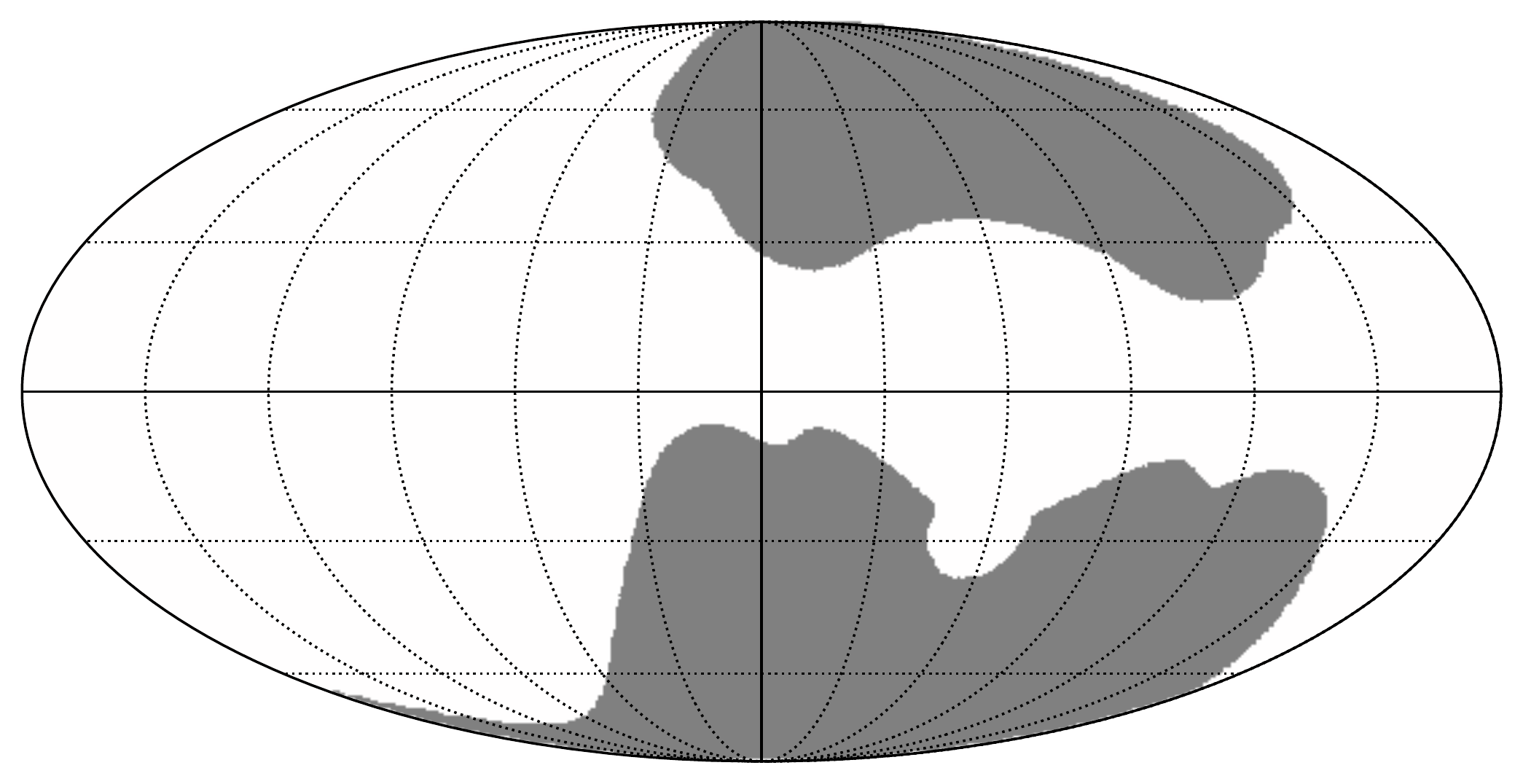}
	\caption{Sky coverage of the LSPE-SWIPE (left) and SO-SAT (right) instruments in Galactic coordinates. The observed region of the sky is shown in grey.}
	\label{fig:patches}
\end{figure*}
\begin{figure*}
\centering	\includegraphics[width=0.48\linewidth]{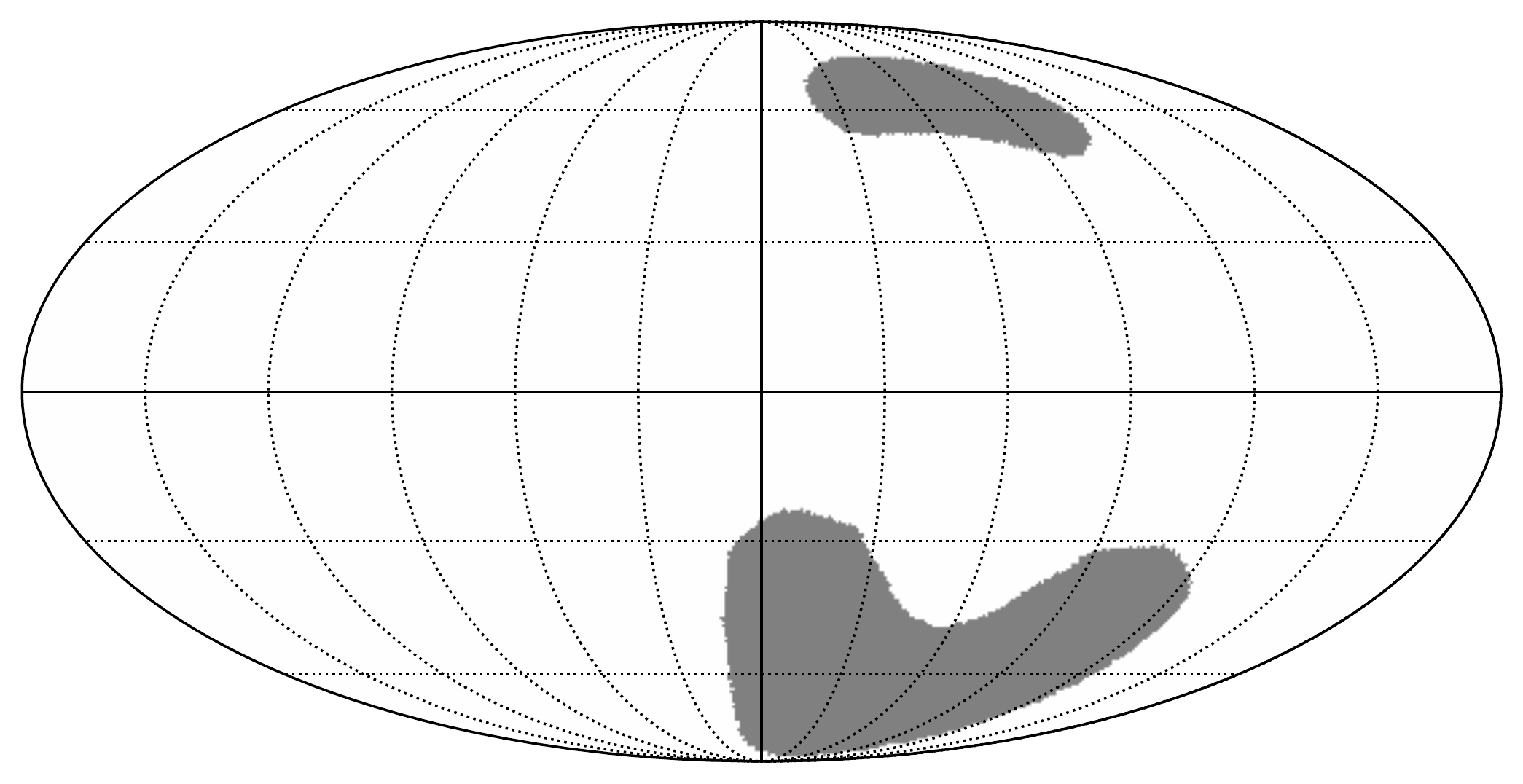}
	\caption{Sky area with the highest signal-to-noise ratio in the SO SAT patch. This is obtained considering pixels with the largest values in the hit counts map for a final sky fraction of $f_{sky}=10\%$. The ILC methods are applied within this patch.}
\label{fig:patch_so}
\end{figure*}

\section{Extension of ILC methods to partial-sky observations}
\label{sec:extensions}
The Internal Linear Combination (ILC) method is one of the most widely used approaches for the reduction of contaminants in CMB observations. It was first adopted in the analysis of the intensity data from the Wilkinson Microwave Anisotropy Probe (WMAP) satellite \citep{2003ApJS..148...97B}. Then, throughout the years, several extensions of the basic algorithm have been proposed. The methods applicable to polarisation data considered in this work are:
\begin{itemize}
    \item ILC,
    \item Polarisation ILC (PILC) \citep{2016MNRAS.459..441F},
    \item Needlet ILC (NILC) \citep{2009A&A...493..835D}.
\end{itemize}
All the above ILC methods consist in linearly combining the $N_{\nu}$ multi-frequency maps of one or multiple CMB experiments with frequency-dependent weights $\omega$:
\begin{equation}
\hat{X} = \sum_{j=1}^{N_{\nu}}\omega_{j}\cdot X_{j},
\label{eq:ILC}
\end{equation}
where $X$ is the $B$-field for ILC, $P=Q+iU$ for PILC and a set of $B$-mode needlet coefficients for NILC (as discussed in the text below). \\
The weights are derived by minimising the variance of $\hat{X}$ to reduce the contamination of Galactic emission and instrumental noise. Therefore, among the component separation techniques, ILC methods require the smallest number of a priori assumptions. Recently, a further variation of these techniques has been proposed, the constrained Moments ILC (cMILC), where the weights estimation is constrained to de-project some moments of the foregrounds emission \citep{2021MNRAS.503.2478R}. In this work, however, we will not consider such an extension. \\
Input maps of Eq. \ref{eq:ILC} have to be smoothed with a common beam that usually corresponds to the one of the channel with the lowest angular resolution. \\
Each frequency map $X_{j}$ can be described as the sum of the CMB signal $s$, foregrounds emission $f_{j}$ and instrumental noise $n_{j}$:
\begin{equation}
X_{j}= a_{j}s + f_{j} + n_{j},
\label{eq:input}
\end{equation}
where $a_{j}$ are the calibration coefficients. Assuming that the observations are calibrated with respect to the component of interest (the CMB) and are expressed in thermodynamic units, we have $a_{j}=1,\ \forall j$. Therefore, to preserve the CMB signal in the output map, the weights must satisfy the further constraint $\sum_{j=1}^{N_{\nu}}\omega_{j}=1$. This last condition, together with Eqs. \ref{eq:ILC} and \ref{eq:input}, enables us to demonstrate that the output solution contains the full CMB signal and some foregrounds and noise residuals:
\begin{equation}
\hat{X} = \sum_{j=1}^{N_{\nu}}\omega_{j}\cdot (s + f_{j} + n_{j})=s+\sum_{j=1}^{N_{\nu}}\omega_{j}\cdot (f_{j} + n_{j}).
\label{eq:ILC_new}
\end{equation}
In the case of simulated data-sets, the estimation of foregrounds and noise residuals is straightforward, while for actual data they have to be estimated through Monte Carlo simulations based on our prior knowledge of the contamination in the input multi-frequency data-set. \\
The weights which minimise the output variance and whose sum is equal to unity can be analytically estimated for ILC, PILC and NILC (see \citealt{2003ApJS..148...97B},\citealt{2016MNRAS.459..441F},\citealt{2009A&A...493..835D}). The only assumption of these methods is that the CMB has a known emission law (black-body spectrum) and no correlations with foregrounds or noise. \\
NILC, which is the main algorithm considered in this work, represents a refinement of the ILC method \citep{2009A&A...493..835D}, since the effectiveness of variance minimisation is enhanced when performed in the needlet domain. \emph{Needlets} are a particular wavelet system that guarantees simultaneous localisation in harmonic and pixel space. They have been introduced into the statistical literature by \citet{2006math......6599B} and have been applied for the first time to CMB data in \citet{2006PhRvD..74d3524P}. \\
In practise, the needlet coefficients of an input $B$-mode map at frequency $i$, $\beta_{j}^{i}$, are obtained by filtering its harmonic coefficients $a_{\ell m}^{i}$ with a weighting function $b_{j}(\ell)$ that selects modes at different angular scales for each needlet scale $j$:
\begin{equation}
\beta_{j}^{i}(\hat{\gamma})
=\sum_{\substack{\ell,m}} \left(a_{\ell m}^{i}\cdot b_{j}(\ell)\right)\cdotp Y_{\ell m}(\hat{\gamma}),
\label{eq:needlets_map}
\end{equation}
where $\hat{\gamma}$ is a direction in the sky. This procedure in harmonic space is equivalent to performing a convolution of the map in real domain. \\
The shape of the needlet bands is defined by the choice of the harmonic function $b$ whose width is set by a parameter $B$: lower values of $B$ correspond to a tighter localisation in harmonic space (fewer multipoles entering into any needlet coefficient), whereas larger values result in wider harmonic bands. Commonly adopted constructions of the harmonic function $b(\ell)$ are the \emph{standard} \citep{doi:10.1137/040614359,2008MNRAS.383..539M}, the \emph{cosine} \citep{2012MNRAS.419.1163B} and the \emph{mexican} needlets \citep{2008arXiv0811.4440G}. \\
The input needlet maps are then linearly combined in such a way as to obtain a minimum variance map $\beta_{j}^{NILC}$ at each scale $j$:
\begin{equation}
\beta_{j}^{NILC}(\hat{\gamma}) = \sum_{i=1}^{N_{\nu}}\omega_{i}^{j}(\hat{\gamma})\cdot \beta_{j}^{i}(\hat{\gamma})=\sum_{\substack{\ell,m}} a_{\ell m,j}^{NILC} \cdotp Y_{\ell m}(\hat{\gamma}).
\label{eq:NILC}
\end{equation}
The pixel-dependent weights are computed as follows:
\begin{equation}
\omega_{i}^{j}(\hat{\gamma})=\frac{\sum_{k}C_{ik}^{j}(\hat{\gamma})^{-1}}{\sum_{ik}C_{ik}^{j}(\hat{\gamma})^{-1}},
\label{eq:NILC_weights} 
\end{equation}
where the covariance matrix $C_{ik}^{j}(\hat{\gamma})=\langle \beta_{j}^{i}\cdot \beta_{j}^{k} \rangle$ for scale $j$ at pixel $\hat{\gamma}$ is estimated as the  average of the product of needlet coefficients in some space domain $\mathcal{D}$. This domain is usually a Gaussian window function centred at $\hat{\gamma}$, whose width varies with the considered needlet scale $j$. \\
The final NILC map is then reconstructed by filtering again the harmonic coefficients $a_{\ell m,j}^{NILC}$ in Eq. \ref{eq:NILC} and summing them all for each $\ell$ and $m$:
\begin{equation}
X_{NILC}(\hat{\gamma})=\sum_{\substack{\ell,m}} a_{\ell m}^{NILC} \cdotp Y_{\ell m}(\hat{\gamma})
=\sum_{\substack{\ell,m}}\Bigg(\sum_{j} a_{\ell m,j}^{NILC}\cdot b_{j}(\ell)\Bigg)\cdotp Y_{\ell m}(\hat{\gamma}).
\label{eq:needlets_alm_final}
\end{equation}
Minimising the variance (and hence the contamination) separately on different needlet scales leads to a more effective cleaning. On large scales, diffuse Galactic foregrounds dominate over the other components and are better removed in NILC with respect to ILC. The same is expected to happen for the instrumental noise on smaller scales (larger multipoles). \\
We note that any error on the CMB calibration in Eq. \ref{eq:input} could have a relevant impact on the CMB reconstruction, as discussed in \citet{2010MNRAS.401.1602D}. Furthermore, as noted in \citet{2009LNP...665..159D} and further discussed in \citet{2009A&A...493..835D}, one of the main limitations of ILC methods is the generation of empirical correlations between the CMB and the contaminants \citep{2009A&A...493..835D} induced by the departure of
the empirical correlation matrix $C_{ik}$ from its ensemble average due to the
finite size of the sample over which it is estimated. \\
This can lead to a negative bias in the reconstructed CMB power spectrum, especially at low multipoles, due to the cancellation of CMB modes projected in the foregrounds and noise sub-space. This effect has to be taken into account when estimating the CMB power spectrum. The NILC bias problem in this analysis is further discussed in the Appendix \ref{app:nilc_bias}. \\
In the case of ground-based and balloon-borne experiments, ILC methods cannot be applied straightforwardly to $B$-mode data. Indeed, three main complications arise:
\begin{itemize}
    \item the E-B leakage: some CMB $E$ modes are estimated as $B$ modes in the E-B decomposition performed on partial-sky Q and U maps (see Sect. \ref{sec:leak}). It affects the ILC and NILC pipelines.
    \item for NILC, needlet filtering of a partial-sky signal can lead to a loss of modes and power, especially close to the borders of the mask (see Sect. \ref{sec:bands}).
    \item degrading cut-sky polarisation maps to a common angular resolution, analogously to needlet filtering, can lead to errors in the reconstruction of the signal in the pixels close to the border of the observed patch (see Sect. \ref{sec:smooth}). It impacts the application of ILC, PILC and NILC.
\end{itemize}
The E-B leakage effect is associated to the fact that the E-B decomposition of partial-sky Stokes parameters is not exact: some modes (\emph{ambiguous modes}) satisfy both the $E$- and $B$-conditions simultaneously. When we split the polarisation field into an $E$- and a $B$-part, these ambiguous modes can go into either component \citep{2003PhRvD..67b3501B}. In the case of CMB, where $E$ modes are much brighter than $B$ modes, this leads to an over-estimation of the power in $B$ modes, especially on large scales. \\
The needlet filtering, on the other hand, represents a convolution of the data. If $B$ modes are observed only in a portion of the sky, some modes (on large scales) can be lost in the process if the configuration of the needlet bands is not carefully chosen, because the convolution mixes the signal of the pixels close to the border with the null one of the unobserved region. \\
Both these effects are relevant especially on those angular scales which are most sensitive to the amplitude of primordial tensor perturbations, and thus they must be properly analysed and treated to have a correct reconstruction of the primordial tensor CMB $B$-mode signal. \\
Finally, the ILC algorithms require input maps with a common angular resolution (which usually is the one of the frequency channel with the largest beam). The convolution for a beam, in analogy with the needlet filtering, is not trivial if one deals with observations only on a portion of the sky due to the leakage of null signal of the unobserved region within the patch. \\ 
The relevance and possible correction of such issues are explored in Sects. \ref{sec:leak}, \ref{sec:bands} and \ref{sec:smooth} by comparing the angular power spectra (evaluated in the footprint of the considered experiments) of leakage-corrected, needlet-filtered, and smoothed CMB simulations with those of the input exact $B$-mode signal, reconstructed with a full-sky B-decomposition of input Q and U maps. In such analysis, we consider the entire footprint observed by SO, given in the right panel of Fig. \ref{fig:patches}, since this is the region
over which the $B$ modes will be reconstructed from Q and U maps.
Specifically, we use the estimator of the pseudo-angular power spectrum, which indicates the rotationally invariant variance of the harmonic coefficients of a map: 
\begin{equation}
    {C}_{\ell}=\frac{1}{2\ell +1}\sum_{m=-\ell}^{m=\ell}\Big|a_{\ell m}\Big|^{2}. 
\label{eq:Cl}
\end{equation}
It is customary to report an analogous quantity: 
${D}_{\ell}=\frac{\ell(\ell+1)}{2\pi}{C}_{\ell}$, which would be constant on large scales for a scale-invariant primordial density perturbations spectrum. In this work, the estimation of the power spectra of the $B$-mode maps is performed with the \emph{NaMaster} Python package\footnote{\url{https://namaster.readthedocs.io/en/latest/}} that extends the estimator in Eq. \ref{eq:Cl} taking into account the effects of masking and beam convolution (see \citealt{2019MNRAS.484.4127A}). \\

\subsection{Correction of the E-B leakage}
\label{sec:leak}
The polarisation field of the CMB is a spin-2 quantity that can be described in terms of the Stokes parameters $Q$ and $U$ as follows:
\begin{equation}
    P_{\pm}(\hat{\gamma}) = Q(\hat{\gamma}) \pm iU(\hat{\gamma}),
\label{eq:P}
\end{equation}
where $\hat{\gamma}$ denotes the position in the sky. This field can be expanded in spin-weighted spherical harmonics:
\begin{equation}
    P_{\pm}(\hat{\gamma}) = \sum_{\ell m}a_{\pm2,\ell m}\ {}_{\pm2}Y_{\ell m}(\hat{\gamma}).
\label{eq:P_lm}
\end{equation}
\begin{figure*}
	\centering
	\includegraphics[width=0.70\textwidth]{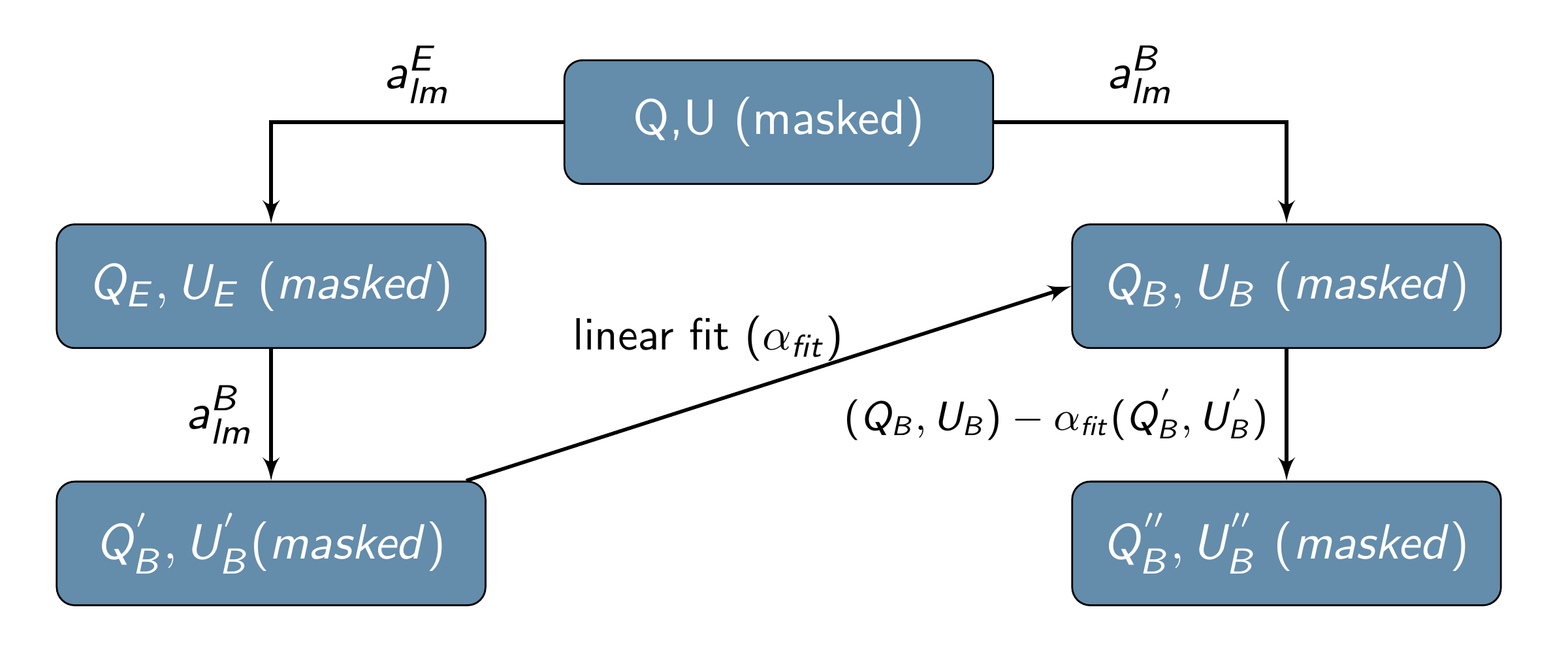}
	\caption{Recycling method for E-B leakage correction}
	\label{fig:rec_method}
\end{figure*}
From the harmonic coefficients in Eq. \ref{eq:P_lm}, it is possible to build the polarisation fields \emph{E} (a scalar map of even-parity) and \emph{B} (a pseudo-scalar field of odd-parity):
\begin{equation}
    \begin{aligned}
        E_{\ell m}=& -\frac{1}{2}\left(a_{2,\ell m}+a_{-2,\ell m}\right) \\
        B_{\ell m}=& -\frac{1}{2i}\left(a_{2,\ell m}-a_{-2,\ell m}\right). \\
    \end{aligned}
\label{eq:EB_lm}
\end{equation}
The construction of maps from $E_{\ell m}$ and $B_{\ell m}$ represents the E- and B-decomposition of a set of Q and U maps. \\
For partial-sky observations, we do not have information on Q and U on the entire sky and, therefore, Eqs. \ref{eq:P}, \ref{eq:P_lm} and \ref{eq:EB_lm} lead to an incorrect reconstruction of $E$ and $B$ maps. The E-B decomposition on the cut-sky is indeed not unique due to the non-locality of the transformation and, therefore, some $E$ modes will be interpreted as $B$ modes and vice versa (\emph{ambiguous modes}) (see  \citealt{2001PhRvD..65b3505L}).
The CMB $E$-mode power is much greater at all multipoles than that of $B$ modes and therefore the issue of leakage of the $E$ modes into the $B$ modes is the most relevant for partial-sky observations, leading to an over-estimate of the power in the reconstructed CMB $B$-mode map. \\
Several different methods have been proposed to address the leakage correction problem \citep{2001PhRvD..65b3505L,2003PhRvD..67b3501B,2010A&A...519A.104K,2019PhRvD.100b3538L,2010PhRvD..82b3001Z,2011A&A...531A..32K,2021JCAP...02..036G}. In this work, we consider: 
\begin{itemize}
    \item the recycling method introduced in \citet{2019PhRvD.100b3538L}
    \item the ZB method presented in \citet{2010PhRvD..82b3001Z}.
\end{itemize}
We test these techniques on $200$ CMB-only simulations that include lensing from $E$ modes and primordial tensor perturbations with $r=0.01$ for the cases of the SWIPE and SO SAT patches (shown in Fig. \ref{fig:patches}). We have chosen this value of the tensor-to-scalar ratio because it is close to the upper bound targeted at the $95\%$ confidence level (CL) by LSPE in the case of no detection \citep{2021JCAP...08..008A} and represents the amplitude expected to be observed at high significance by SO \citep{2019JCAP...02..056A}. \\
The performance of these methods has been evaluated by comparing the angular power spectra, $D_{\ell}^{BB}$ and $D_{\ell,in}^{BB}$, of leakage-corrected and exact $B$-mode maps. The exact $B$ modes are reconstructed with a full-sky B-decomposition of CMB Q and U simulations. Furthermore, we consider an effective tensor-to-scalar ratio:
\begin{equation*}
   r_{leak} = \frac{\left | D_{\ell,in}^{BB}-D_{\ell}^{BB}\right |}{D_{\ell,r=1}^{BB}}, 
\end{equation*}
 which quantifies the absolute error we can make in the estimation of the amplitude of primordial tensor perturbations due to the presence of leakage residuals at the different angular scales. This parameter accounts for the error in the reconstruction of both the tensor modes and the lensing signal. \\ 
We detail the recycling and ZB methods in the following sections. 
\begin{figure*}
	\centering
	\includegraphics[width=0.45\textwidth]{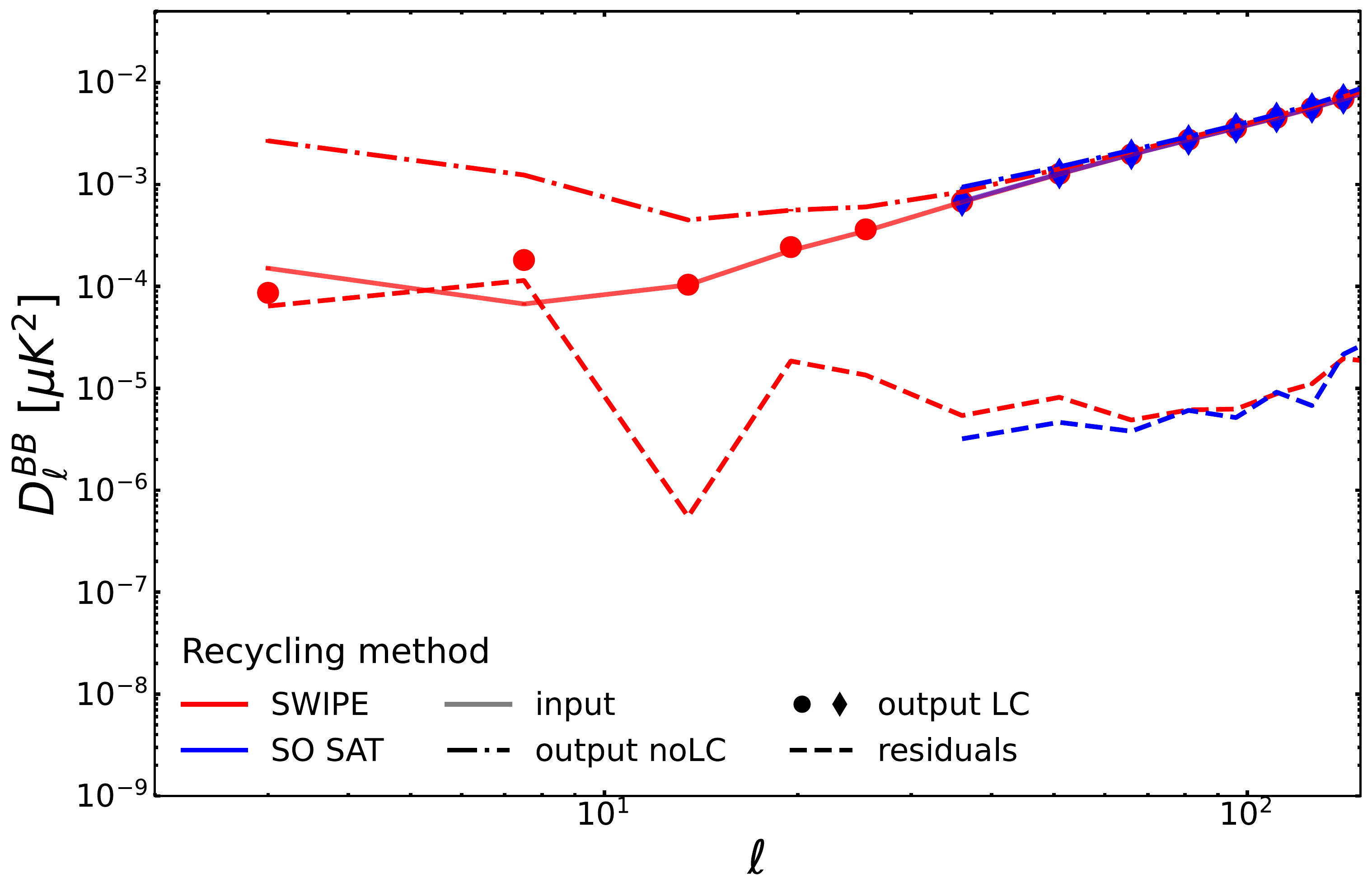}
	\hspace{0.5 cm}
	\includegraphics[width=0.455\textwidth]{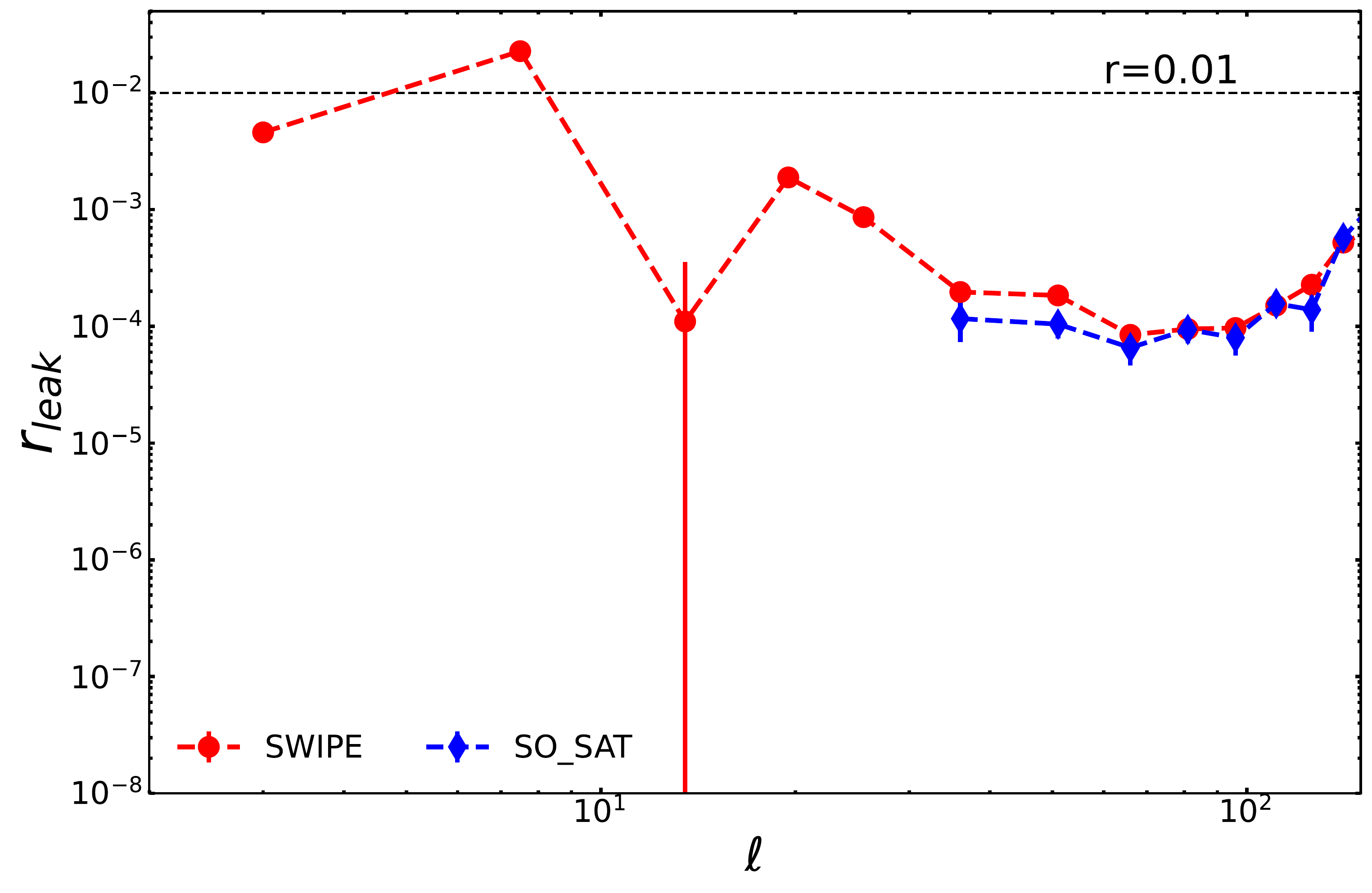}
	\includegraphics[width=0.45\textwidth]{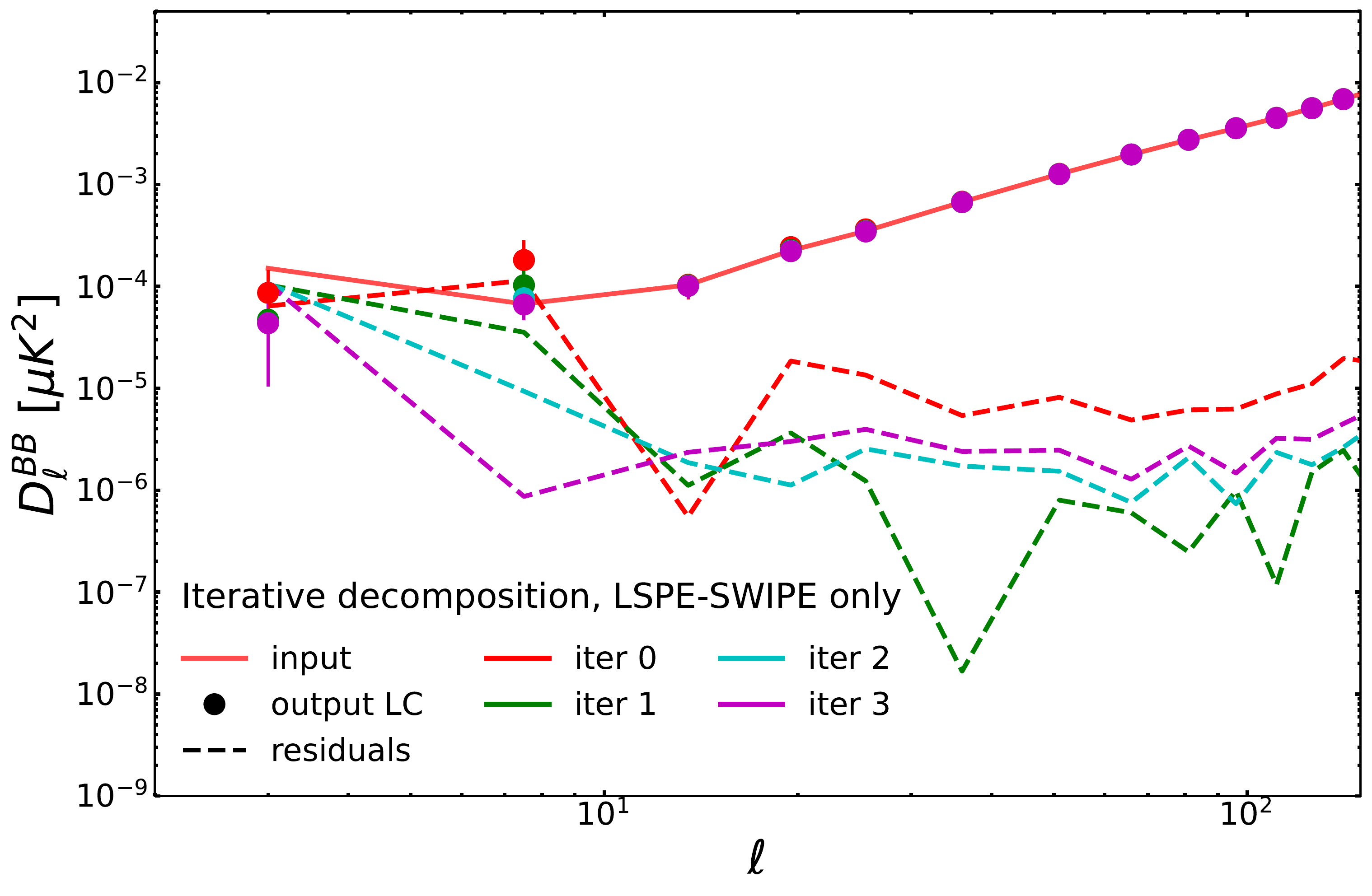}
	\hspace{0.5 cm}
	\includegraphics[width=0.45\textwidth]{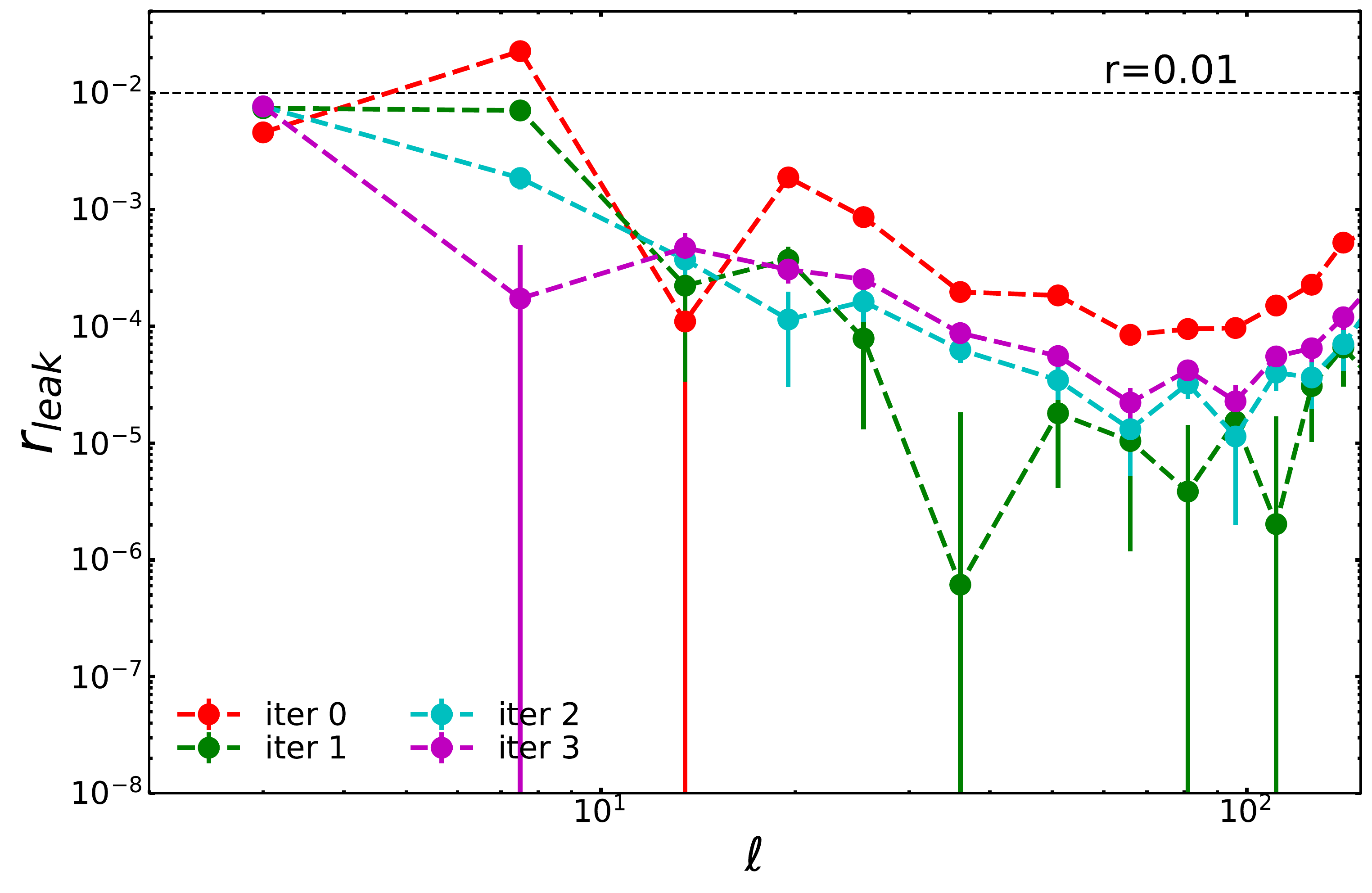}
	\includegraphics[width=0.45\textwidth]{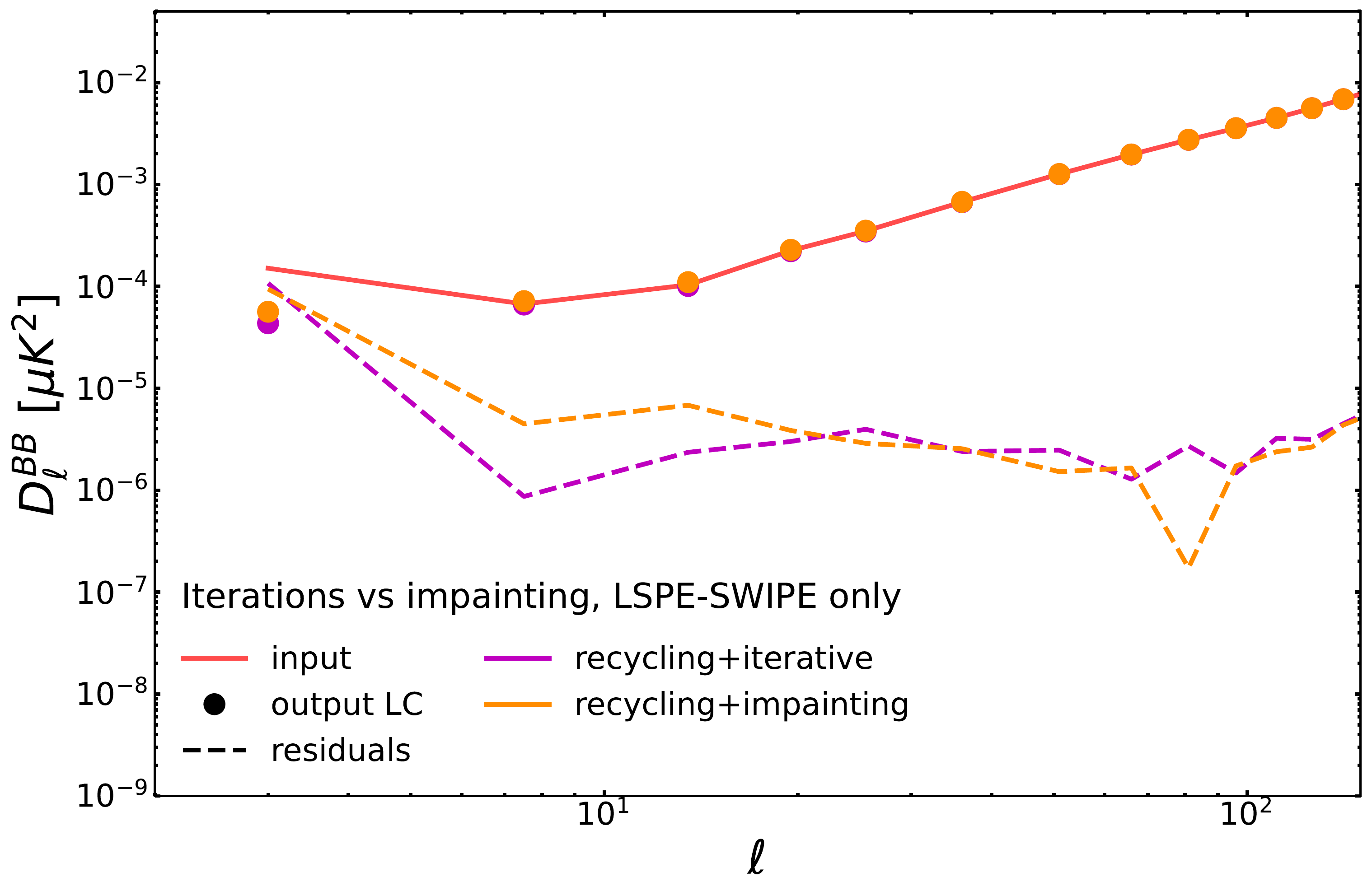}
	\hspace{0.5 cm}
	\includegraphics[width=0.45\textwidth]{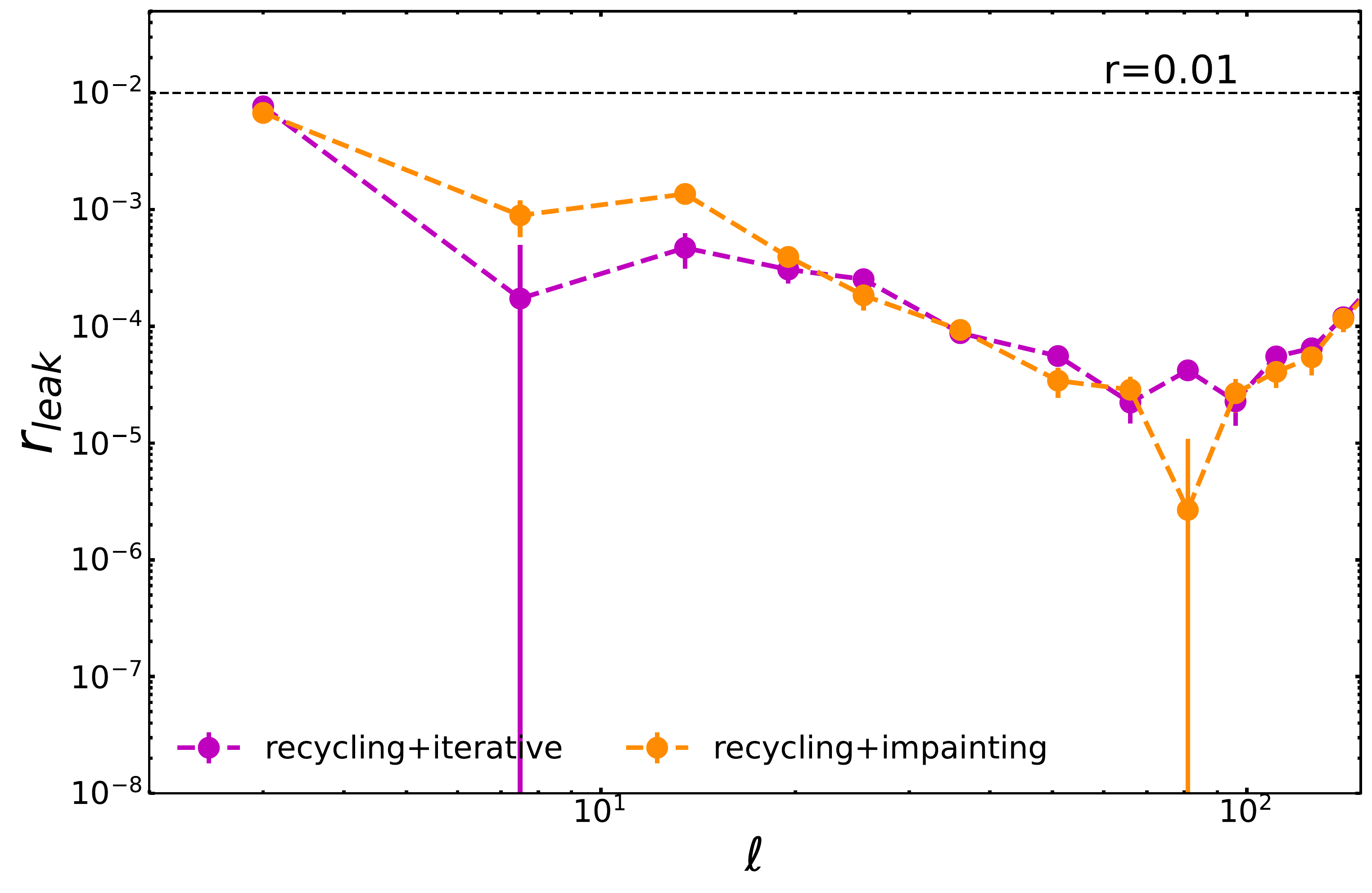}
	\caption{On the left: mean angular power spectrum over $N_{sims}=200$ CMB-only simulations that include lensing and a primordial tensor signal with $r=0.01$. Solid lines represent the power spectrum of exact $B$-maps (input), reconstructed with a full-sky B-decomposition of Q and U and then masked for power spectrum estimation; circles and diamonds that of leakage-corrected maps (output LC) for SWIPE and SO SAT, respectively; the dashed lines the leakage residuals after the correction; the dashed-dotted lines (if present) the CMB $B$ modes without any leakage correction (output noLC). On the right: the effective tensor-to-scalar ratio associated to the absolute difference between leakage-corrected and exact angular power spectra. Top panel shows the results when the \emph{recycling method} is applied to CMB $B$-mode maps for the SWIPE (red) and SO SAT (blue) footprints; middle panel those obtained when iterative B-decomposition with no (red), 1 (green), 2 (cyan) or 3 (magenta) iterations is performed; in the bottom panel, we compare the performance of iterative decomposition with three iterations (magenta) and diffusive inpainting (orange) in correcting the residual leakage. The middle and bottom panels present results only for SWIPE. The adopted binning scheme is $\Delta\ell =3$ for $\ell \leq 4$, $\Delta\ell =6$ for $5 \leq \ell \leq 28$ and $\Delta\ell =15$ for $\ell \geq 29$ for the spectra computed in the LSPE-SWIPE patch, while a constant $\Delta\ell =15$ for those estimated in the SO footprint. The error-bars highlight the uncertainty on the mean at $1\sigma$, estimated from the dispersion of the angular power spectra of the simulations divided by $\sqrt{N_{sims}}$. See text for details.}
	\label{fig:leak_liu_swiso}
\end{figure*}
\begin{figure*}
	\centering
	\includegraphics[width=0.70\textwidth]{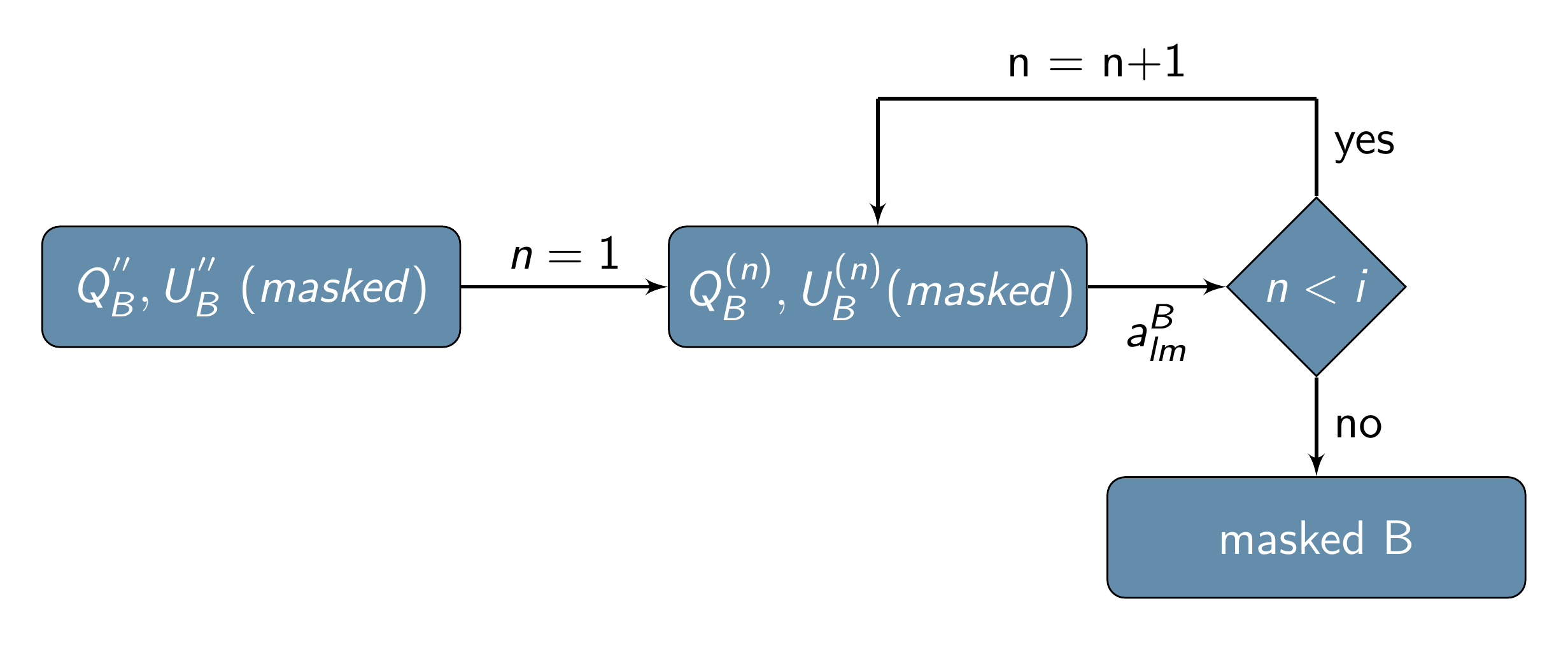}
	\caption{Iterative B-decomposition, extension of the recycling E-B leakage correction method.}
	\label{fig:rec_method_iter}
\end{figure*}
\subsubsection{The recycling method}
\label{sec:leak_liu}
The \emph{recycling method} has been introduced in \citet{2019PhRvD.100b3538L}. The procedure is summarised in Fig. \ref{fig:rec_method}. It consists in
decomposing the masked polarisation field $P=(Q,U)$ into the so-called E- and B-family: $P_{E}=(Q_{E},U_{E})$ and $P_{B}=(Q_{B},U_{B})$, which receive contributions only from $E$- and $B$-mode harmonic coefficients, respectively. For partial-sky observations, $P_{B}$ is largely affected by the presence of ambiguous $E$ modes due to the E-B leakage. To correct for this effect, in the recycling method, $P^{'}_{B}=(Q_{B},U_{B})^{'}$ is constructed through the B-decomposition of masked $P_{E}$ and is then used as a template of the E-B leakage contamination. This template is linearly fitted to $P_{B}$ and removed from it, thus providing a leakage-corrected B-family $P^{''}_{B}=(Q_{B},U_{B})^{''}$. The fit is performed with a simple least-squares linear regression method. We then obtain the final CMB $B$-mode map through a B-decomposition of $P^{''}_{B}$. The leakage correction of the recycling method is not exact, due to our lack of knowledge of the Q and U parameters in the unobserved region of the sky, and therefore some residuals are still present in the reconstructed map. \\ 
The performance of the method on $200$ CMB-only simulations with lensing+r=0.01 for the SWIPE and SO SAT patches is shown in the upper panels of Fig. \ref{fig:leak_liu_swiso}, where the average angular power spectrum of the corrected $B$ maps is compared to that of the exact signal reconstructed with a full-sky B-decomposition of the simulated Q and U maps. As can be seen in the top left panel, if not corrected, the E-B leakage highly contaminates the estimation of the angular power spectrum over most of the multipole range of interest.\\
The leakage correction is performed on the full SWIPE and SO SAT footprints; the angular power spectra are estimated, instead, masking the $4\%$ of pixels closest to the borders, where the residual leakage is expected to still affect the reconstruction of the CMB $B$-mode power spectrum. \\
It is possible to observe that the residuals after the application of the recycling method are negligible for the SO SAT case in the entire multipole range of interest ($30 \lesssim \ell \lesssim 130$), if compared to the upper bound on $r$ targeted by the experiment ($r\lesssim 0.003$ at $1\sigma$, \citealt{2019JCAP...02..056A}). Indeed, the effective tensor-to-scalar ratio associated to the leakage residuals is at the level of $r_{leak} \sim 10^{-4}$ at all angular scales. \\
For SWIPE, the recombination bump can be exactly reconstructed given the sensitivity of the instrument ($\Delta r\lesssim 0.015$ at $95\%$ CL) with $r_{leak} \sim 10^{-4}$, while large angular scales still suffer relevant contamination due to residual leakage. We also observe an increasing trend of $r_{leak}$ for $\ell \geq 100$, which is caused by two main effects: 
\begin{itemize}
    \item the angular resolution of the input CMB maps: on those multipoles, indeed, the instrumental beam begins to have an impact and, therefore, the correction coefficient of the recycling method is less sensitive to the E-B leakage contamination on those smaller angular scales
    \item the CMB lensing $B$-mode signal becomes dominant while the expected primordial tensor spectrum decreases, and even a small relative error in the reconstruction results in a larger value of $r_{leak}$.
\end{itemize}
Given the capability of LSPE to measure primordial $B$ modes on the largest angular scales, it would be desirable to lower the E-B leakage residuals in that multipole range with a further processing of the maps corrected with the recycling method. To this end, we implement two separate extensions:
\begin{itemize}
    \item the iterative B-decomposition
    \item the diffusive inpainting.
\end{itemize}
Both proposed techniques are not applied to the SO SAT data-set, since in this case optimal results in correcting the E-B leakage effect are already obtained with the standard recycling method. \\
\textbf{Iterative B-decomposition} is summarised in Fig. \ref{fig:rec_method_iter} and consists of iteratively reconstructing a B-family, $P^{(n)}_{B}=(Q_{B},U_{B})^{(n)}$, by performing a B-decomposition of the masked $P^{(n-1)}_{B}=(Q_{B},U_{B})^{(n-1)}$. The starting set of Stokes parameters for the iterations, $P^{(1)}_{B}=(Q_{B},U_{B})^{(1)}$, is the leakage-corrected B-family returned by the recycling method, $P^{''}_{B}=(Q_{B},U_{B})^{''}$. With such iterations, the pure $B$ modes in the map are preserved, while the residual ambiguous $E$ modes are progressively de-projected, because interpreted as $E$ in one of the iterations. \\
The results of the iterative B-decomposition are shown in the second row of Fig. \ref{fig:leak_liu_swiso}, where we plot the average angular power spectra of the $200$ CMB-only simulations when either the recycling method alone (iter $0$) or a post-processing with $1,\ 2$ or $3$ iterative B-decompositions  is applied. The plots highlight how, with this additional procedure, we are able to de-project the residual E-B leakage contamination in the $B$-mode CMB maps at most of the angular scales of interest, obtaining, when three iterations are performed, residuals at a much lower level ($r_{leak} \sim 10^{-4}$) than the LSPE targeted sensitivity already for $\ell > 4$. On the other hand, iterative B-decomposition does not lead to an improvement in the reconstruction for $\ell \leq 4$. This does not mean that we have not further removed residual E-B leakage contamination on those scales, but highlights another unavoidable phenomenon of the QU to EB decomposition of partial-sky observations. Both the first (in the standard recycling method) and the following B-decompositions (in the iterations) of the input Q and U maps also suffer from B-E leakage. In each harmonic transformation, some ambiguous modes, which would be identified as $B$ modes in a full-sky analysis, are erroneously estimated as $E$ modes, causing a loss of power in the final $B$-mode map. This effect is particularly relevant on large scales and when E-B leakage residuals are very low. This explains our inability to correctly reconstruct the input power at $\ell \leq 4$ and the larger (but still irrelevant) errors in the map at $\ell \geq 20$ when several iterations are performed with respect to the case with just one. In these cases, the B-E leakage can become the limiting factor in our ability to reconstruct the input CMB $B$-mode signal, possibly leading to an under-estimate of the input tensor-to-scalar ratio. This is particularly meaningful for those experiments (such as LSPE) targeting the observation of the reionisation peak. It is important, then, to determine which multipoles are critically affected by this phenomenon and to exclude them from the cosmological analysis. \\
The effective tensor-to-scalar ratio fitted on the obtained residuals (see the right panels of Fig. \ref{fig:leak_liu_swiso}) highlights that the first three multipoles ($\ell \leq 4$) suffer a significant B-E leakage in the SWIPE case ($r_{leak} \sim 7\cdot 10^{-3}$) and must be excluded. The hexadecupole has been estimated as the largest multipole to reject because including those modes in the second bin of the angular power spectra of the reconstructed CMB maps in Fig. \ref{fig:leak_liu_swiso} leads to a significant under-estimate of its value if compared to the input one. \\
The applicability of iterative B-decomposition may present a drawback if the use of a large number of iterations leads to a considerable power loss due to B-E leakage along with the subtraction of residual ambiguous $E$-mode contamination. To assess whether this is the case, we perform the iterative B-decomposition on two separate masked initial B-families: $P_{B}^{onlyB}$ and $P_{B}^{leak}$. The former is the set of Stokes parameters reconstructed with a full-sky B-decomposition of input Q and U and represents the ideal goal of a leakage correction method. $P_{B}^{leak}$, instead, is composed of Q and U maps sourced only by the residual E-B leakage after the application of the recycling method. In Fig. \ref{fig:leak_onlyB_iters}, we compare the angular power spectra of the output $B$-mode maps from the B-decomposition performed on the two families after several iterations with those of the initial $P_{B}^{onlyB}$ and $P_{B}^{leak}$. It is possible to observe that for $\ell \geq 5$, iterations do not de-project modes of the $B$ maps obtained from the initial $P_{B}^{onlyB}$ and at the same time are able to subtract most of the leakage residuals ($60\%$ of the power already after the first iteration). At $\ell < 5$, instead, the iterative B-decomposition suffers loss of power in $P_{B}^{onlyB}$ due to B-E leakage, but these scales would have been excluded anyway from the cosmological analysis of the NILC solution because an analogous significant loss is observed even without any iteration. Therefore, Fig. \ref{fig:leak_onlyB_iters} highlights the benefit of employing the iterative B-decomposition. \\
Given the previous results, we apply the NILC pipeline on multi-frequency $B$-mode maps reconstructed in the SWIPE patch and processed with the recycling method and three iterations of B-decomposition. Hereafter, we refer to it as \emph{rit-NILC}. \\
\begin{figure}
	\centering
	\includegraphics[width=0.45\textwidth]{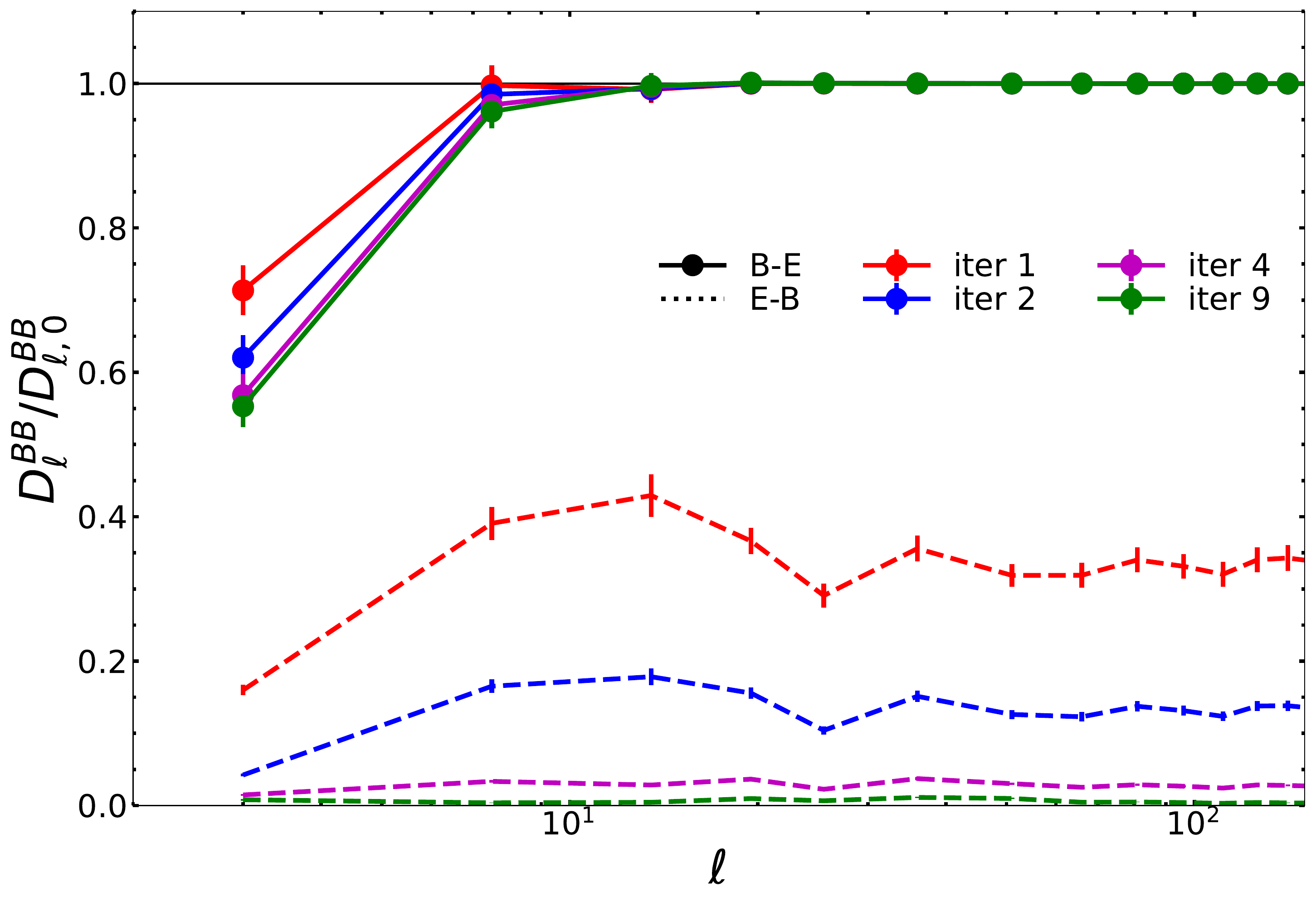}
	\caption{
	LSPE-SWIPE only. Ratio between the mean angular power spectrum of either pure $B$ modes only (solid lines) or E-B leakage residuals (dashed lines) after having applied the iterative B-decomposition correction and that of corresponding input maps. The cases with $1$ (red), $2$ (blue), $4$ (magenta) and $9$ (green) iterations are shown. $200$ different realisations of CMB have been considered, including lensing and a primordial tensor signal with $r=0.01$. The trend of dashed lines highlights the reduction of E-B leakage residuals obtained by performing the iterations. The trend of solid lines, instead, the loss of power suffered due to the B-E leakage when applying the iterative B-decomposition. A binning scheme of $\Delta\ell =3$ for $\ell \leq 4$, $\Delta\ell =6$ for $5 \leq \ell \leq 28$ and $\Delta\ell =15$ for $\ell \geq 29$ is employed. The error-bars highlight the uncertainty on the mean at $1\sigma$, estimated from the dispersion of the angular power spectra of the simulations divided by $\sqrt{N_{sims}}$. See text for details.}
\label{fig:leak_onlyB_iters}
\end{figure}
\begin{figure}
	\centering
	\includegraphics[width=0.45\textwidth]{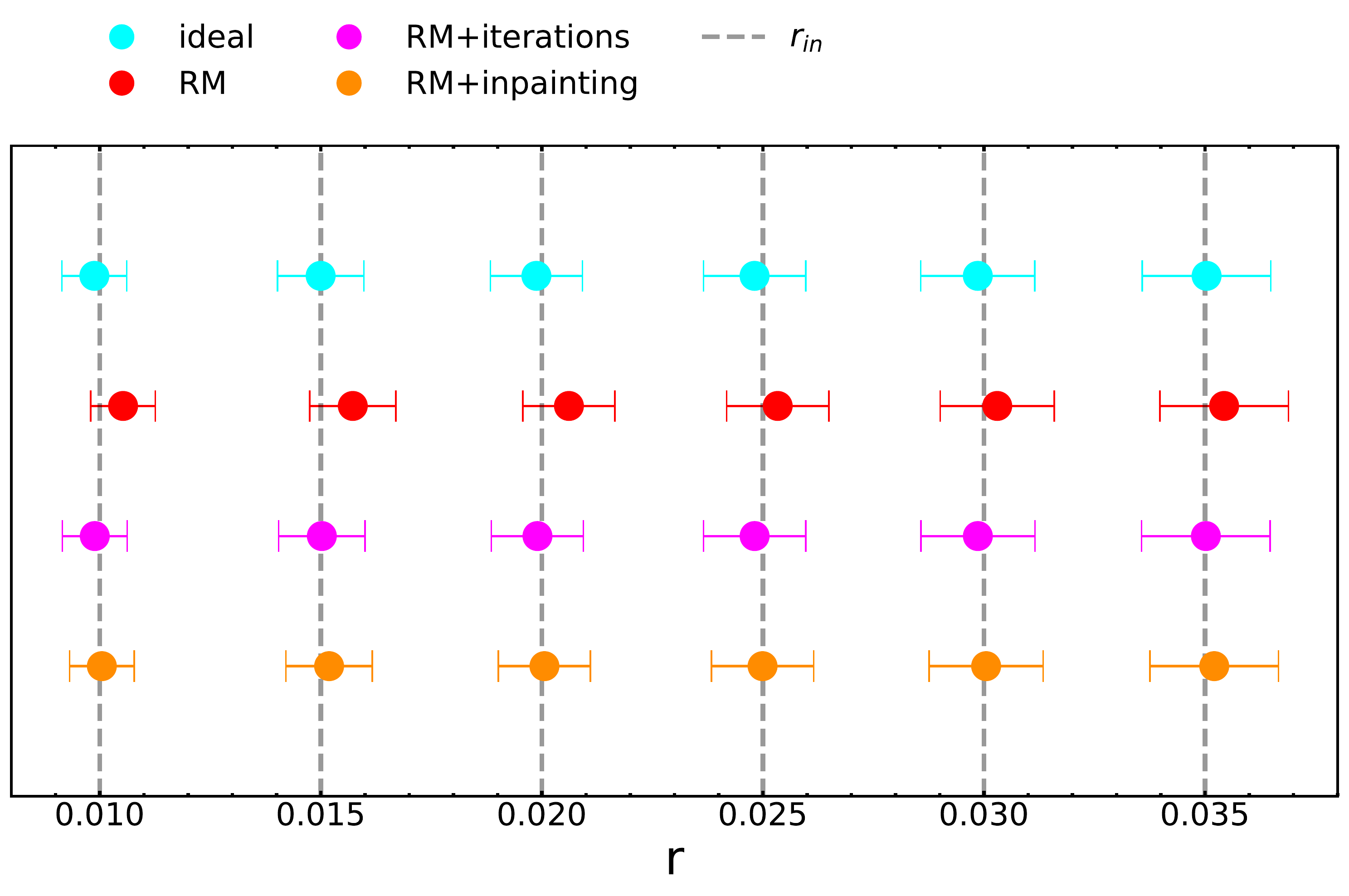}
	\caption{LSPE-SWIPE. Fitted tensor-to-scalar ratio on the mean angular power spectrum over $200$ CMB-only simulations, including lensing and different values of the amplitude of the primordial tensor power spectrum. We consider the cases of $B$-mode maps reconstructed with: the recycling method (RM, red), the iterative recycling method with three iterations (magenta), the inpainting recycling method (orange) and without any E-B leakage contamination (ideal, cyan). 
	The input values of $r$ are represented with vertical dashed grey lines. The used likelihood is shown in Eq. \ref{eq:like_cmb} and the covariance is given only by the input CMB signal (ideal case). A binning scheme of $\Delta\ell =15$ is employed and the first three multipoles ($\ell \leq 4$) are excluded. The error bars indicate the bounds at $1\sigma$ obtained from the posterior distribution.}
\label{fig:leak_iters_diff_rs}
\end{figure}
\textbf{Diffusive inpainting.} The second proposed method to improve leakage removal on large scales for SWIPE is \emph{diffusive inpainting}. This technique has already been introduced in \citet{2019PhRvD.100b3538L} as an alternative to the recycling method. 
It is based on the fact that ambiguous modes $\psi$ satisfy the spherical bi-Laplacian equation:
\begin{equation}
    \nabla^{2}(\nabla^{2} + 2) \psi = 0
\end{equation}
subject to homogeneous Neumann and Dirichlet boundary conditions at the edge of the observed patch. An approximate solution of the equation above can be obtained by replacing the bi-Laplacian with the Laplacian and neglecting the Neumann boundary conditions. In this case, a template of residual ambiguous modes in the recycling CMB $B$-map is given by diffusive inpainting. The procedure consists in imposing as boundary conditions the values of the pixels of the recycling CMB solution at the edges of the patch and replacing iteratively inner sky pixels in the footprint with the average of
their neighbours. The obtained template of the residuals is then subtracted from the input recycling CMB map. \\
The results of the application of the diffusive inpainting on the solutions of the recycling method for $200$ CMB-only simulations with lensing+r=0.01 on the LSPE-SWIPE patch are shown in the bottom panels of Fig. \ref{fig:leak_liu_swiso}. They are compared with those obtained with iterative B-decomposition and three iterations. Even in this case, we are able to significantly reduce the leakage residuals with respect to the standard recycling method for $\ell \geq 5$, while the angular scales at $\ell < 5$ are too affected by the B-E leakage and have to be excluded from the subsequent cosmological analysis of the NILC solution. As for iterative B-decomposition, just including $\ell=4$ in the second bin of the angular power spectra of the reconstructed CMB maps leads to a significant under-estimate of its value with respect to that of input $B$ modes. The comparison of $r_{leak}$ in Fig. \ref{fig:leak_liu_swiso} when diffusive inpainting or iterative B-decomposition is applied highlights that the latter performs better on the largest scales ($5 \leq \ell \leq 20$), while they are equivalent for intermediate and small ones ($\ell > 20$). \\
Given these results, the NILC method is also applied to $B$-maps reconstructed in the SWIPE patch and corrected with the recycling method and the inpainting technique. Hereafter we refer to it as \emph{rin-NILC}. \\
Finally, Fig. \ref{fig:leak_iters_diff_rs} shows that, independently of the amplitude of the input tensor perturbations, iterative B-decomposition with three iterations and diffusive inpainting allow us to perfectly reconstruct the $B$-mode CMB anisotropies for $\ell \geq 5$. We generate CMB simulations with different input amplitudes of tensor perturbations and obtain leakage-corrected $B$-mode maps with the three techniques described above: the standard recycling method, iterative recycling, and inpainting recycling. Then, a tensor-to-scalar ratio is fitted on the average binned angular power spectrum ($C_{\ell_{b}}^{BB}$) of the output CMB $B$-mode maps for all different cases and excluding the first three multipoles ($\ell \le 4$). The fit of the tensor-to-scalar ratio is performed with a Gaussian likelihood \citep{2008PhRvD..77j3013H, 2020FrP.....8...15G}:
\begin{equation}
    -2\log\mathcal{L}(r)=\sum_{\ell_{b},\ell'_{b}}\Big(C_{\ell_{b}}^{BB}-rC_{\ell_{b}}^{r=1}-C_{\ell_{b}}^{lens}\Big)M_{\ell_{b}\ell'_{b}}^{-1}\Big(C_{\ell_{b}}^{BB}-rC_{\ell_{b}}^{r=1}-C_{\ell_{b}}^{lens}\Big),
\label{eq:like_cmb}
\end{equation}
where $C_{\ell_{b}}^{r=1}$ is the binned angular power spectrum of primordial tensor CMB $B$ modes for a tensor-to-scalar ratio $r=1$, $C_{\ell_{b}}^{lens}$ the binned BB lensing angular power spectrum and $M_{\ell_{b}\ell'_{b}}$ the covariance matrix associated to $C_{\ell_{b}}^{BB}$. A binning scheme of $\Delta_{\ell}=15$ is adopted to make the angular power spectrum Gaussianly distributed.\\
We have considered an input $r$ that varies between the sensitivity of LSPE and the current upper bounds reported in \citet{2022PhRvD.105h3524T}. In every case, with both iterative recycling and inpainting recycling methods, we are able to recover the same results obtained in an ideal case without any E-B leakage contamination in the maps. Instead, with the standard recycling method, a bias is observable, especially for lower values of the input tensor-to-scalar ratio, because of the residual E-B leakage contamination on large scales. 

\subsubsection{The ZB method}
\label{sec:leak_zhao}
The second technique we consider for the correction of E-B leakage is the \emph{ZB method} introduced in \citet{2010PhRvD..82b3001Z}. In this case, two different but related definitions for scalar and pseudo-scalar polarisation fields are employed:
\begin{equation}
    \begin{aligned}
    \mathcal{E}(\hat{\gamma}) =& -\frac{1}{2}\left[\Bar{\eth}_{1}\Bar{\eth}_{2}P_{+}(\hat{\gamma}) +\eth_{1}\eth_{2}P_{-}(\hat{\gamma})\right] \\
    \mathcal{B}(\hat{\gamma}) =& -\frac{1}{2i}\left[\Bar{\eth}_{1}\Bar{\eth}_{2}P_{+}(\hat{\gamma}) -\eth_{1}\eth_{2}P_{-}(\hat{\gamma})\right] ,
    \end{aligned}
\label{eq:EB_zhao}
\end{equation}
where $\Bar{\eth}_{s}$ and $\eth_{s}$ are the spin-lowering and raising operators:
\begin{equation}
    \begin{aligned}
        \Bar{\eth}_{s} = -(\sin{\theta})^{-s}\Bigg(\frac{\partial}{\partial\theta}-\frac{i}{\sin{\theta}}\frac{\partial}{\partial\phi}\Bigg)[\sin^{s}\theta] \\
        \eth_{s} = -(\sin{\theta})^{-s}\Bigg(\frac{\partial}{\partial\theta}+\frac{i}{\sin{\theta}}\frac{\partial}{\partial\phi}\Bigg)[\sin^{s}\theta].
    \end{aligned}
\label{eq:d_spin}
\end{equation}
The harmonic coefficients $\mathcal{E}_{\ell m}$ and $\mathcal{B}_{\ell m}$ of $\mathcal{E}$ and $\mathcal{B}$ are related to $E_{\ell m}$ and $B_{\ell m}$ of Eq. \ref{eq:EB_lm} by a $\ell$-dependent numerical factor $N_{\ell}$:
\begin{equation}
    \mathcal{E}_{\ell m}=N_{\ell}E_{\ell m},\quad \mathcal{B}_{\ell m}=N_{\ell}B_{\ell m},\quad N_{\ell}=\sqrt{(\ell+2)!/(\ell-2)!}.
\end{equation}
In the case of partial-sky observations, the spin operators have to be applied on $P_{+}$ and $P_{-}$ fields multiplied by a window function $W(\hat{\gamma})$ (non-zero only in the observed patch of the sky), and the recovered quantities $\Tilde{\mathcal{E}}$ and $\Tilde{\mathcal{B}}$ will be affected by the E-B leakage:
\begin{equation}
    \begin{aligned}
    \Tilde{\mathcal{E}}(\hat{\gamma}) =& -\frac{1}{2}\left[\Bar{\eth}_{1}\Bar{\eth}_{2}(P_{+}(\hat{\gamma})W(\hat{\gamma})) +\eth_{1}\eth_{2}(P_{-}(\hat{\gamma})W(\hat{\gamma}))\right] \\
    \Tilde{\mathcal{B}}(\hat{\gamma}) =& -\frac{1}{2i}\left[\Bar{\eth}_{1}\Bar{\eth}_{2}(P_{+}(\hat{\gamma})W(\hat{\gamma})) -\eth_{1}\eth_{2}(P_{-}(\hat{\gamma})W(\hat{\gamma}))\right] .
    \end{aligned}
\label{eq:EB_zhao_mask}
\end{equation}
In \citet{2010PhRvD..82b3001Z}, the correction term to reconstruct the pure pseudo-scalar field $\mathcal{B}$ in the observed region of the sky has been derived as:
\begin{equation}
    \mathcal{B} = \Tilde{\mathcal{B}}W^{-1} + ct\cdot W^{-2},
\label{eq:EB_zhao_ct}
\end{equation}
where:
\begin{equation}
    \begin{aligned}
    ct =& U[3\cot{\theta}WW_{x} + W(W_{xx}-W_{yy})-2(W_{x}^{2}-W_{y}^{2})] \\
    &- Q[2\cot{\theta}WW_{y} + 2WW_{xy}-4W_{x}W_{y}] \\ 
    &-2W_{y}[(QW)_{x}+(UW)_{y}]+2W_{x}[(UW)_{x}-(QW)_{y}],
    \end{aligned}
\label{eq:zhao_ct_1}
\end{equation}
with $F_{x}=\frac{\partial F}{\partial\theta},\ F_{y}=\frac{\partial F}{\sin{\theta}\partial\phi},\ F_{xx}=\frac{\partial^{2} F}{\partial\theta^{2}},\  F_{yy}=\frac{\partial F^{2}}{\sin^{2}{\theta}\partial\phi^{2}}$ and $F_{xy}=\frac{\partial F^{2}}{\sin{\theta}\partial\phi\partial\theta}$. \\
\begin{figure*}
	\centering
	\includegraphics[width=0.45\textwidth]{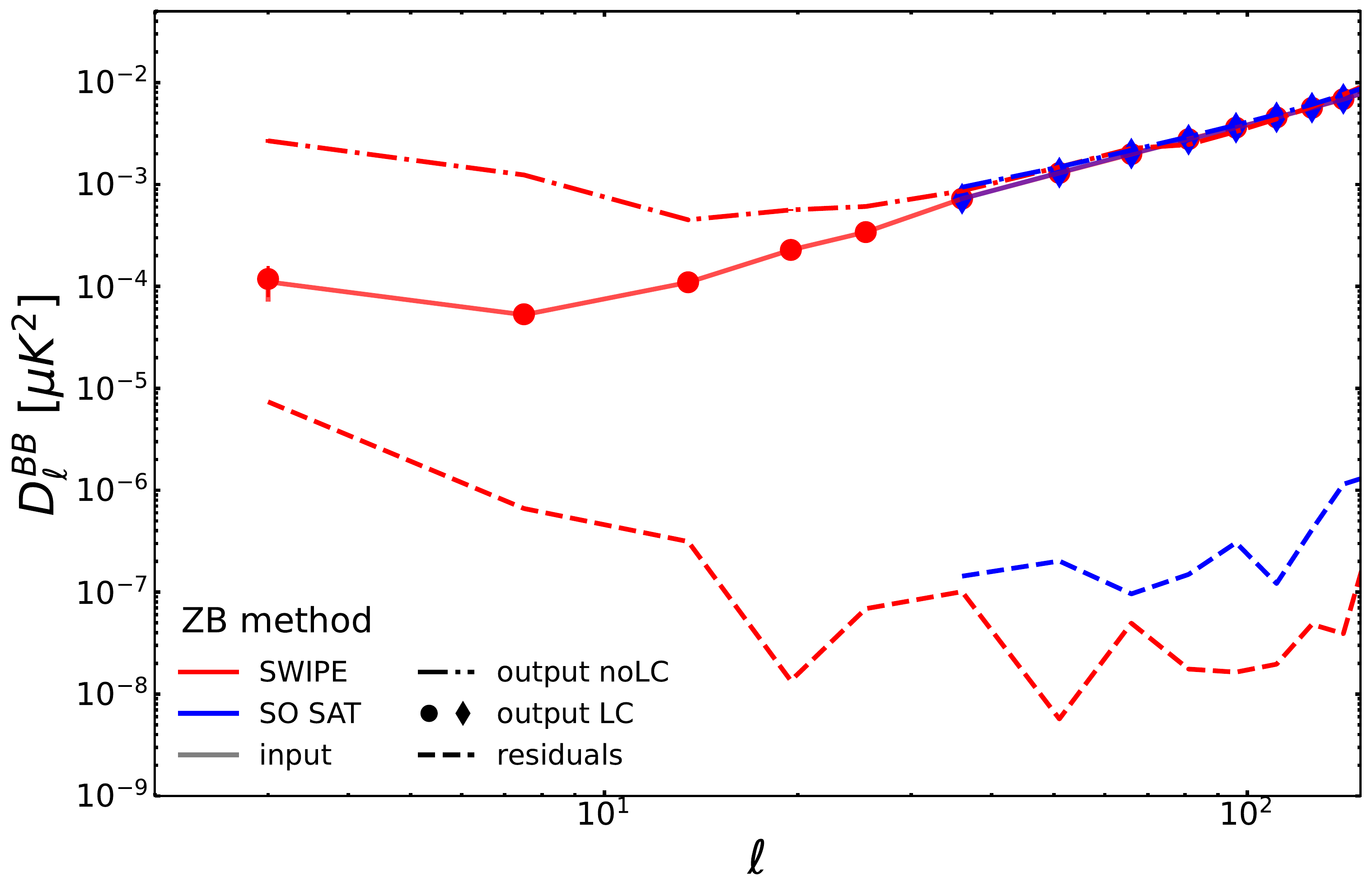} 
	\hspace{0.5 cm}
	\includegraphics[width=0.45\textwidth]{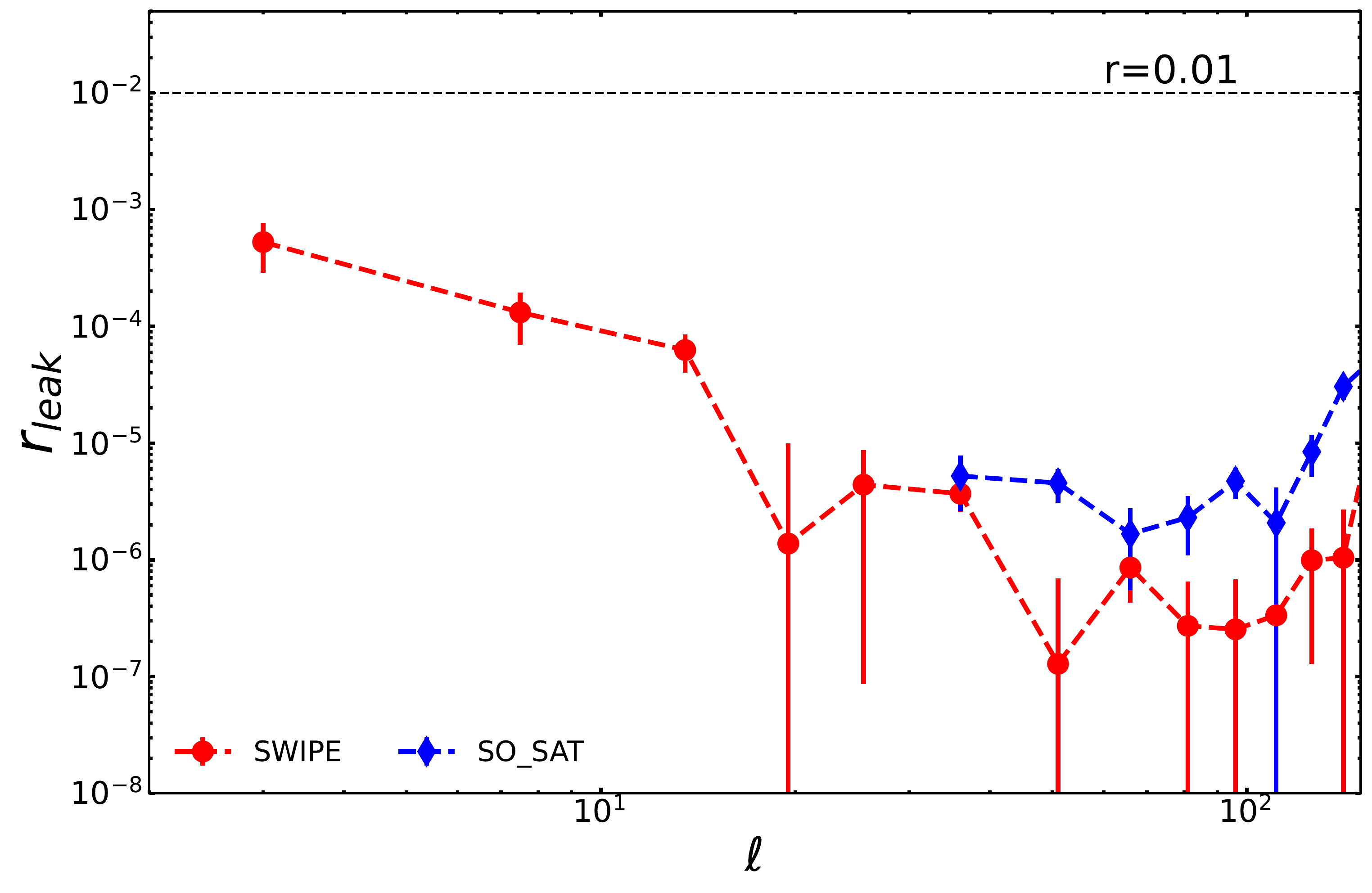}
	\caption{On the left: mean angular power spectrum over 200 CMB-only simulations including lensing and a primordial tensor signal with $r=0.01$. Solid lines represent the power of exact $B$-maps reconstructed with full-sky EB-decomposition of Q and U; circles and diamonds that of leakage-corrected maps (output LC) with the \emph{ZB method}; the dashed lines the leakage residuals after the correction; the dashed-dotted lines the CMB $B$ modes without any leakage correction (output noLC). The LSPE-SWIPE (red) and SO SAT (blue) footprints have been considered. On the right: the effective tensor-to-scalar ratio associated to the absolute difference between leakage-corrected and exact angular power spectra is shown for SWIPE (red) and SO SAT (blue). A binning scheme of $\Delta\ell =3$ for $\ell \leq 4$, $\Delta\ell =6$ for $5 \leq \ell \leq 28$ and $\Delta\ell =15$ for $\ell \geq 29$ is employed for the spectra computed in the LSPE-SWIPE patch, while a constant $\Delta\ell =15$ for those estimated in the SO footprint. The error-bars highlight the uncertainty on the mean at $1\sigma$. See text for details.}
	\label{fig:leak_corr_zhao}
\end{figure*}
In this framework, as window functions, we adopt the footprints of SWIPE and SO (shown in Fig. \ref{fig:patches}) apodised with the "$C^{1}$" scheme, where any observed pixel is multiplied by a weighting function $f$:
\begin{equation}
    f=\begin{cases} x-\sin(2\pi x)/(2\pi) & x<1\\ 1 & {\rm otherwise} \end{cases}
\label{eq:mask_C1}
\end{equation}
with $x=\sqrt{(1-\cos\theta)/(1-\cos(\theta_*))}$, being $\theta_*$ the apodisation scale and $\theta$ the angular separation between the pixel itself and the closest unobserved one. Such an apodisation scheme ensures that the total uncertainty of the recovered power spectra is (nearly) minimal (see \citealt{2009PhRvD..79l3515G}). In our analysis, we adopt an apodisation length of $5\degree$ for both SWIPE and SO. In the ZB method, masks have to be apodised because spinorial derivatives of the polarisation field are badly estimated close to the borders of the patch if one adopts a binary mask. \\ 
The reconstruction of the pure $\mathcal{B}$ field with this procedure is exact. However, the ill-behaved nature of $W^{-1}$ and $W^{-2}$ may complicate it, especially in the proximity of the unobserved region. Therefore, in this analysis, we have excluded all pixels where $W \leq 0.01$ with a loss of sky fraction that corresponds to almost $1\%$ for both footprints. \\
In practise, $\Tilde{\mathcal{E}}$ and $\Tilde{\mathcal{B}}$ are obtained by computing $E_{\ell m}$ and $B_{\ell m}$ of Eq. \ref{eq:EB_lm} with an harmonic decomposition of masked Q and U maps and then filtering them with $N_{\ell}$. \\
Fig. \ref{fig:leak_corr_zhao} compares the mean angular power spectrum of $200$ leakage-corrected CMB-only simulations with lensing+$r=0.01$ for the SWIPE and SO patches with that of the exact signal reconstructed with a full-sky E-B decomposition of the simulated Q and U maps. In these cases, all angular power spectra have been computed employing masks obtained from the exclusion of pixels with values lower than $0.01$ in the original apodised patches and apodising them with a $7\degree$ apodisation scale, leading to final sky fractions of $f_{sky}=32\%$ and $29\%$, respectively, for SWIPE and SO. In both footprints, the power of the residuals is lower across the entire multipole range of interest than the sensitivities targeted by the experiments, with associated values of the effective tensor-to-scalar ratio: $10^{-7} \lesssim r_{leak}\lesssim 10^{-3}$ (see the right panel in Fig. \ref{fig:leak_corr_zhao}). With the ZB method we are even able to confidently reconstruct the primordial tensor $B$-mode power spectrum at $\ell < 5$. This is motivated by the fact that, in this case, in contrast to the application of the recycling method, the correction term in Eq. \ref{eq:zhao_ct_1} is exact and we can employ apodised masks which reduce both the E-B and the B-E leakage effects. Thus, these large angular scales are included in the cosmological analysis of the NILC CMB solutions when the ZB method is adopted. \\
\begin{figure*}
	\centering
	\hspace{-0.6 cm}
	\includegraphics[width=0.45\textwidth]{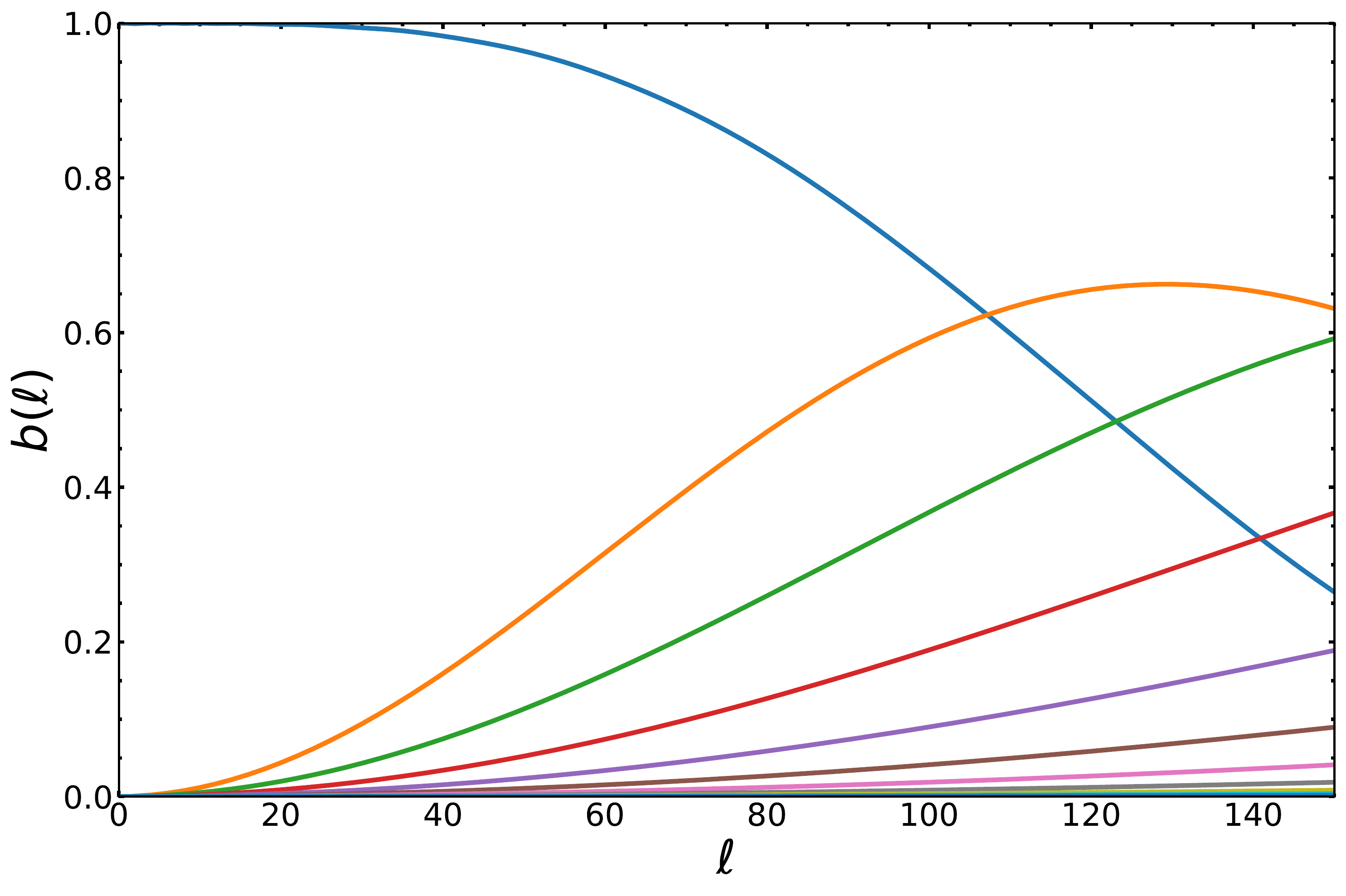} 
	\hspace{0.5 cm}
	\includegraphics[width=0.45\textwidth]{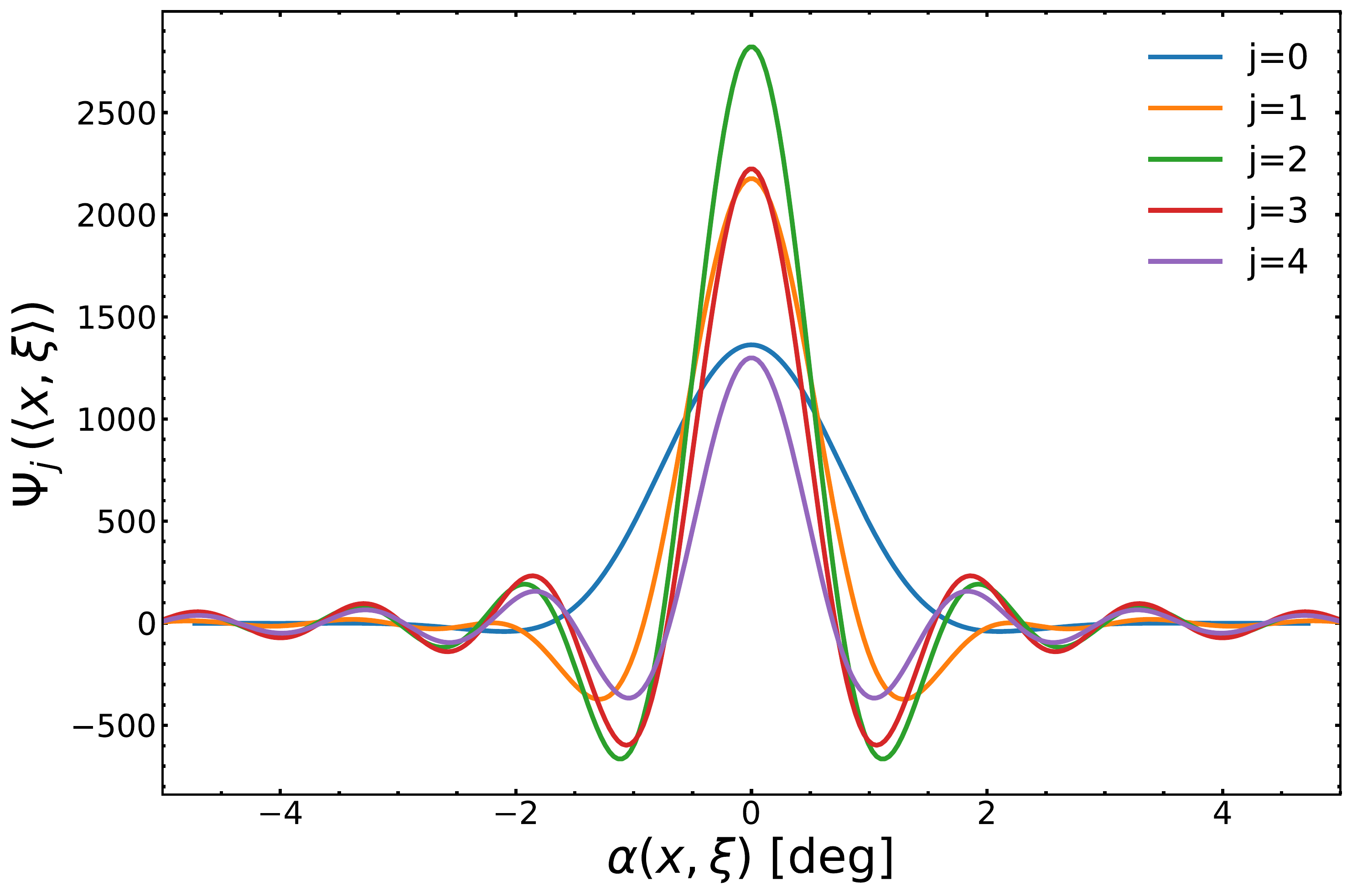}
	\caption{Left panel: mexican needlets bands with $B=1.5$ in the harmonic domain. Right: corresponding needlet filters in real space of the first five bands, plotted with the same colours of the associated harmonic bands.}
	\label{fig:bands}
\end{figure*}
\begin{figure*}
	\centering
	\includegraphics[width=0.45\textwidth]{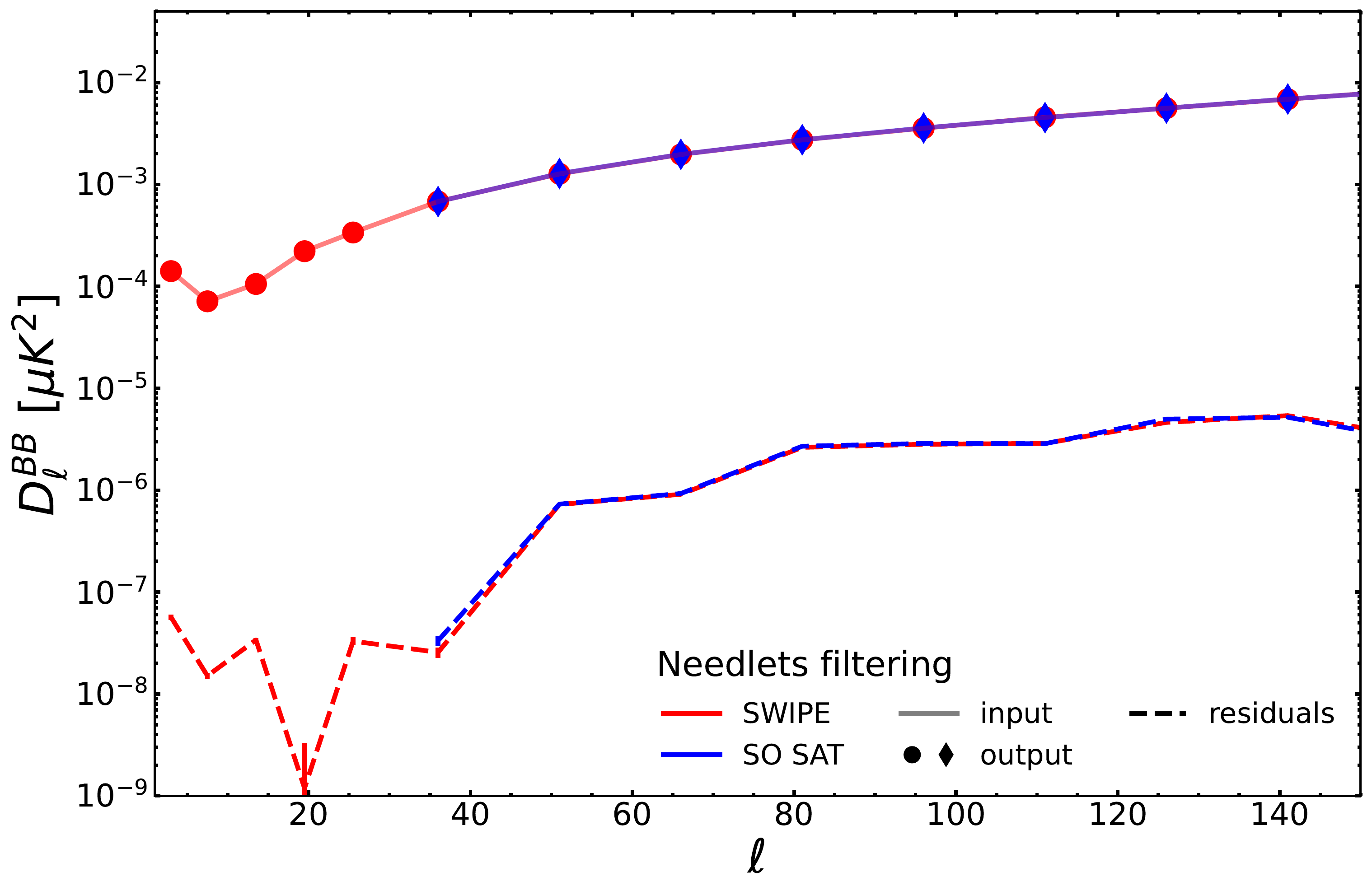}
	\hspace{0.5 cm}
	\includegraphics[width=0.45\textwidth]{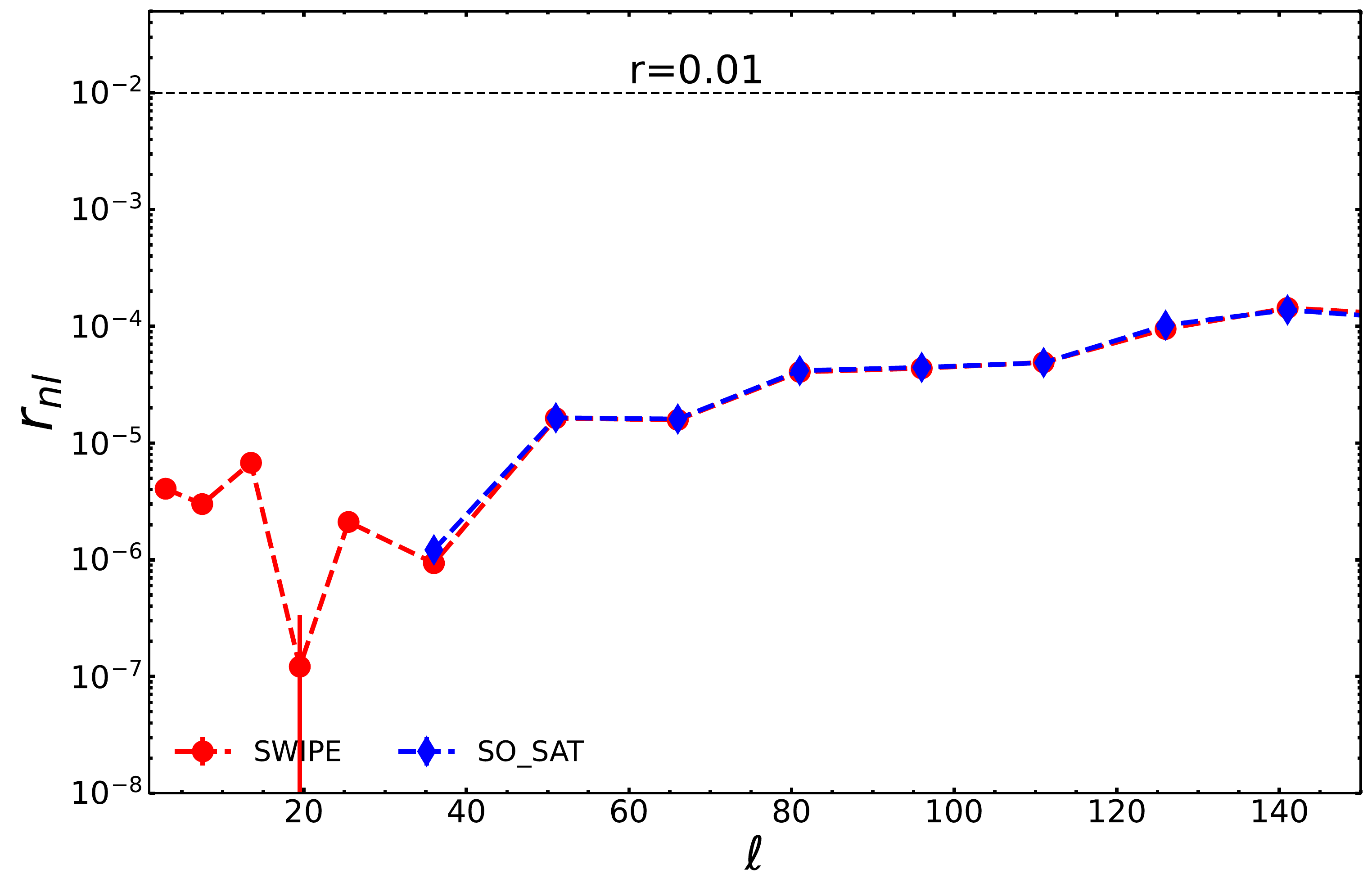}
	\includegraphics[width=0.45\textwidth]{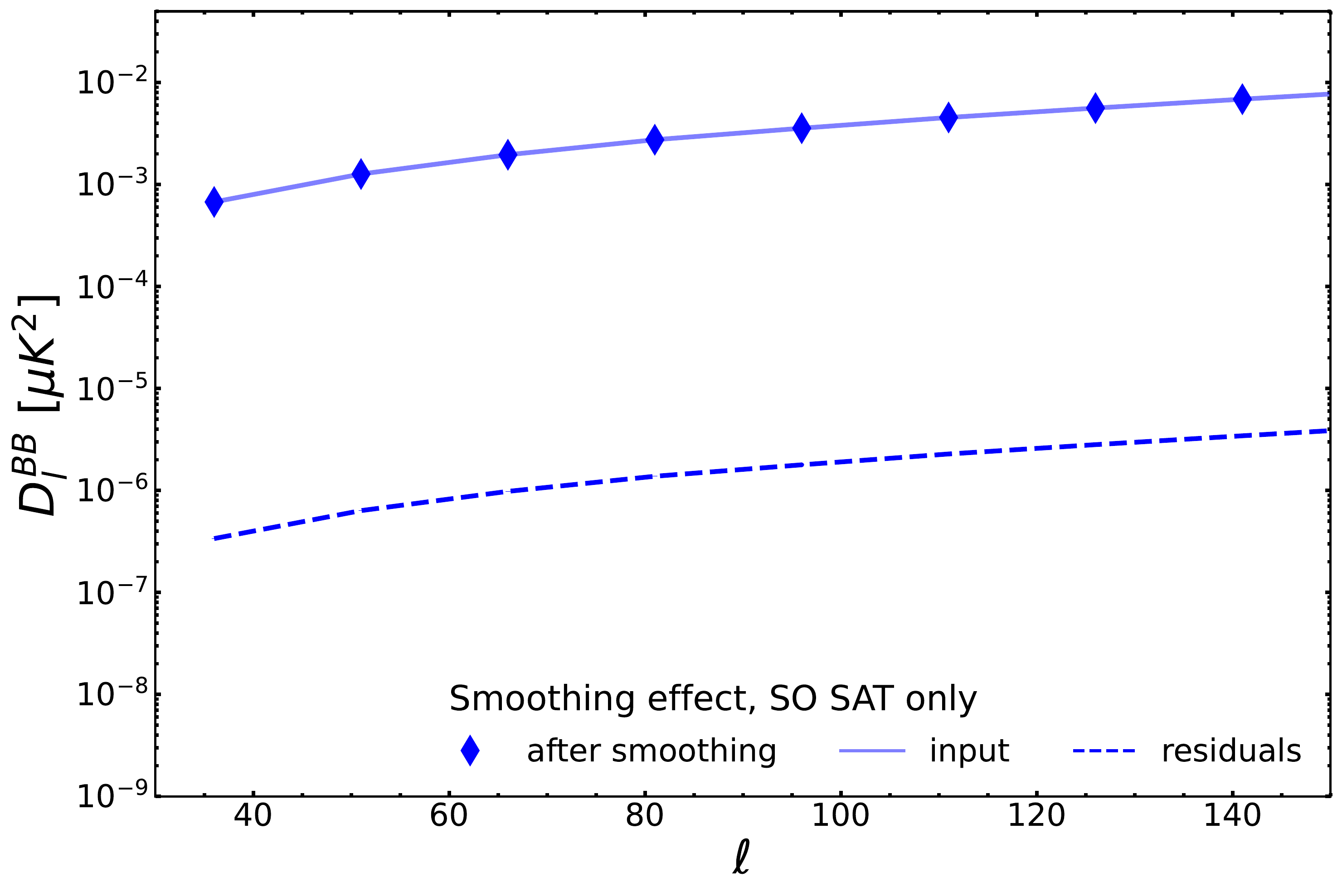}
	\hspace{0.5 cm}
	\includegraphics[width=0.45\textwidth]{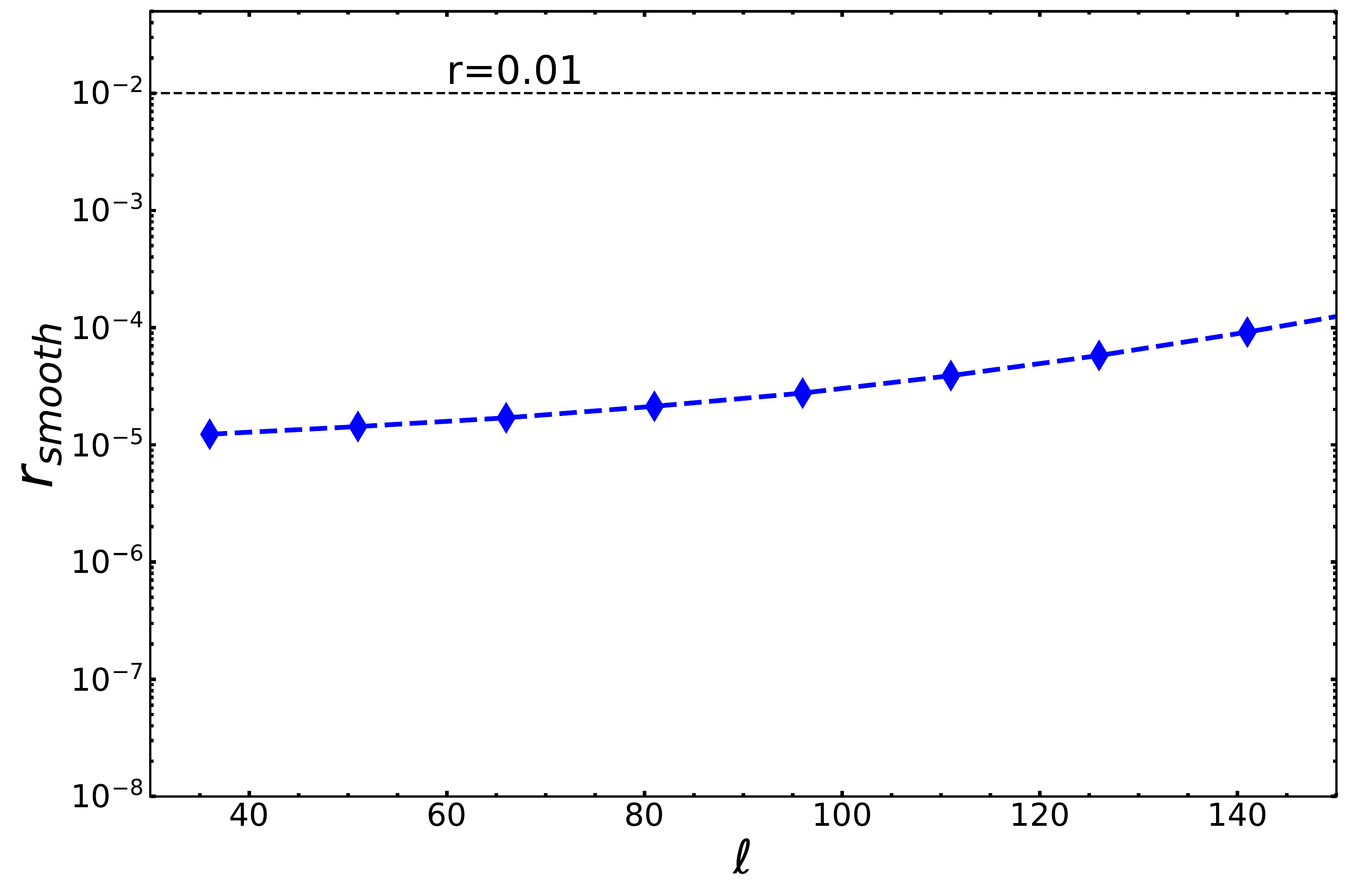}
	\caption{Goodness of the reconstruction of the CMB $B$-mode signal (lensing+$r=0.01$) after needlet filtering (top) in the SWIPE (red) and SO (blue) patches or after smoothing CMB maps with input angular resolution of FWHM$=9'$ to bring them to FWHM=$91'$ in the SO footprint (bottom). On the left: the mean angular power spectra (over 200 CMB-only simulations) of the input exact $B$-maps reconstructed with a full-sky EB-decomposition of Q and U (solid lines); those of the reconstructed maps after needlet filtering or beam convolution (dots and diamonds); the corresponding residuals (dashed lines). The plot on the right represents the effective tensor-to-scalar ratios associated to the absolute differences between output power spectra after needlet filtering and beam convolution and exact input ones in the SWIPE (red) and SO (blue) patches. A binning scheme of $\Delta\ell =3$ for $\ell \leq 4$, $\Delta\ell =6$ for $5 \leq \ell \leq 28$ and $\Delta\ell =15$ for $\ell \geq 29$ is employed for the spectra computed in the LSPE-SWIPE patch, while a constant $\Delta\ell =15$ for those estimated in the SO footprint. The error-bars highlight the uncertainty on the mean of the simulations at $1\sigma$.}
	\label{fig:leak_nl_corr_1}
\end{figure*}

\subsection{Needlet filtering of partial-sky maps}
\label{sec:bands}
The second step in the pre-processing of $B$-mode data which could introduce distortions in the CMB reconstruction before applying a NILC minimisation is the needlet filtering of partial-sky observations. \\
Needlet coefficients, indeed, are computed by convolving a map with a filter in pixel-space. If observations only cover a partial fraction of the sky, the signal in the pixels close to the borders of the observed patch is mixed with the null values of the unobserved region by the convolution procedure. Thus, the estimation of needlet coefficients can be highly affected by this operation. The wider the needlet band is in pixel space, the larger this effect will be. \\
To assess the impact of needlet filtering, we generate $200$ CMB $B$-mode maps from a full-sky E-B decomposition of Q and U, simulated with Planck 2018 best-fit parameters and lensing+r=0.01. Then, we apply a forward and an inverse needlet transformation to these $B$-mode maps, properly masked to recover the SWIPE and SO footprints. We assess the goodness of the reconstruction by comparing the angular power spectra of maps obtained after a forward and inverse needlet transform, $D_{\ell,nl}^{BB}$, and those of the input ones, $D_{\ell,in}^{BB}$. \\
In analogy to the analysis of E-B leakage correction, we consider an effective tensor-to-scalar ratio $r_{nl}$:
\begin{equation*}
    r_{nl} = \left | D_{\ell,in}^{BB}-D_{\ell,nl}^{BB}\right |/D_{\ell,r=1}^{BB},
\end{equation*} 
which quantifies the absolute error we can make in the partial-sky reconstruction of both the tensor modes and the lensing signal due to the needlet filtering. \\
In this analysis and in the rest of the paper, we adopt mexican needlets \citep{2008arXiv0811.4440G}, which are characterised by the following Gaussian-related weighting function:
\begin{equation}
b_{m}\Bigg(\frac{\ell}{B^{j}}\Bigg)=\Bigg(\frac{\ell}{B^{j}}\Bigg)^{p}\cdot exp\Bigg(-\frac{1}{2}\cdot\Bigg(\frac{\ell}{B^{j}}\Bigg)^{2}\Bigg).
\label{eq:b_mex}
\end{equation}
We have set $B=1.5$ and $p=1$. The localisation properties of these wavelets in the real domain are better than those of standard and cosine needlets, because their tails are Gaussian, and therefore the convolution procedure is expected to be less affected by border effects. \\
The NILC CMB reconstruction can be distorted, especially on the largest scales, by two phenomena:
\begin{itemize}
\item if the first needlet harmonic band includes only few low multipoles, the localisation of the filter would be very poor and the needlet coefficients estimation would be largely impacted by the null values of the unobserved pixels;
\item if only few modes are sampled by the first needlet band, the empirical covariance matrix of Eq. \ref{eq:NILC_weights} would be highly uncertain and chance correlations between the CMB and other components could lead to the loss of some CMB power in the NILC reconstruction \citep{2009A&A...493..835D}.
\end{itemize}
To avoid these effects, we add together the first $11$ bands of the employed mexican needlets in the following way:
\begin{equation}
	b^{\textit{new}}(\ell) =\sqrt{\sum_{j=j_{min}}^{j_{max}}b_{j}^{2}(\ell)},
\label{eq:b_merge}
\end{equation}
leading to the configuration of needlet filters shown in Fig. \ref{fig:bands}. \\
Top panels of Fig. \ref{fig:leak_nl_corr_1} show that filtering masked $B$-mode maps with such a needlet configuration does not introduce any significant error in the reconstruction of the CMB signal on a partial sky. Indeed, the residuals between the input maps and those obtained after a forward and an inverse needlet transform are at the level of $r_{nl} \lesssim 10^{-4}$ for both SWIPE and SO at all angular scales (see the upper right panel of Fig. \ref{fig:leak_nl_corr_1}). 
\subsection{Smoothing effect}
\label{sec:smooth}
All ILC methods require input frequency maps with a common angular resolution, which is usually that of the channel with the largest beam.
As anticipated in Sect. \ref{sec:maps}, we consider two data-sets in this analysis: LSPE-SWIPE complemented with Planck frequencies (SWIPE+PLANCK) and SO-SAT. \\
For SWIPE+Planck it suffices to bring the full-sky Planck maps to the lower SWIPE angular resolution of FWHM=$85\arcmin$. \\
For SO SAT, instead, where frequency channels have different angular resolutions, such an operation is not trivial. The convolution for a beam of partial-sky maps can lead to an error in the CMB reconstruction, as in the case of needlet filtering, because the values of the pixels close to the border of the patch may be influenced by the null values outside the observed region. \\
To assess the relevance of this effect, we generate $200$ CMB $B$-mode simulated maps with lensing+r=0.01 at the highest angular resolution of SO (FWHM =$9^{\prime}$) and compare the average angular power spectrum, computed in the SO SAT patch, of these maps when they are brought to the lowest resolution of SO (FWHM=$91^{\prime}$) full-sky ($D_{\ell,in}^{BB}$) or after having been masked ($D_{\ell,smooth}^{BB}$). \\
The comparison between these spectra is shown in the bottom panels of Fig. \ref{fig:leak_nl_corr_1} together with an assessment of the residuals through an effective tensor-to-scalar ratio:
\begin{equation*}
    r_{smooth} = \left | D_{\ell,in}^{BB}-D_{\ell,smooth}^{BB}\right |/D_{\ell,r=1}^{BB}.
\end{equation*}
It is possible to observe that the most aggressive convolution, which has to be applied to SO maps, does not considerably affect the $B$-mode CMB reconstruction with values of $r_{smooth}$ of the order of or below $10^{-4}$ at all the angular scales of interest for $B$-mode science. 
\section{NILC performance: analysis setup}
\label{sec:results_ideal}
After having applied a leakage correction method and needlet filtering to input partial-sky $B$-mode multi-frequency maps, the NILC application leads to the best blind estimate of the CMB $B$-mode signal in the considered patch. Such a solution, as shown in Eq. \ref{eq:ILC_new}, is contaminated by residuals of the polarised foregrounds emission, especially close to the Galactic plane. For this reason, when computing the angular power spectrum for a cosmological analysis, the expected most contaminated regions are masked, lowering the effective sky fraction. The masking strategies adopted in this work are described in Sect. \ref{sec:masks}. \\
In this work, where we analyse simulated data, we can precisely determine the expected level of foregrounds and noise residual contamination in the NILC CMB solution by combining the input foregrounds and noise simulations at different frequencies with the NILC weights. Therefore, it is possible to easily assess the impact of these residuals on the estimate of the tensor-to-scalar ratio following the approach detailed in Sect. \ref{sec:like}.\\
In addition to the residual foregrounds contamination, the other main issue of the NILC CMB reconstruction is the loss of modes induced by the empirical correlation of the reconstructed CMB with other components, especially on large scales, which might lead to a negative bias in the estimated power spectrum. However, we have verified that this effect is negligible in our analysis, given the adopted configuration of needlet bands (see Appendix \ref{app:nilc_bias}).

\subsection{Foregrounds masks}
\label{sec:masks}
In order to perform a cosmological analysis of a NILC CMB solution, masking the most foregrounds contaminated regions is usually needed before estimating the angular power spectrum. We have adopted two distinct and alternative methods to generate such masks:
\begin{enumerate}
    \item computing the polarised intensity $P^{2}=Q^{2}+U^{2}$ of simulated foregrounds maps at a CMB frequency channel of the considered data-set, smoothing it with a Gaussian beam of FWHM=$12.5\degree$, thresholding it to obtain masks with decreasing $f_{sky}$;
    \item considering directly the average foreground residuals of the method under study, thresholding this map to obtain binary masks with decreasing $f_{sky}$, smoothing them with a Gaussian beam of FWHM=$8\degree$, finally thresholding the smoothed masks to get binary masks with $f_{sky}$ equal to the original ones.
\end{enumerate}
In both cases, the considered SO SAT patch is shown in Fig. \ref{fig:patch_so}, and corresponds, as already discussed in Sect. \ref{sec:maps}, to the region where the ILC methods are applied. The rationale of the first masking strategy is to have common masks for all the considered methods (NILC, ILC and PILC). In such a way, we can assess and compare their foregrounds residuals in the same patch of the sky. These masks are obtained by assuming to have perfect knowledge of the polarised Galactic emission across the sky at a CMB channel. This is not possible when dealing with real data. However, similar masks can be obtained from foreground templates built by re-scaling with appropriate SEDs the Q and U maps of the considered data-set at the lowest (for synchrotron) and highest (for thermal dust) frequencies. The smoothing procedure is performed on the map to be thresholded in order to attenuate the small-scale fluctuations, which in real data are mainly induced by noise and CMB, and thus to avoid masks with a patchy structure. \\
Instead, the second masking approach is more method-specific. It is based on the assumption to be able in the future to better predict the distribution and intensity of foregrounds residuals in the sky even thanks to future surveys devoted to the studies of Galactic emission at microwave frequencies. This masking strategy is adopted to compute the angular power spectra of NILC CMB solutions and of its residuals when the E-B leakage is corrected with the recycling or ZB method. \\
When the ZB method is applied for the E-B leakage correction, the mask used for the power spectrum computation is apodised. Indeed, when dealing with second spinorial derivatives of a field on the sphere, the use of a binary mask would lead to an ill-behaved estimation of the power spectrum. We adopt the "$C^{1}$" scheme (see Eq. \ref{eq:mask_C1}) with an apodisation length of $12\degree$ and $2\degree$ for SWIPE and SO, respectively. The choice of the apodisation lengths in the two cases is influenced by the minimum multipole ($\ell=2$ for SWIPE+Planck, $\ell=30$ for SO SAT) that we consider for the computation of the angular power spectra. \\
In the case of the NILC application on SWIPE+Planck $B$-mode maps corrected with the recycling method, we exclude the $4\%$ of the pixels closest to the patch border, where some residual leakage could affect the estimate of the power spectra. This step, instead, is not needed for SO SAT, because the reduced patch of Fig. \ref{fig:patch_so} already excludes the borders of the original footprint. \\
In this second masking strategy, the sky fractions for each case have been determined by comparing the level of the foregrounds residuals with the primordial tensor signal targeted by the experiment. 

\subsection{Estimates of the residuals contamination}
\label{sec:like}
When NILC is applied on simulated data-sets, we have access to its foregrounds and noise residuals in the $B$-mode solution. Therefore, it is possible to compare their angular power spectra with a primordial BB power spectrum of tensor modes with $r$ equal to the target of the experiment under consideration. \\
If cross correlations among components are negligible (as proved in the Appendix \ref{app:nilc_bias}), the power spectrum of the cleaned map ($C_{\ell}^{out}$) can be written as follows:
\begin{equation}
    C_{\ell}^{out} = C_{\ell}^{cmb} + C_{\ell}^{fgds} + C_{\ell}^{noi},
\end{equation}
where $C_{\ell}^{fgds}$ and $C_{\ell}^{noi}$ are the angular power spectra of residual foregrounds and noise. The noise bias, $C_{\ell}^{noi}$, can be subtracted at the power spectrum level by estimating its contribution through Monte Carlo simulations or by computing cross-spectra between uncorrelated splits of the data (\textit{e.g.} maps from subsets of detectors or from partial-mission observations). Hence, the main systematic bias to the final angular power spectrum will be given by the residual Galactic contamination. \\
To assess the impact of this residual term on the estimation of $r$ from $C_{\ell}^{\textrm{out}}$, we fit $r$ on binned $C_{\ell_{b}}^{fgds}$ assuming a Gaussian likelihood \citep{2008PhRvD..77j3013H,2020FrP.....8...15G}:
\begin{equation}
    -2\log\mathcal{L}(r)=\sum_{\ell_{b},\ell'_{b}}\Big(C_{\ell_{b}}^{fgds}-rC_{\ell_{b}}^{r=1}\Big)M_{\ell_{b}\ell'_{b}}^{-1}\Big(C_{\ell'_{b}}^{fgds}-rC_{\ell'_{b}}^{r=1}\Big),
\label{eq:like}
\end{equation}
where $C_{\ell_{b}}^{r=1}$ is the primordial tensor CMB $B$-mode angular power spectrum for a tensor-to-scalar ratio $r=1$, and $M_{\ell_{b}\ell'_{b}}$ is the covariance matrix for a fiducial cosmological model with $r=0$. \\
An aggressive binning makes the angular power spectrum gaussianly distributed and mitigates correlations among different modes (induced by the use of a mask). In this work, we adopt a constant binning of $\Delta_{\ell}=15$. We have tested the robustness of the obtained constraints by both varying the binning scheme and even considering an inverse-Wishart likelihood function.\\
The covariance matrix is estimated from the angular power spectra of $200$ NILC simulated solutions as:
\begin{equation}
M_{\ell_{b}\ell'_{b}}=\mathrm{Cov}\Big(C_{\ell_{b}}^{out}-(1-A_{L})\cdot C_{\ell_{b}}^{lens} ,C_{\ell'_{b}}^{out}-(1-A_{L})\cdot C_{\ell_{b}}^{lens} \Big),    
\label{eq:Mllb}
\end{equation}
where $A_{L}$ quantifies our ability to de-lens the output map with $A_{L}=1$ for no de-lensing and $A_{L}=0$ for full de-lensing. Eq. \ref{eq:Mllb} accounts for the cosmic variance associated to the residual lensing signal, the sample variance of the residual foreground and noise angular power spectra and all the cross-terms. In this work, we assume $A_{L}=0.5$ (which is the level that can be achieved with multi-tracers techniques; see, e.g., \citealt{delensing}). \\
\begin{figure*}
	\centering
	\includegraphics[width=0.45\textwidth]{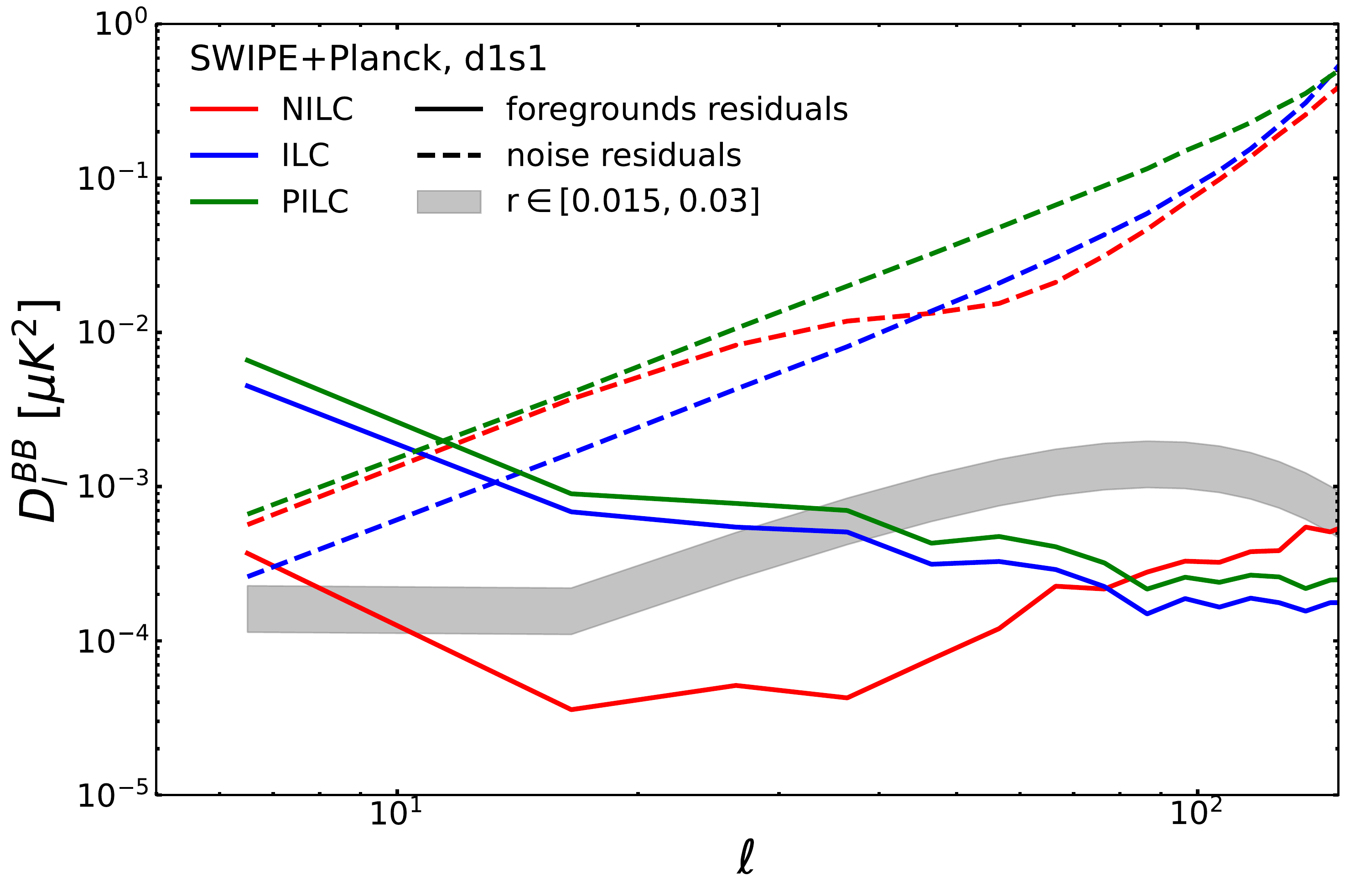} 
	\hspace{0.5 cm}
	\includegraphics[width=0.442\textwidth]{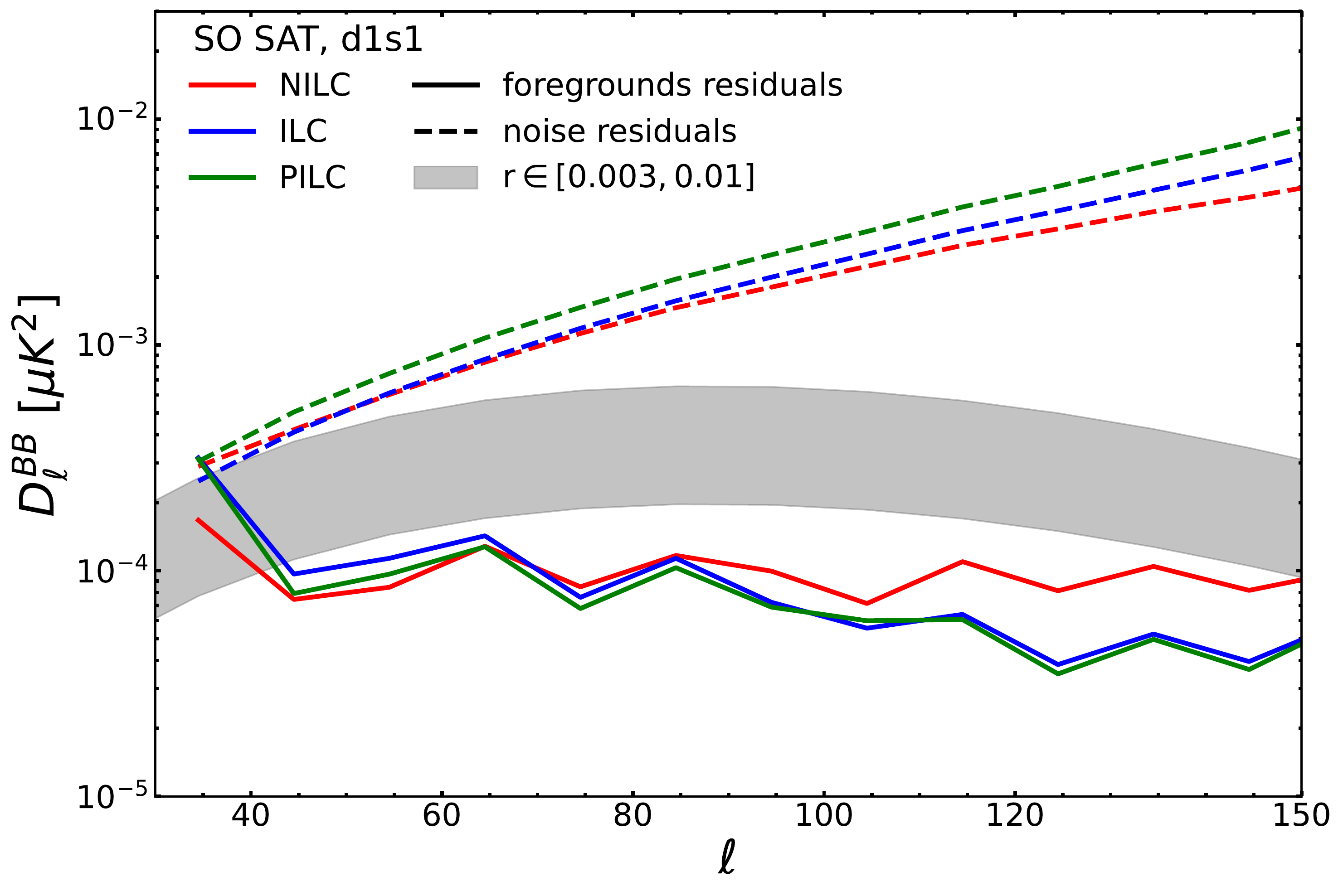} 
	\caption{Mean angular power spectra (over 200 simulations) of foregrounds (solid lines) and noise (dashed lines) residuals when PILC (green), ILC (blue) or NILC (red) are applied on $B$-mode data (reconstructed on the full-sky and then masked) of SWIPE+Planck (left) or Simons Observatory (right) frequency maps. The spectra have been computed considering the $15\%$ (for SWIPE) and $7.5\%$ (for SO) of the sky which is expected to be less contaminated by Galactic emission in the CMB frequency channels of SWIPE and SO. The input foregrounds emission is simulated with the \texttt{d1s1} sky model. A binning scheme of $\Delta\ell =10$ is employed.}
	\label{fig:pilc_ilc}
\end{figure*}
\section{Results}
\label{sec:results}
In the processing of future ground-based and balloon-borne CMB polarisation experiments, NILC represents an effective alternative to the commonly used parametric methods, given the very low number of assumptions on which the algorithm is based. \\
In Sect. \ref{sec:nilc_perf}, we assess the ability of NILC to recover the input CMB $B$-mode signal in an ideal case where we assume to be able to perfectly reconstruct the needlet and pixel coefficients of $B$-mode data even for partial-sky observations. We compare these results with those obtained with the application of ILC and PILC. \\
Once we have verified that it is worth applying NILC on $B$ modes, we proceed to report in Sect. \ref{sec:real_results} the results of the application of NILC to a more realistic case with input maps leakage-corrected with the recycling or ZB method. In this case, also the needlet filtering and the beam convolution are performed on partial-sky maps. \\
In all the following analyses, the input CMB $B$-mode signal has no tensor modes and only lensing. We have verified that the same conclusions hold when injecting a primordial inflationary tensor signal. \\
We consider two different data-sets:
\begin{itemize}
    \item the three channels of SWIPE, complemented with the seven polarised Planck frequency maps (SWIPE+Planck)
    \item the six simulated frequency maps of the Small Aperture Telescope of the Simons Observatory (SO SAT)
\end{itemize}
The elements of the covariance matrix of Eq. \ref{eq:NILC_weights} associated to the Planck channels are computed considering only the observed SWIPE region (shown in Fig. \ref{fig:patches}). For SO, the considered patch is shown in Fig. \ref{fig:patch_so} and represents a faithful approximation of the SO SAT sky coverage under the assumption of homogeneous noise. \\
For the application of NILC, the used configuration of needlet bands in all cases is the same of the test performed in Sect. \ref{sec:bands}: mexican needlets with $B=1.5$ and the first eleven bands added together.

\subsection{Do we need NILC for B-modes?}
\label{sec:nilc_perf}
At the moment, PILC (see Sect. \ref{sec:extensions}) is the only blind method available to easily subtract foregrounds emission from ground-based and balloon-borne CMB polarisation data. However, it works with polarisation intensity data, which account for both $E$ and $B$ modes; therefore, the application of ILC and NILC algorithms could represent a valid alternative, since they are sensitive only to Galactic contamination in $B$ modes. \\
We compare the performance of PILC, ILC and NILC by looking at the angular power spectra of their foregrounds and noise residuals in an ideal case, where we assume to be able to perfectly recover the $B$-mode signal in needlet and pixel space for partial-sky observations. This is explicitly obtained by reconstructing from simulated full-sky Q and U maps full-sky $B$-mode maps, which are then smoothed to a common angular resolution and needlet filtered, and only at the end they are masked according to the patch of the considered experiments. In all cases, a model of foregrounds with spatially varying spectral indices is assumed for polarised dust and synchrotron emissions (\texttt{d1s1}). \\
This ideal case with a full-sky reconstruction of $B$-mode maps is considered just to compare the performance of the different blind methods and to assess the masking strategy needed to reach the scientific targets of the experiments under study. Then, in Sects. \ref{sec:rNILC} and \ref{sec:ZBNILC} we report the results for a realistic approach with a reconstruction of $B$-mode maps from partial-sky Q and U observations. \\
In Fig. \ref{fig:pilc_ilc}, the angular power spectra of foregrounds and noise residuals from the application of PILC, ILC and NILC are shown. They have been computed using Galactic masks retaining the $15\%$ (for SWIPE) and $7.5\%$ (for SO) of the sky and generated with the first procedure described in Sect. \ref{sec:masks}. \\
We start by considering the SWIPE data-set complemented with Planck frequencies (SWIPE+Planck) as introduced in Sect. \ref{sec:maps}. A method which minimises the variance directly of $B$ modes (ILC) instead of polarised intensity (PILC) leads to lower residuals of Galactic emission and instrumental noise at all angular scales. The CMB, indeed, is much brighter in $P=\sqrt{Q^2 + U^2}$ (due to the presence of the $E$-mode signal) than in $B$ and therefore the PILC weights are less capable of tracing and subtracting the frequency-dependent contaminants in the $B$-mode sky. Furthermore, extending the procedure of variance minimisation to the needlet domain (NILC) yields significantly lower
foregrounds residuals, especially on large scales, with respect to ILC and PILC at the price of a mild increase of the noise residuals. This justifies the implementation of the NILC algorithm for cut-sky observations given the sensitivity to the tensor-to-scalar ratio in that multipole range. \\
\begin{figure*}
	\centering
	\includegraphics[width=0.45\textwidth]{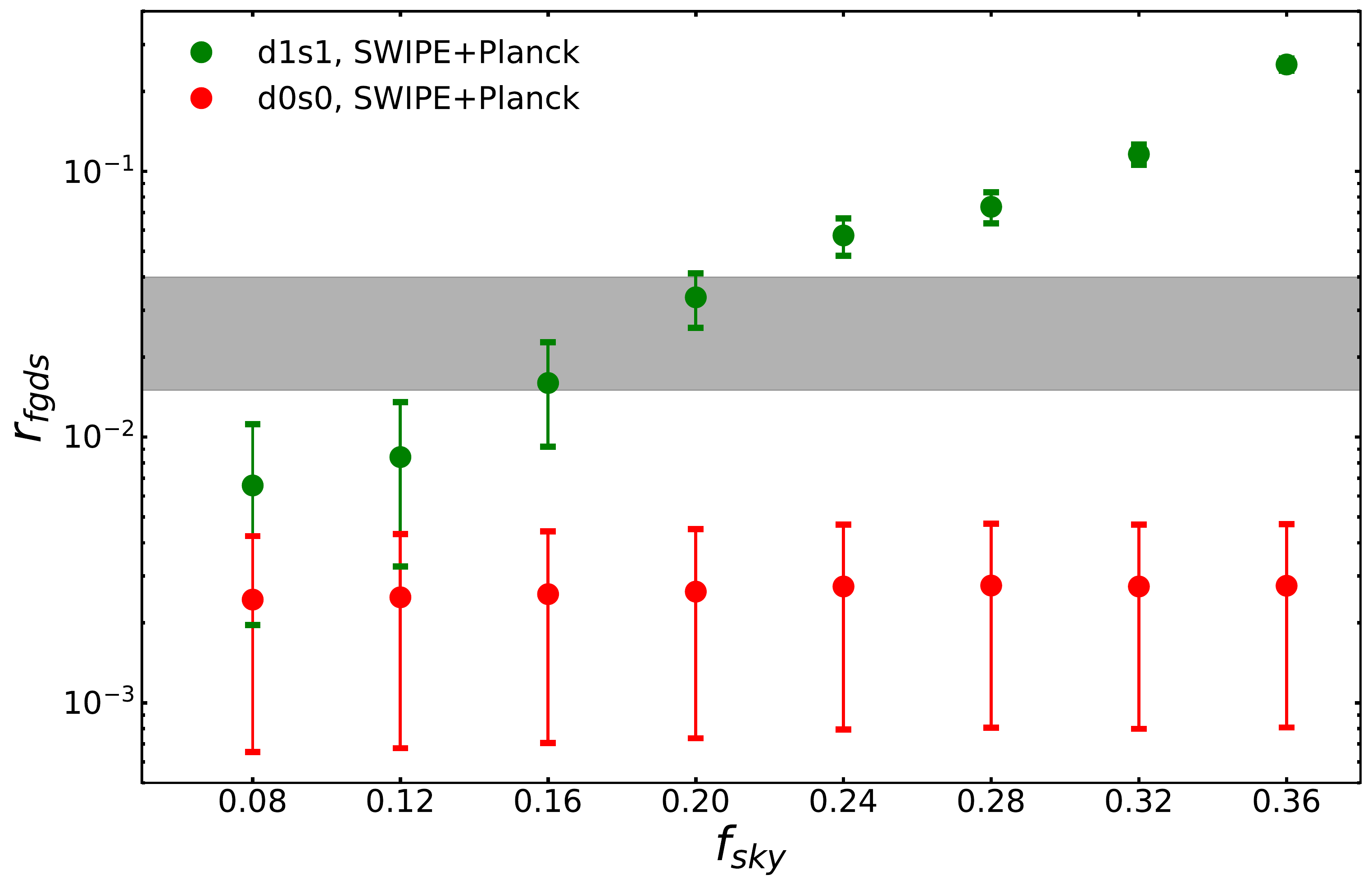} 
	\hspace{0.5 cm}
	\includegraphics[width=0.445\textwidth]{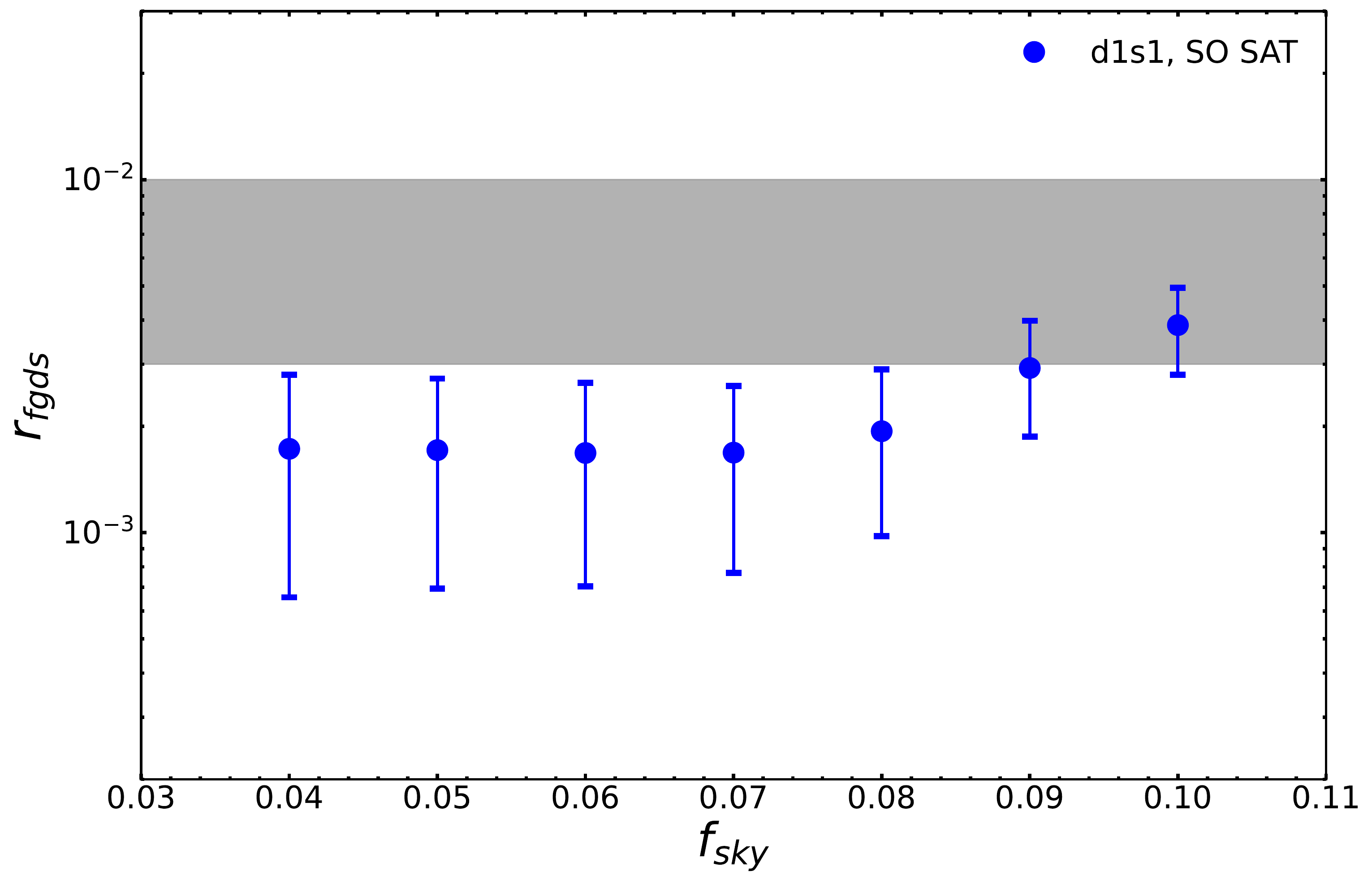} 
	\caption{Trend of the tensor-to-scalar ratio $r_{fgds}$ fitted on the NILC foregrounds residuals angular power spectra with the likelihood of Eq. \ref{eq:like} when the sky fraction of the employed mask varies. The adopted masking strategy is the second one described in Sect. \ref{sec:masks}: we mask the most contaminated regions by foregrounds residuals. In this case, the input multi-frequency $B$-mode maps are constructed full-sky and then masked. The considered cases are SWIPE+Planck (left) with both \texttt{d0s0} (red) and \texttt{d1s1} (green) Galactic models and SO SAT data-set (right) assuming only the \texttt{d1s1} sky model. The grey shaded areas show the targeted sensitivity of the LSPE and SO experiments. The angular power spectra in the likelihood are binned with $\Delta\ell =15$. The error bars indicate the bounds at $2\sigma$ (left) and $1\sigma$ (right) obtained from the posterior distributions.}
	\label{fig:r_fgds_masks}
\end{figure*}
The NILC effectiveness is particularly relevant for SWIPE because we have access to information at very low multipoles. For SO, which will observe only modes at $\ell \geq 30$, indeed, NILC foregrounds residuals are comparable with those of ILC and PILC. Indeed, the range of multipoles where NILC performs better in subtracting Galactic contamination is highly dependent on the sensitivity and angular resolution of the considered data-set. However, as expected, the power of total residuals is lower for NILC than for ILC and PILC across all multipoles, because the output variance is separately minimised at different needlet scales. \\
The above results support the effort to extend the application of NILC on $B$-mode data for ground-based and balloon-borne experiments. 
\subsection{NILC results for realistic SWIPE and SO case-studies}
\label{sec:real_results}
A realistic application of NILC on partial-sky $B$-mode maps can be summarised with the following steps:
\begin{itemize}
    \item Correct the E-B leakage effect in the multi-frequency data-set, which include CMB, noise and Galactic foregrounds
    \item where needed, bring all the frequency maps to the same resolution
    \item filter the maps with needlet bands
    \item perform the variance minimisation independently on the different needlet scales
    \item reconstruct a cleaned CMB map in real space.
\end{itemize}
Depending on the technique employed to correct the E-B leakage, we can identify several distinct pipelines:
\begin{itemize}
    \item[$\bullet$] \emph{r-NILC}, for maps corrected with the recycling method,
    \item[$\bullet$] \emph{rit-NILC}, when the recycling method and iterative B-decomposition are applied,
    \item[$\bullet$] \emph{rin-NILC}, for maps corrected with the recycling method and diffusive inpainting,
    \item[$\bullet$] \emph{ZB-NILC}, when the ZB technique is performed.
\end{itemize}
We have verified that all the leakage-correction methods presented in Sect \ref{sec:leak} perform analogously in the correction of the E-B leakage even in the presence of instrumental noise and Galactic foregrounds. \\
We will report separately the NILC results obtained when recycling (and its extensions) and ZB methods are applied, for both SWIPE+Planck and SO SAT, in Sect. \ref{sec:rNILC} and \ref{sec:ZBNILC}, respectively. Figs. \ref{fig:results_liu_nilc_swi_d0s0}, \ref{fig:results_liu_nilc_swi_d1s1} and
\ref{fig:results_liu_nilc_so_d1s1} provide a detailed comparison of the two approaches. \\
We analyse simulated data of:
\begin{itemize}
    \item SWIPE+Planck with \texttt{d0s0} Galactic model
    \item SWIPE+Planck with \texttt{d1s1} Galactic model
    \item SO SAT with \texttt{d1s1} Galactic model
\end{itemize}
The case with constant spectral indices for polarised thermal dust and synchrotron emission is considered to have a direct comparison with the recent forecast of the sensitivity on $r$ for the LSPE experiment published in \citet{2021JCAP...08..008A}. \\
Considering the results obtained in Sect. \ref{sec:leak} regarding the effectiveness of the different methods in correcting E-B leakage in the two footprints, we apply rit-NILC, rin-NILC, and ZB-NILC to the SWIPE+Planck data-set, while r-NILC and ZB-NILC to SO SAT simulated maps. In the case of the application of rit-NILC and rin-NILC to SWIPE+Planck, the minimum considered multipole is $\ell = 5$ both in the plots and for the cosmological analysis, given the significant loss of power on larger angular scales due to B-E leakage, as found in Sect. \ref{sec:leak_liu}. In the SO SAT analysis, the leakage correction methods are performed within the full SO footprint (shown in the right panel of Fig. \ref{fig:patches}), since CMB data will be available in that region. The NILC pipelines, instead, are applied considering the patch of Fig. \ref{fig:patch_so}, which retains $10\%$ of the sky and takes into account the highly inhomogeneous scanning strategy of the telescope. \\
The reference sky fraction for the cosmological analysis in each case is established by looking at the results in Fig. \ref{fig:r_fgds_masks}. In this analysis, we perform a fit of $r$ on the power spectra of NILC foregrounds residuals (see Sect. \ref{sec:like}) in the ideal case considered in Sect. \ref{sec:nilc_perf} adopting the second masking approach described in Sect. \ref{sec:masks} and varying the sky fraction. \\
It is possible to observe that when NILC is applied on \texttt{d0s0} simulations of the SWIPE+Planck data-set, no masking of the foregrounds residuals is needed to achieve the scientific goal of the mission in terms of sensitivity on the tensor-to-scalar ratio $r$. 
When a more realistic Galactic model (\texttt{d1s1}) is instead considered for the same data-set, an aggressive masking strategy is needed. The computation of the power spectrum in approximately the $12 \%$ less contaminated fraction of the sky allows us to obtain an upper bound at $95\%$ CL on $r$ lower than the sensitivity targeted by the experiment. Considering larger portions of the sky with such a blind method would lead to a bias on $r$ at the level of $1.5 \cdot 10^{-2}$ or greater. When, instead, NILC is applied on ideal simulated $B$-mode data of SO SAT, with a sky fraction of $8\%$, the foregrounds contamination is sensibly lower than the primordial tensor signal targeted by the experiment, even assuming a Galactic foreground model with varying spectral indices across the sky. \\
\begin{figure*}
	\centering
	\includegraphics[width=0.45\textwidth]{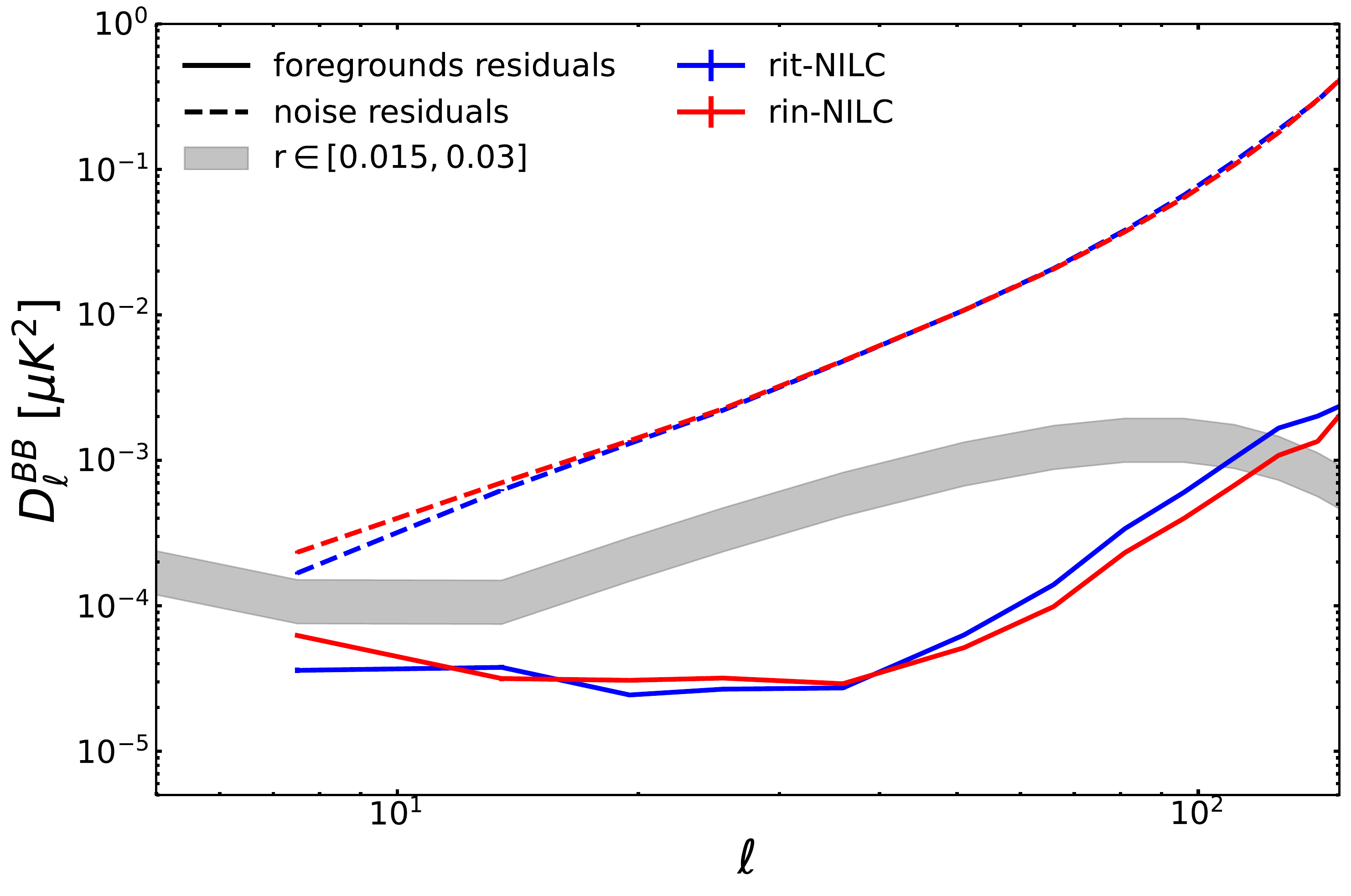}
	\hspace{0.5 cm}
	\includegraphics[width=0.45\textwidth]{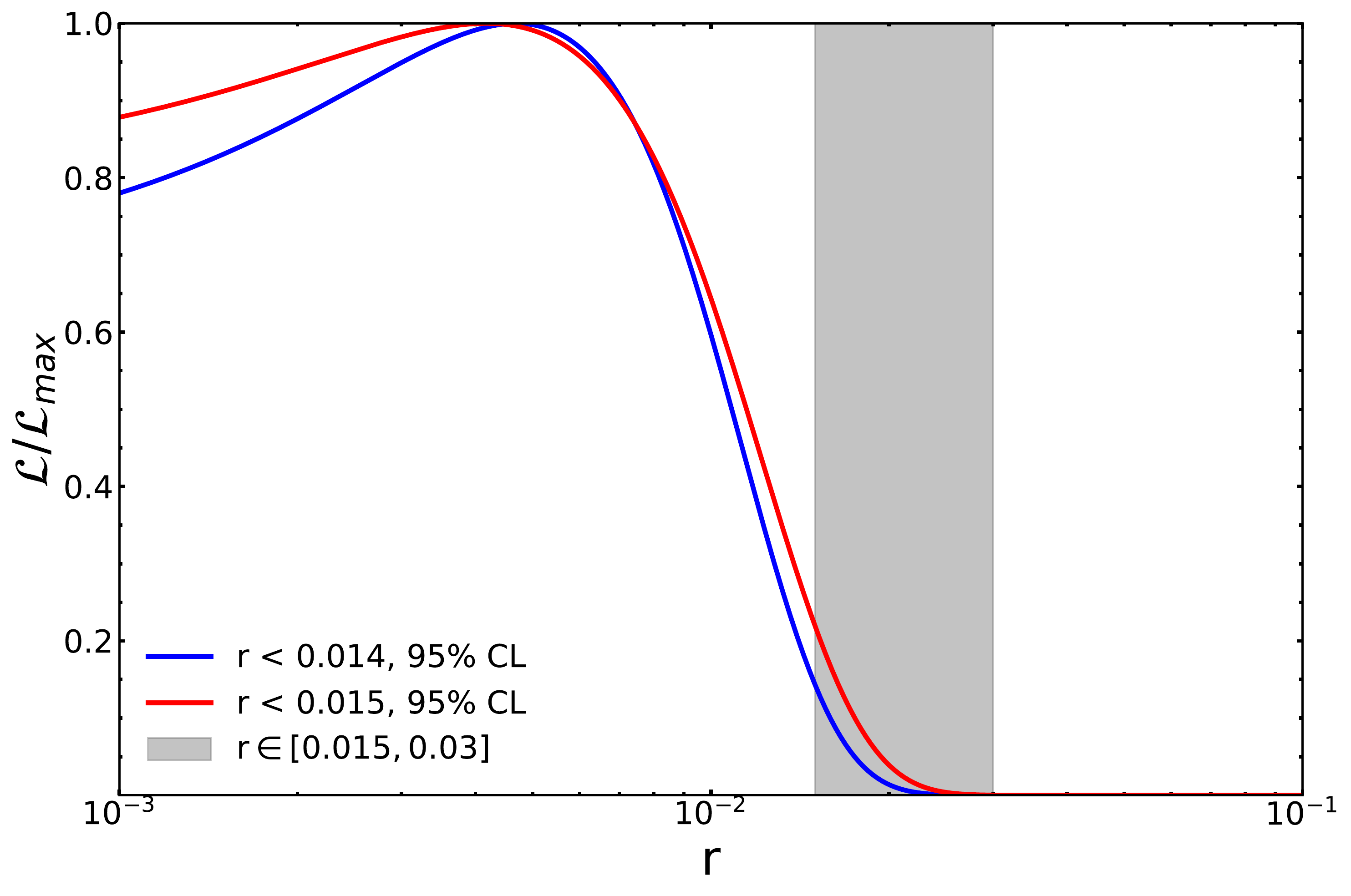}
	\includegraphics[width=0.45\textwidth]{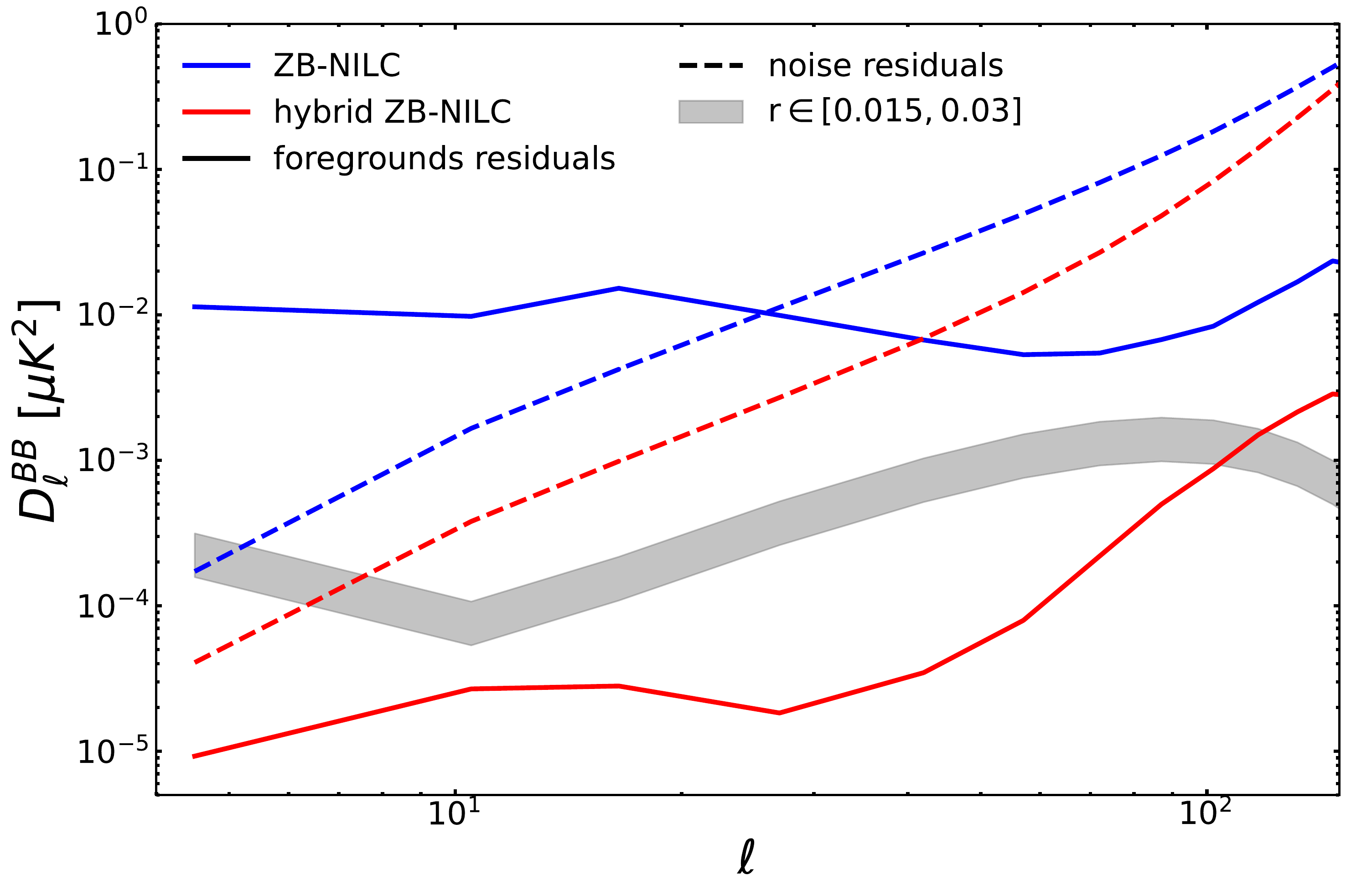}
	\hspace{0.5 cm}
	\includegraphics[width=0.45\textwidth]{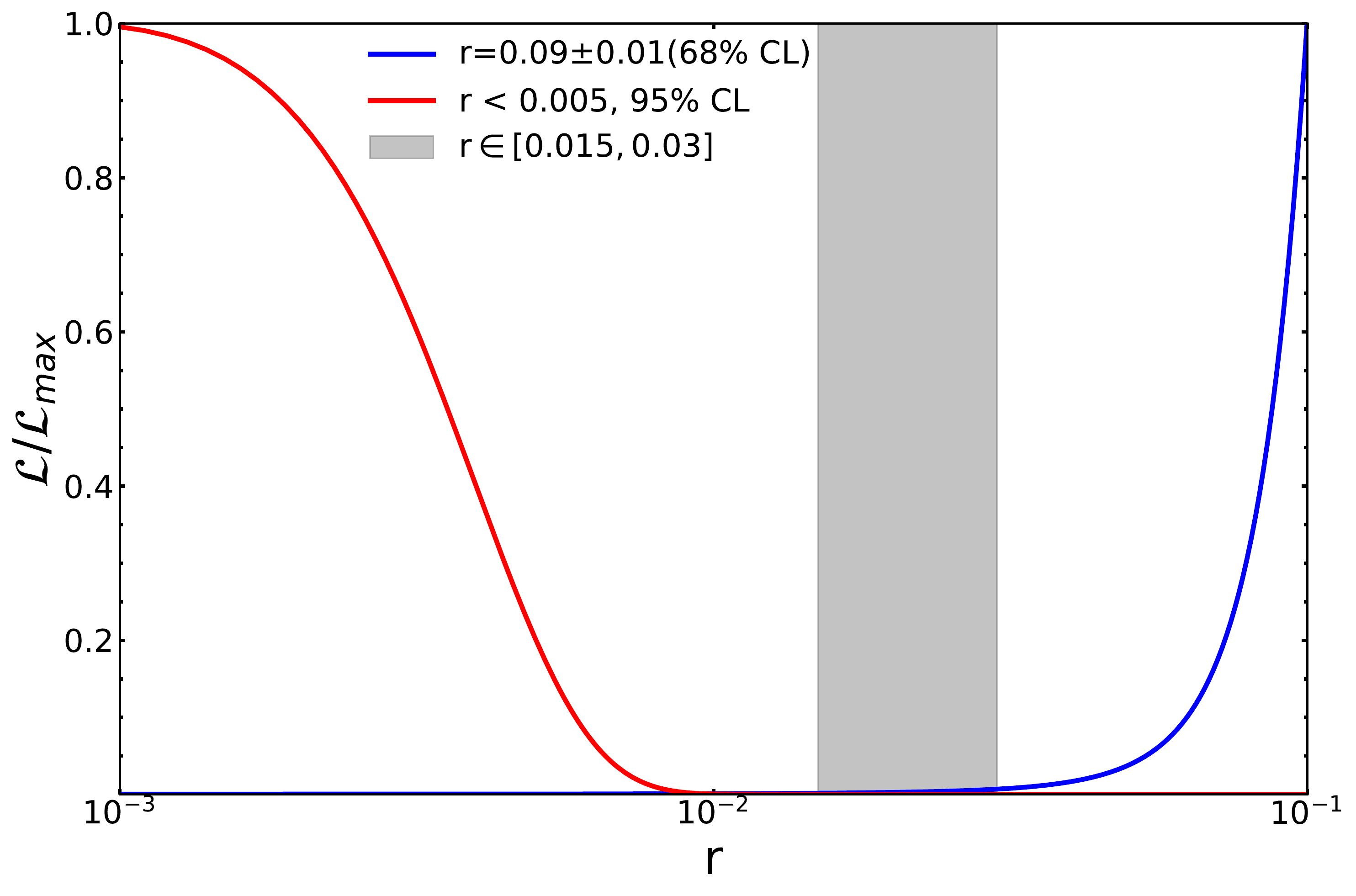}
	\caption{SWIPE+Planck data-set with the \texttt{d0s0} Galactic model. On the left: the mean angular power spectra of NILC foregrounds (solid) and noise (dashed) residuals over $200$ simulations.
	On the right: the posterior distributions of an \textit{effective} tensor-to-scalar ratio fitted on the foregrounds residuals in the case of half de-lensing ($A_{L}=0.5$). For the estimation of the posterior a binning scheme of $\Delta\ell =15$ has been used to make the angular power spectrum gaussianly distributed (see Sect. \ref{sec:like}). 
	On the top: blue and red lines represent the results when E-B leakage is corrected with, respectively, the iterative and inpainting recycling methods (rit-NILC and rin-NILC). On the bottom: results when the ZB- (blue) and hybrid ZB-NILC (red) methods are applied. The adopted binning scheme is $\Delta\ell =6$ for $5 \leq \ell \leq 28$ and $\Delta\ell =15$ for $\ell \geq 29$ on the top left, while $\Delta\ell =6$ for $2 \leq \ell \leq 19$ and $\Delta\ell =15$ for $\ell \geq 20$ on the bottom left. In both cases the chosen binning strategy is just for visualisation purposes. Everywhere the grey region represents the primordial tensor BB angular power spectrum for $r\in [0.015,0.03]$. The lower bound represents the targeted upper limit at $95\%$ CL by LSPE in case of no detection, the upper one the targeted detection at $99.7\%$ CL. The spectra have been estimated without any masking of the regions most contaminated by foregrounds. The final $f_{sky}$ is $32\%$ in the case on top, because $4\%$ of the pixels closest to the borders are masked to avoid residual E-B leakage effects, and $28\%$ in the case on the bottom due to the needed apodisation of the mask.}
	\label{fig:results_liu_nilc_swi_d0s0}
\end{figure*}
\begin{figure*}
	\centering
	\includegraphics[width=0.45\textwidth]{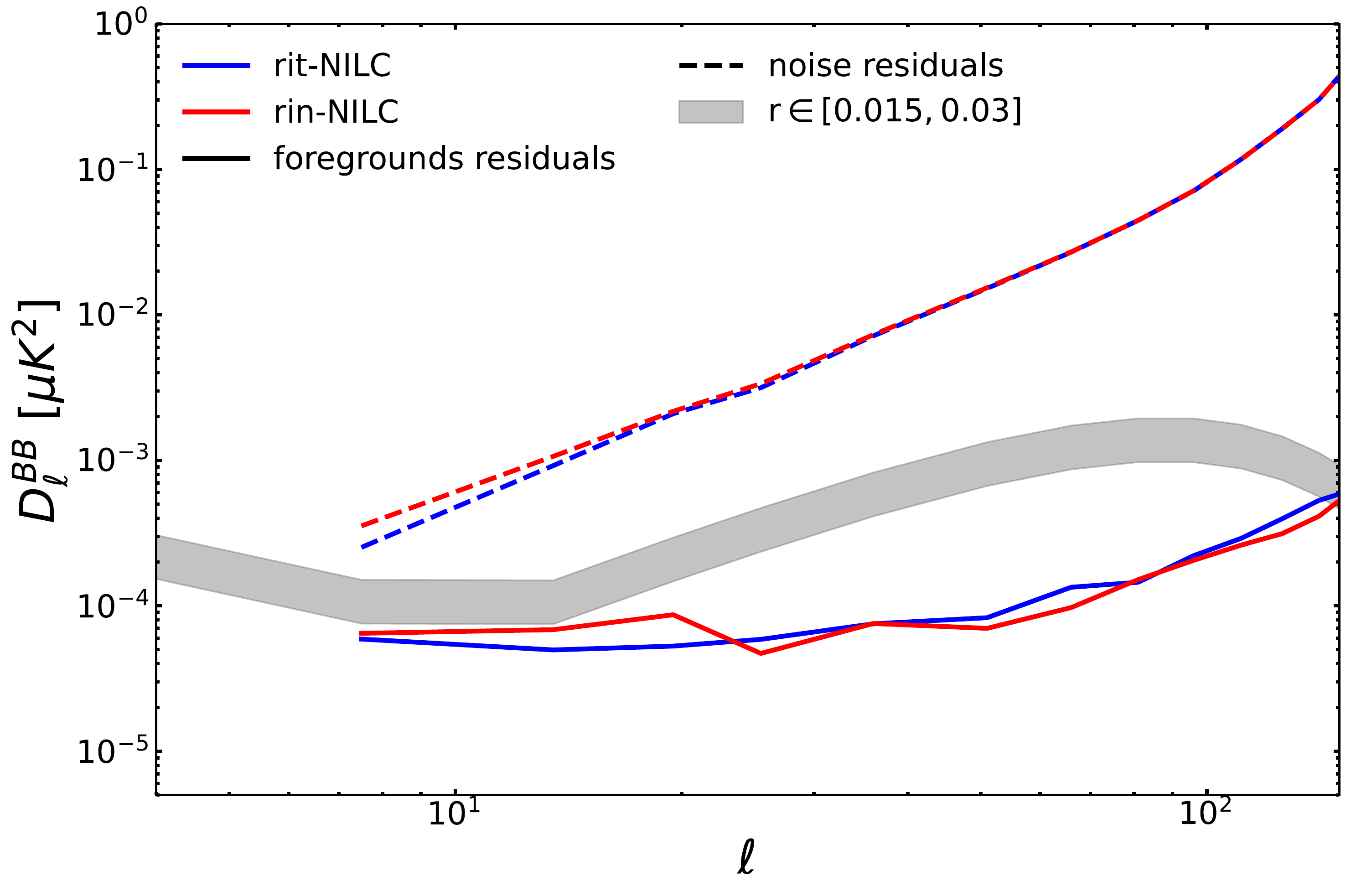}
	\hspace{0.5 cm}
	\includegraphics[width=0.45\textwidth]{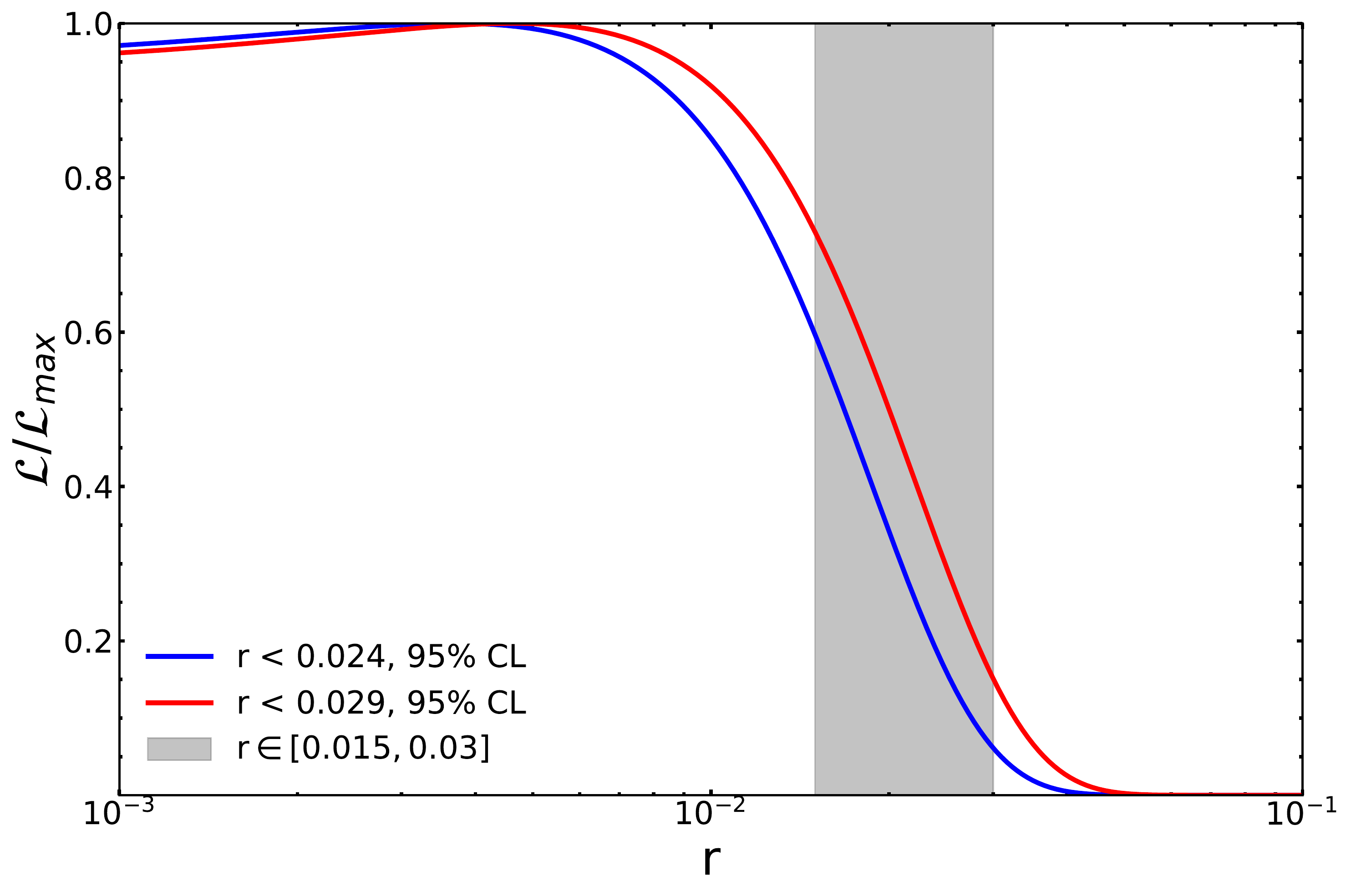}
	\includegraphics[width=0.45\textwidth]{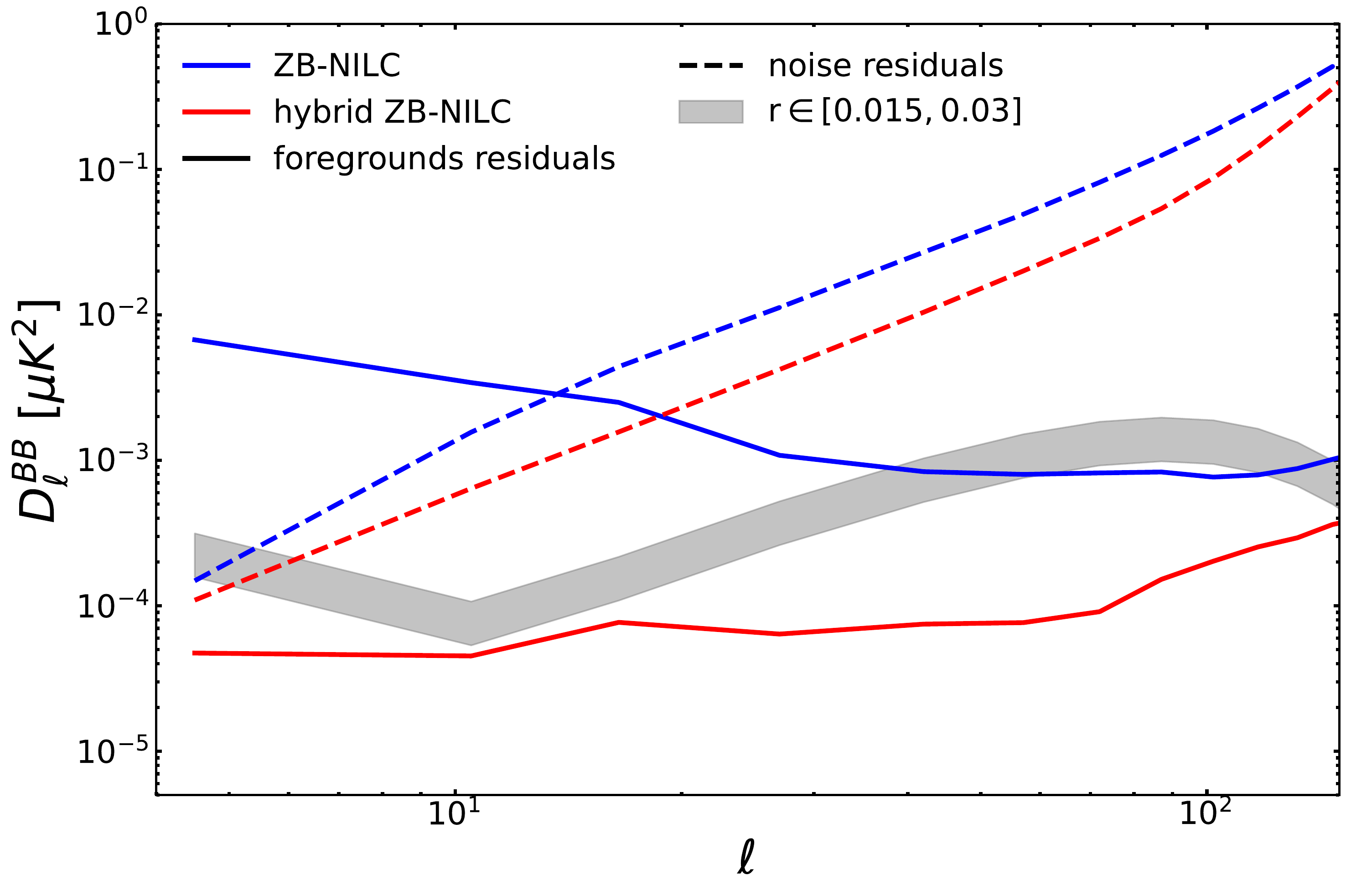}
	\hspace{0.5 cm}
	\includegraphics[width=0.45\textwidth]{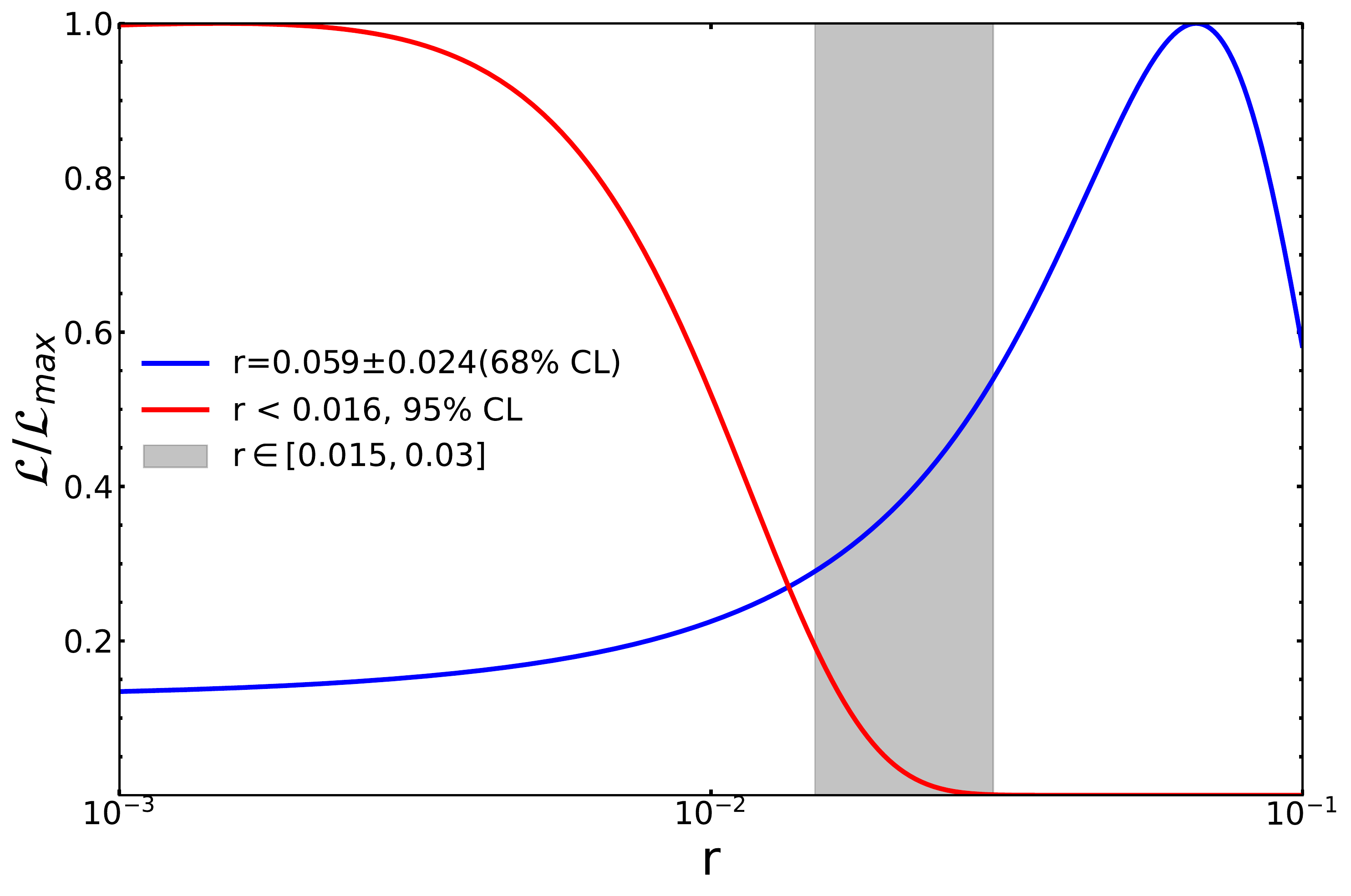}
	\caption{SWIPE+Planck data-set with the \texttt{d1s1} Galactic model. On the left: the mean angular power spectra of NILC foregrounds (solid) and noise (dashed) residuals over $200$ simulations.
	On the right: the posterior distributions of an \textit{effective} tensor-to-scalar ratio fitted to the foregrounds residuals in the case of half de-lensing ($A_{L}=0.5$). For the estimation of the posteriors, a binning scheme of $\Delta\ell =15$ has been used to make the angular power spectrum gaussianly distributed (see \ref{sec:like}). 
	On the top: blue and red lines represent the results when E-B leakage is corrected with, respectively, the iterative and inpainting recycling methods (rit-NILC and rin-NILC). On the bottom: results when the ZB- (blue) and hybrid ZB-NILC (red) methods are applied. The adopted binning scheme is $\Delta\ell =6$ for $5 \leq \ell \leq 28$ and $\Delta\ell =15$ for $\ell \geq 29$ on the top left, while $\Delta\ell =6$ for $2 \leq \ell \leq 19$ and $\Delta\ell =15$ for $\ell \geq 20$ on the bottom left. In both cases the chosen binning strategy is just for visualisation purposes. Everywhere the grey region represents the primordial tensor BB angular power spectrum for $r\in [0.015,0.03]$. The lower bound represents the targeted upper limit at $95\%$ CL by LSPE in case of no detection, the upper one the targeted detection at $99.7\%$ CL. The spectra have been estimated masking the regions most contaminated by foregrounds for a final $f_{sky}$ of $12\%$.}
	\label{fig:results_liu_nilc_swi_d1s1}
\end{figure*}
\begin{figure*}
	\centering
	\includegraphics[width=0.45\textwidth]{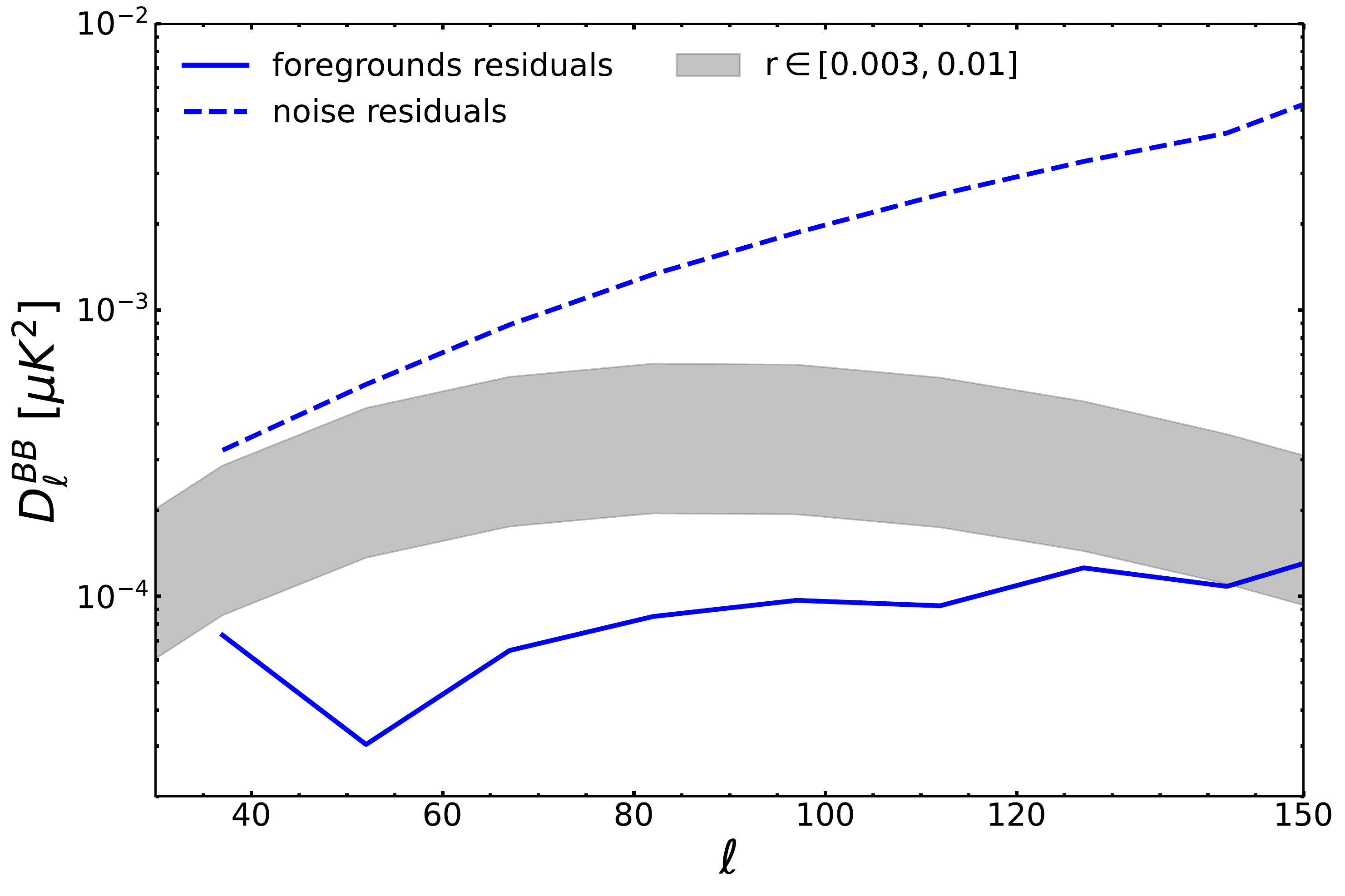}
	\hspace{0.5 cm}
	\includegraphics[width=0.45\textwidth]{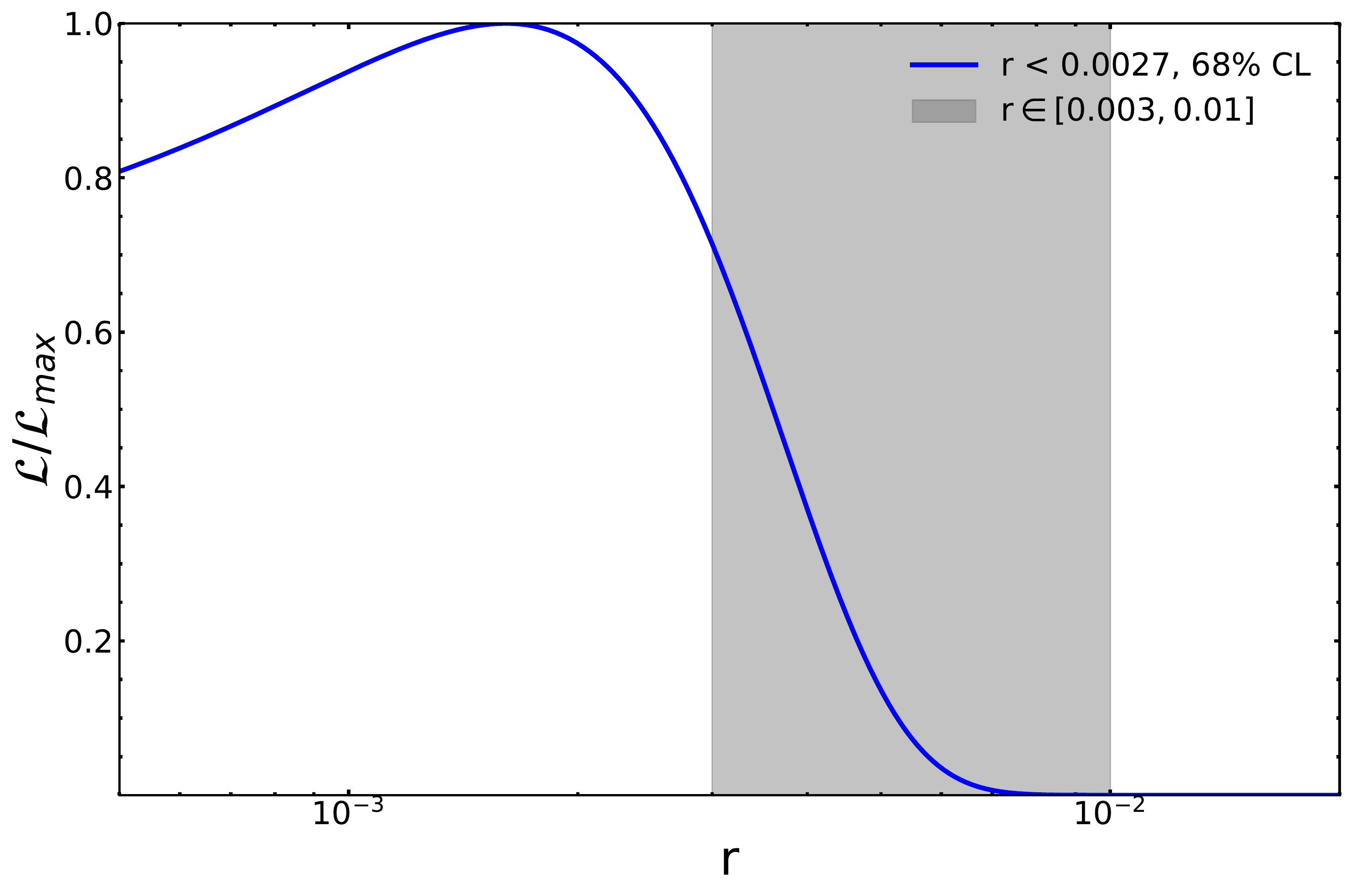}
	\includegraphics[width=0.45\textwidth]{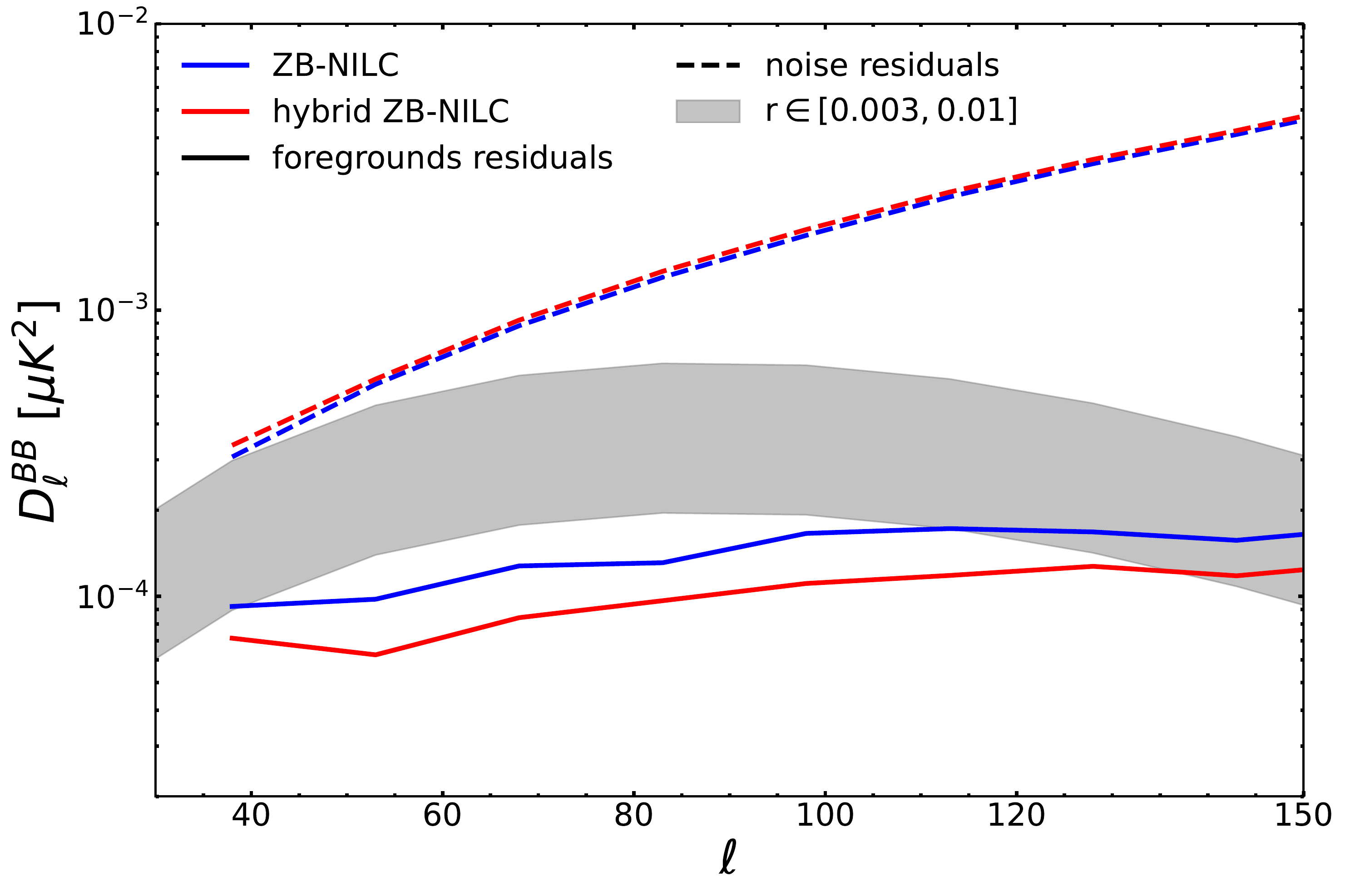}
	\hspace{0.5 cm}
	\includegraphics[width=0.45\textwidth]{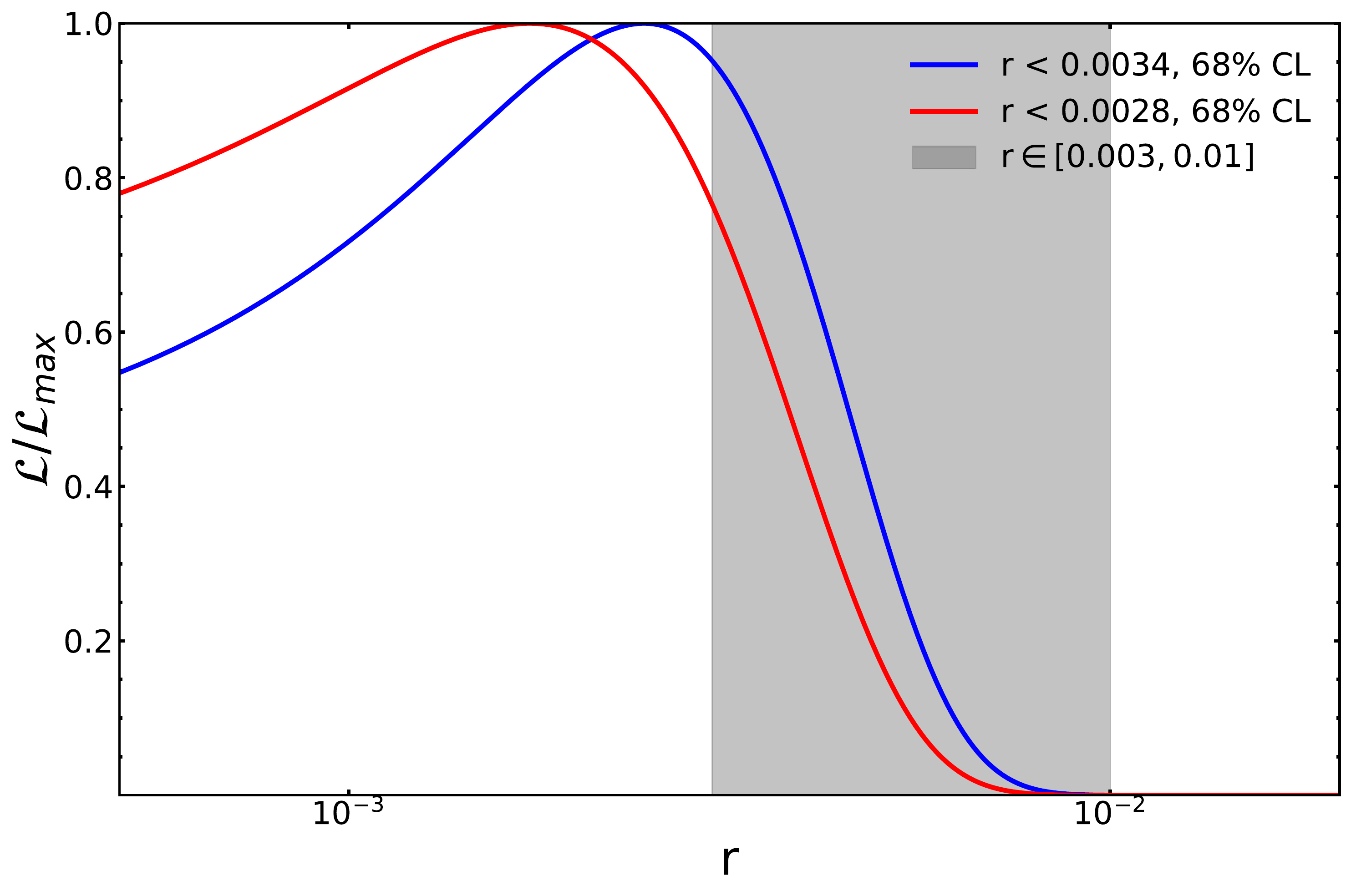}
	\caption{SO SAT data-set with the \texttt{d1s1} Galactic model. On the left: the mean angular power spectra of NILC foregrounds (solid) and noise (dashed) residuals over $200$ simulations.
	On the right: the posterior distributions of an \textit{effective} tensor-to-scalar ratio fitted to the foregrounds residuals in the case of half de-lensing ($A_{L}=0.5$). Both for the plots of the angular power spectra and the estimation of the posterior a binning scheme of $\Delta\ell =15$ has been used. 
	On the top: results when E-B leakage is corrected with the recycling method (r-NILC). On the bottom: results when the ZB- (blue) and hybrid ZB-NILC (red) methods are applied. Everywhere the grey region represents the primordial tensor BB angular power spectrum for $r\in [0.003,0.01]$. The lower bound represents the targeted upper limit at $68\%$ CL by SO in case of no detection, the upper one the targeted detection at few $\sigma$ significance. The spectra have been estimated masking the regions most contaminated by foregrounds for a final $f_{sky}$ of $8\%$.}
	\label{fig:results_liu_nilc_so_d1s1}
\end{figure*}
In the following, for all cases, we show the posterior distributions of an effective tensor-to-scalar ratio fitted on the foregrounds residuals average power spectrum, as described in Sect. \ref{sec:like}. When $r_{fgds}=0$ is within $2\sigma$, we report upper bounds on $r_{fgds}$ at $95\ \%$ CL (for SWIPE) or $68\ \%$ CL (for SO) to be compared with those in the official forecasts of the two experiments (see \citealt{2021JCAP...08..008A} and \citealt{2019JCAP...02..056A})..
\subsubsection{Recycling NILC (r-NILC)}
\label{sec:rNILC}
In this section we report the results obtained when the NILC algorithm is applied to multi-frequency $B$-mode maps that include CMB, noise and foregrounds, and have been corrected with the recycling method (r-NILC). We note that, in the case of the LSPE+Planck data-set, the maps have also been post-processed either with the iterative B-decomposition with three iterations (rit-NILC) or with the diffusive inpainting (rin-NILC). \\
Recently, updated forecasts on $r$ for the LSPE experiment have been published in \citet{2021JCAP...08..008A} considering a Galactic model with constant spectral indices for polarised thermal dust and synchrotron. Following the same approach, we first test the recycling NILC on the SWIPE+Planck data-set adopting the \texttt{d0s0} model. \\
The mean angular power spectrum of foregrounds and noise residuals in this case are shown in the upper left panel of Fig. \ref{fig:results_liu_nilc_swi_d0s0}. The adopted binning scheme is $\Delta\ell =6$ for $5 \leq \ell \leq 28$ and $\Delta\ell =15$ for $\ell \geq 29$. The foregrounds residuals are confidently below both the reionisation and recombination bumps for $r=0.015$ (the LSPE targeted upper bound at $2\sigma$ in case of no detection). The angular power spectra have been estimated without any masking of the most contaminated regions by residuals of the Galactic emission. \\ 
The plot on the upper right-hand side of Fig. \ref{fig:results_liu_nilc_swi_d0s0} shows that the effective tensor-to-scalar ratio associated to foregrounds residuals has a posterior distribution with an upper limit at $95\%$ CL of $\sim 1.5\cdot 10^{-2}$, which is fully in accordance with the LSPE target. This constraint has been obtained by excluding the multipoles $\ell \leq 4$ where a significant negative bias is observed in the reconstructed CMB $B$-mode power spectrum due to the B-E leakage effect (see Sect. \ref{sec:leak_liu}). \\
When a more complicated sky model is considered (\texttt{d1s1} with anisotropic spectral indices for dust and synchrotron), the NILC weights try to trace locally several different SEDs at the same time, leading to a more noisy and less effective removal of the Galactic emission in each pixel. Therefore, in this case, a more aggressive masking strategy (based on the second approach described in Sect. \ref{sec:masks}) is needed to exclude the most contaminated regions. We have found in the analysis of Fig. \ref{fig:r_fgds_masks} that a sky fraction of $f_{sky}=12\%$ is enough to avoid biases in the estimate of the tensor-to-scalar ratio. \\
The angular power spectra of the residuals are shown in the upper left panel of Fig. \ref{fig:results_liu_nilc_swi_d1s1}. Both the rit-NILC and rin-NILC cleaning methods still allow one to possibly detect both peaks in the $B$-mode primordial spectrum. \\
The posterior distribution of an \textit{effective} tensor-to-scalar ratio fitted on the foregrounds residuals has an upper limit of around $r\lesssim 2.5 \cdot 10^{-2}$ at $2\sigma$, which is mainly sourced by the contribution of the noise residuals to the covariance matrix of Eq. \ref{eq:Mllb} and the exclusion of the first three multipoles from the cosmological analysis. \\
The very same pipeline can be applied to SO SAT simulations. In this case, the standard recycling method is adopted, given its effectiveness in removing E-B leakage contamination at the angular scales of interest for SO (see Sect. \ref{sec:leak_liu}). The target of SO is the observation of only the recombination bump, as the experiment is not sensitive to multipoles $\ell < 30$ due to atmospheric contamination. Therefore, the needlet bands and the cleaning algorithm do not take into account modes on angular scales larger than almost $6\degree$. \\
For SO SAT we have considered only a Galactic emission simulated with the \texttt{d1s1} model. The residuals and the posterior distribution of the tensor-to-scalar ratio are shown in the top panels of Fig. \ref{fig:results_liu_nilc_so_d1s1}. The goal of SO is a detection at high significance of $r=0.01$. The application of r-NILC leads to foregrounds residuals that are below such a level in all the multipole range of interest, yielding an upper bound in the posterior distribution of $r<0.0027$ ($68\%$ CL). 
This result is fully in agreement with the constraint targeted by the SO Collaboration ($\delta r \sim 0.003$, \citealt{2019JCAP...02..056A}). However, in this analysis, we have considered input simulations with simple white and isotropic noise, since we simply aim to validate a partial-sky extension of the NILC pipeline to $B$-mode data from future experiments. A proper SO forecast should also consider 
a $1/\textrm{f}$ noise component, which is, however, expected to have only a mild impact on the effectiveness of the cleaning method and on the width of the posterior distribution. \\
The above results for the SO case have been obtained considering the cleanest $8 \%$ of the sky, as suggested by the analysis reported in Fig. \ref{fig:r_fgds_masks}. 
\subsubsection{ZB-NILC}
\label{sec:ZBNILC}
ZB-NILC consists of implementing the variance minimisation in needlet domain after having applied the ZB leakage correction method on multi-frequency $B$-mode maps that include CMB, noise and foregrounds. Therefore, the cleaning algorithm is applied to $\mathcal{B}$ maps of Eq. \ref{eq:EB_zhao} instead of the usual $B$ field. The standard $B$-mode signal is recovered only at the end when the angular power spectra are computed. \\
We consider the same data-sets and cases of the r-NILC analysis to closely compare the results of the two methodologies by looking at Figs. \ref{fig:results_liu_nilc_swi_d0s0}, \ref{fig:results_liu_nilc_swi_d1s1} and
\ref{fig:results_liu_nilc_so_d1s1}. 
The mean angular power spectra (over $200$ simulations) of the foregrounds and noise residuals are shown with blue solid and dashed lines, respectively. \\
In the bottom left panel of Fig. \ref{fig:results_liu_nilc_swi_d0s0}, the results for the Planck+SWIPE data-set with the \texttt{d0s0} Galactic model are reported. It is possible to observe that the ZB-NILC foregrounds residuals are much higher than those of the recycling NILC. This is caused by the fact that, in ZB-NILC, the input maps are convolved for the harmonic filter $N_{\ell}=\sqrt{(\ell+2)!/(\ell-2)!}$ (see Sect. \ref{sec:leak} for details), which boosts the power on smaller scales. Therefore, even the weights of the first needlet band are sensitive much more to contamination on intermediate scales (where noise has a not negligible contribution) than to that on the largest ones (where diffused foregrounds dominate). As a consequence, ZB-NILC weights are much less capable of removing Galactic polarised emission in $\mathcal{B}$ maps. \\
To obtain an acceptable level of foregrounds residuals even in the case of the application of the NILC pipeline to ZB-corrected maps, we have implemented the \emph{Hybrid ZB-NILC} (HZB-NILC). HZB-NILC consists of correcting the E-B leakage effect in frequency maps with the ZB method, but then combining them using the NILC weights estimated from the $B$-mode maps processed with the standard recycling method. \\
The foregrounds and noise residuals of the HZB-NILC are shown with solid and dashed red lines in the bottom panels of Figs. \ref{fig:results_liu_nilc_swi_d0s0}, \ref{fig:results_liu_nilc_swi_d1s1} and
\ref{fig:results_liu_nilc_so_d1s1}. With such an implementation, we can recover foregrounds and noise residuals at a level comparable to recycling NILC, but at the same time we can exploit the greater leakage correction capabilities of the ZB technique on the largest angular scales ($\ell \leq 4$). \\
In this case, for SWIPE+Planck with the \texttt{d0s0} model, we obtain an upper bound on an \textit{effective} tensor-to-scalar ratio fitted on the foregrounds residuals of $5\cdot 10^{-3}$ at $2\sigma$ that is fully compatible with the LSPE target (see bottom right panel of Fig. \ref{fig:results_liu_nilc_swi_d0s0}). The improved constraint on $r$ with respect to r-NILC is motivated by the inclusion in this analysis of the lowest multipoles $\ell \leq 4$. \\
The upper limit on the tensor-to-scalar ratio obtained with HZB-NILC is sensibly lower than that obtained in \citet{2021JCAP...08..008A} even though the same Galactic model is assumed. Such a difference can be explained taking into account three key factors: i) in this work we do not take into account the scanning strategy and, thus, the anisotropic distribution of the instrumental noise, which may play a role in the performance of the component separation; ii) we assume to be able to de-lens our $B$-mode power spectrum at a $50\%$ level \citep{delensing}, thus reducing the impact of lensing $B$ modes to the total uncertainty in the $r$ reconstruction; iii) for this simplistic Galactic model the minimisation of the variance across the different needlet scales can in principle outperform the parametric methods, which fit spectral parameters in different regions of the sky. \\
The ZB and hybrid ZB-NILC methods are then applied to the LSPE+Planck data-set assuming the more realistic \texttt{d1s1} model of the Galaxy. The corresponding results are shown in the lower panels of Fig. \ref{fig:results_liu_nilc_swi_d1s1}. We again observe different capabilities of the two methods in subtracting Galactic contamination (in favour of the HZB-NILC). However, in this case, the differences between the power spectra are less evident because the considered sky components are intrinsically more difficult to subtract. With a more aggressive Galactic mask with respect to the \texttt{d0s0} analysis, which retains the $12 \%$ of the sky, both the reionisation and recombination peaks with $r=0.015$ would be observed. The upper bound on an \textit{effective} tensor-to-scalar ratio fitted on the foregrounds residuals is $r<0.016$ ($95\%$ CL) for the Hybrid ZB technique, which is in accordance with the LSPE target in case of no detection. As above, the lower constraint with respect to r-NILC pipelines is given by the possibility of including the first three multipoles in the likelihood. \\
The same trends are obtained when ZB-NILC and Hybrid ZB-NILC have been applied to the simulated SO SAT data-set with Galactic emission simulated with the \texttt{d1s1} model (see the bottom panel of Fig. \ref{fig:results_liu_nilc_so_d1s1}). 
The angular power spectrum of the HZB-NILC Galactic residuals is below the curve of the primordial tensor signal targeted by SO ($r=0.01$) for $\ell > 30$, if we observe the cleanest $8 \%$ of the sky. The HZB-NILC foregrounds residuals lead to an upper bound in the posterior distribution of the effective tensor-to-scalar ratio of $r<0.0028$ ($68\%$ CL) which is compatible with the SO target ($r \leq 3 \cdot 10^{-3}$ at $1 \sigma$). \\
We observe that for SO SAT the difference in the amplitudes of foregrounds residuals on large scales between ZB- and HZB-NILC is much less evident than the one for SWIPE+Planck. This is due to the fact that, without modes with $\ell \leq 30$, the calculation of HZB-NILC weights in the first needlet band is more affected by noise contamination on intermediate scales. This condition is close to that experienced by the ZB-NILC pipeline.
\section{Conclusions}
\label{sec:conclusion}
The main goal of future CMB experiments will be the detection of polarisation $B$ modes generated by primordial tensor perturbations, to definitely confirm the inflationary scenario. \\
Given its low number of assumptions, NILC represents an effective alternative for the analysis of future $B$-mode data to the commonly used parametric component separation methods, which are more subject to systematic biases in the case of mis-modelling of the foreground properties. However, most future surveys will be balloon-borne or ground-based and will observe only a portion of the sky, whereas, at present, NILC has been successfully applied to full-sky data from either Planck or WMAP. \\
In this work, we explore the possibility to extend the NILC formalism to future $B$-mode partial-sky observations, specifically addressing the complications that such an application yields: the E-B leakage, needlet filtering, and beam convolution (see Sect. \ref{sec:extensions}).
Their impact on the reconstruction of CMB $B$ modes has been assessed in Monte Carlo simulations for
two complementary experiments: LSPE-SWIPE and the Small Aperture Telescope of Simons Observatory (SO-SAT). The former aims at a detection with high significance of both reionisation and recombination peaks with $r=0.03$ or to set an upper bound of $r=0.015$ at $95\%$ confidence level; the latter, instead, is designed to be able to measure a signal at the $r = 0.01$ level at a few $\sigma$ significance, or exclude it at similar significance, using the amplitude of the $B$ modes around the recombination bump. In the case of LSPE-SWIPE, realistic simulated Planck maps have been included in the component separation pipeline to have a broader frequency coverage and better trace the spectral properties of $B$-mode foregrounds. \\
When dealing with partial-sky observations, the estimation of $B$ modes from CMB polarisation measurements is challenging due to the mixing of $E$ and $B$ modes. In order to correct for this effect, we implement two different techniques to be applied in pixel space: the \emph{recycling} \citep{2019PhRvD.100b3538L} and the \emph{ZB} \citep{2010PhRvD..82b3001Z} method. We have tested their ability to recover the CMB $B$-mode angular power spectrum on Monte Carlo simulations, also quantifying a possible bias on the estimate of the tensor-to-scalar ratio. We find that:
\begin{itemize}
    \item[$\bullet$] the recycling method is able to reduce the E-B leakage contamination in the $B$-mode maps at a negligible level for SO at all the angular scales of interest $(\ell \geq 30)$, while for LSPE-SWIPE only at $\ell \gtrsim 15$ (see Fig. \ref{fig:leak_liu_swiso}).    
    \item[$\bullet$] the ZB method, instead, allows us to recover the input CMB signal at all angular scales for both the considered experiments (see Fig. \ref{fig:leak_corr_zhao}). However, this technique yields some complications: the need to work with the $\mathcal{B}$ field of Eq. \ref{eq:EB_zhao} and to apodise the mask.
\item[$\bullet$] the E-B leakage residuals of the recycling method on large angular scales ($\ell \lesssim 15$) in the SWIPE patch can be further reduced for $\ell \geq 5$ by applying either of the following two post-processing prescriptions: 
\begin{enumerate}
    \item the diffusive inpainting \citep{2019PhRvD.100b3538L}
    \item the iterative B-decomposition, which we introduce for the first time in this paper. Ambiguous modes, still present in the map, are filtered out by iteratively applying the B-decomposition of the Q and U CMB maps reconstructed with the recycling method.
\end{enumerate}
\item[$\bullet$] the reconstruction of the CMB $B$ modes obtained with the recycling method and its extensions fails at $\ell < 5$ due to the unavoidable loss of modes caused by the leakage of $B$ modes into $E$ modes. Therefore, when this method is applied, the first three multipoles have to be excluded from the final cosmological analysis.
\end{itemize}
Needlet filtering and beam convolution performed on incomplete sky observations can also potentially affect the reconstruction of CMB $B$ modes, as values of the pixels close to the border of the observed patch are influenced by the null ones of the unobserved regions. However, for the experiments under consideration, we have verified that these operations have a negligible impact. \\
We summarise the ability to recover the angular power spectrum of CMB $B$ modes at different angular scales after applying the E-B leakage correction methods, needlet filtering, and beam convolution in Table \ref{tab:leak}. \\
Once the above pre-processing steps have been tested, the performance of the NILC cleaning method has been assessed on simulated multi-frequency $B$-mode data-sets which include CMB with only lensing, instrumental noise, and Galactic foreground emission. For LSPE-SWIPE we have considered two different foreground models: one in which the spectral indices of the synchrotron and dust emissions are assumed to be position independent (\texttt{d0s0}), and another in which they vary across the sky (\texttt{d1s1}). For SO-SAT, instead, we generate simulations only with the \texttt{d1s1} model, as in \citet{2019JCAP...02..056A}. \\
The purpose of this paper is to introduce and validate a complete extension of the NILC pipeline to be applied on partial-sky $B$-mode data from future experiments. A detailed forecast of the performance of the considered CMB experiments should take into account some additional realistic effects, such as the scanning strategy of the instrument or the $1/f$ noise contamination. \\ 
\begin{table*}[htbp!]
\centering
\setlength{\tabcolsep}{3pt}
\renewcommand{\arraystretch}{1.3}
\begin{tabular}{|c|c|c|c|c|}
\hline
 & \textbf{\boldmath $\ell < 5$ (SWIPE)} & \textbf{\boldmath $5 \leq \ell \lesssim 15$ (SWIPE)} & \textbf{\boldmath $ \ell \gtrsim 15$ (SWIPE)} & \textbf{\boldmath $\ell > 30$ (SO SAT)} \\
\hline\hline
\text{\textbf{recycling}} &  \xmark &  \xmark &  \cmark &  \cmark  \\
\hline
\text{\textbf{recycling + iterative}}  & \xmark &  \cmark &  \cmark &   \\
\hline
\text{\textbf{recycling + inpainting}} & \xmark &  \cmark &  \cmark &   \\
\hline
\text{\textbf{ZB}} & \cmark &  \cmark &  \cmark & \cmark  \\
\hline
\text{\textbf{needlet filtering}} & \cmark &  \cmark &  \cmark & \cmark  \\
\hline
\text{\textbf{beam convolution}} & & &  & \cmark  \\
\hline

\end{tabular}
\medskip
\caption{Capability to reconstruct the input angular power spectrum of CMB B-modes with lensing and primordial tensor perturbations at different angular scales for the two experiments considered in this work (SWIPE and SO-SAT). We report the results for the different E-B leakage correction methods, for needlet filtering and beam convolution. The latter is relevant only for SO-SAT for which maps of the different frequency channels need to be brought at the same resolution.}
\label{tab:leak}
\end{table*}
\begin{table*}[htbp!]
\centering
\setlength{\tabcolsep}{3pt}
\renewcommand{\arraystretch}{1.3}
\hspace{0.1 cm}
\begin{tabular}{|c|c|c|c|}
\hline
 & \text{\textbf{SWIPE+Planck, \texttt{d0s0}}} & \text{\textbf{SWIPE+Planck, \texttt{d1s1}}} & \text{\textbf{SO SAT, \texttt{d1s1}}}  \\
\hline\hline
\textbf{Targets} &  $r < 0.015\  (95\%\ \textrm{CL})$ & $r < 0.015\  (95\%\ \textrm{CL})$
& $r < 0.003\ (68\%\ \textrm{CL})$ \\
\hline
\textbf{r-NILC} &  &  & $r <  0.0027\ (68 \%\ \textrm{CL})$ \\
\hline
\textbf{rit-NILC} & $r < 0.014\ (95 \%\ \textrm{CL})$ & $r < 0.024\ (95 \%\ \textrm{CL})$ &  \\
\hline
\textbf{rin-NILC} & $r < 0.015\ (95 \%\ \textrm{CL})$ & $r < 0.029\ (95 \%\ \textrm{CL})$ &   \\
\hline
\textbf{HZB-NILC} & $r < 0.005\ (95 \%\ \textrm{CL})$ & $r < 0.016\ (95 \%\ \textrm{CL})$ & $r < 0.0028\ (68 \%\ \textrm{CL})$  \\
\hline
\end{tabular}
\medskip
\caption{Upper bounds and constraints of an \textit{effective} tensor-to-scalar ratio fitted on the average angular power spectrum of foregrounds residuals given by the application of NILC for the listed cases.}
\label{tab:rs}
\end{table*}
Simulated maps have been corrected for the E-B leakage with:
\begin{itemize}
    \item recycling + iterative decomposition (\emph{rit-NILC}) for SWIPE+Planck
    \item recycling + diffusive inpainting (\emph{rin-NILC}) for SWIPE+Planck
    \item recycling (\emph{r-NILC}) for SO-SAT
    \item ZB (\emph{ZB-NILC}) for both SO-SAT and SWIPE+Planck.
\end{itemize}
For all cases, we quantify the ability of the pipeline to mitigate foreground contamination in terms of an \textit{effective} tensor-to-scalar ratio, which is fitted on the angular power spectrum of foregrounds residuals through a Gaussian likelihood (see Eq. \ref{eq:like}) after the application of an appropriate masking strategy. The derived upper bounds are summarised in Table \ref{tab:rs}. They are reported, respectively, at $95\%$ and $68\%$ CL for LSPE-SWIPE and SO, as in the official forecasts of the two experiments (see \citealt{2021JCAP...08..008A} and \citealt{2019JCAP...02..056A}). \\
We find that SO-SAT bounds are within the goal of the experiment regardless of the adopted E-B leakage correction approach. In this work, as mentioned above, we did not
incorporate a 1/f noise component in the SO SAT simulations, which is necessary for an accurate description of the experiment. However, this component is expected to only mildly affect the performance of the method and the obtained constraints, as also proved in \citet{2019JCAP...02..056A}. 
For SWIPE+Planck with the \texttt{d1s1} foregrounds model, when the recycling method is applied to correct the E-B leakage, the obtained upper bound on $r$ is greater than the target of the experiment. This result is caused by the need to exclude the first three multipoles where the B-E leakage has a significant impact on the CMB power spectrum reconstruction. For all other cases, instead, the constraints are in agreement with the expected sensitivity also for LSPE-SWIPE. \\
When the ZB-NILC pipeline is applied (see Sect. \ref{sec:ZBNILC}), the amplitude of the foregrounds residuals is much greater than that obtained with r-NILC. This can be explained by the fact that in this case the input maps are convolved with the harmonic filter $N_{\ell}=\sqrt{(\ell+2)!/(\ell-2)!}$ (see Sect. \ref{sec:leak_zhao} for details), which boosts the power on small scales where noise is dominant. This reduces our capability to trace and remove Galactic contamination at low multipoles. To solve this problem, we have implement the \emph{Hybrid ZB-NILC} (HZB-NILC), where the E-B leakage effect is corrected with the ZB method, while the NILC weights are estimated on the $B$-mode maps corrected with the recycling technique. \\
We note that the upper limit on the tensor-to-scalar ratio obtained with the HZB-NILC pipeline and assuming the Galactic model \texttt{d0s0} is sensibly lower than the one obtained in \citet{2021JCAP...08..008A}. 
This difference can be motivated by taking into account several factors: i) in this work, we do not consider the scanning strategy and, thus, the anisotropic distribution of the instrumental noise, which may play a role in the performance of the component separation; ii) we assume to be able to de-lens our $B$-mode power spectrum at a $50\%$ level \citep{delensing}, thus reducing the impact of lensing $B$ modes to the total uncertainty in the $r$ reconstruction; iii) for the simplistic \texttt{d0s0} Galactic model, the minimisation of the variance in needlet domain can in principle outperform parametric methods, which fit spectral parameters in different regions of the sky.\\
In this paper, we presented the development and validation of a pipeline, based on the NILC component separation method, for the analysis of future CMB $B$-mode data collected by experiments that will observe the microwave sky from the ground and balloons. Taking into account real-world issues due to the incomplete sky coverage and state-of-the-art simulations of Galactic foregrounds, our results demonstrate the effectiveness of NILC, which emerges as a valid alternative to parametric component separation techniques for these kinds of experiments. Furthermore, this study permits to easily extend the application of other ILC based techniques, \textit{e.g.} cMILC \citep{2021MNRAS.503.2478R} and MC-NILC \citep{2022arXiv221204456C}, to multi-frequency $B$-mode data from ground-based and balloon-borne experiments.

\begin{acknowledgements}
      We thank Luca Pagano, Francesco Piacentini and Giuseppe Puglisi for helpful comments. MM and NV acknowledge support by ASI/COSMOS grant n. 2016-24-H.0 and ASI/LiteBIRD grant n. 2020-9-HH.0. This work has been supported by the ASI-INFN Grant Agreement N. 2021-43-HH.0. This research used resources of the National Energy Research Scientific Computing Center (NERSC), a U.S. Department of Energy Office of Science User Facility located at Lawrence Berkeley National Laboratory. Part of this work was also supported by the InDark and LiteBIRD INFN projects. D.M. acknowledges support from the MIUR Excellence Project awarded to the Department of Mathematics, Università di Roma Tor Vergata, CUP E83C18000100006.
\end{acknowledgements}

\bibliography{lspe} 
\bibliographystyle{aa} 

\begin{appendix}

\section{The NILC bias}
\label{app:nilc_bias}
\begin{figure*}[t]
	\includegraphics[width=0.45\textwidth]{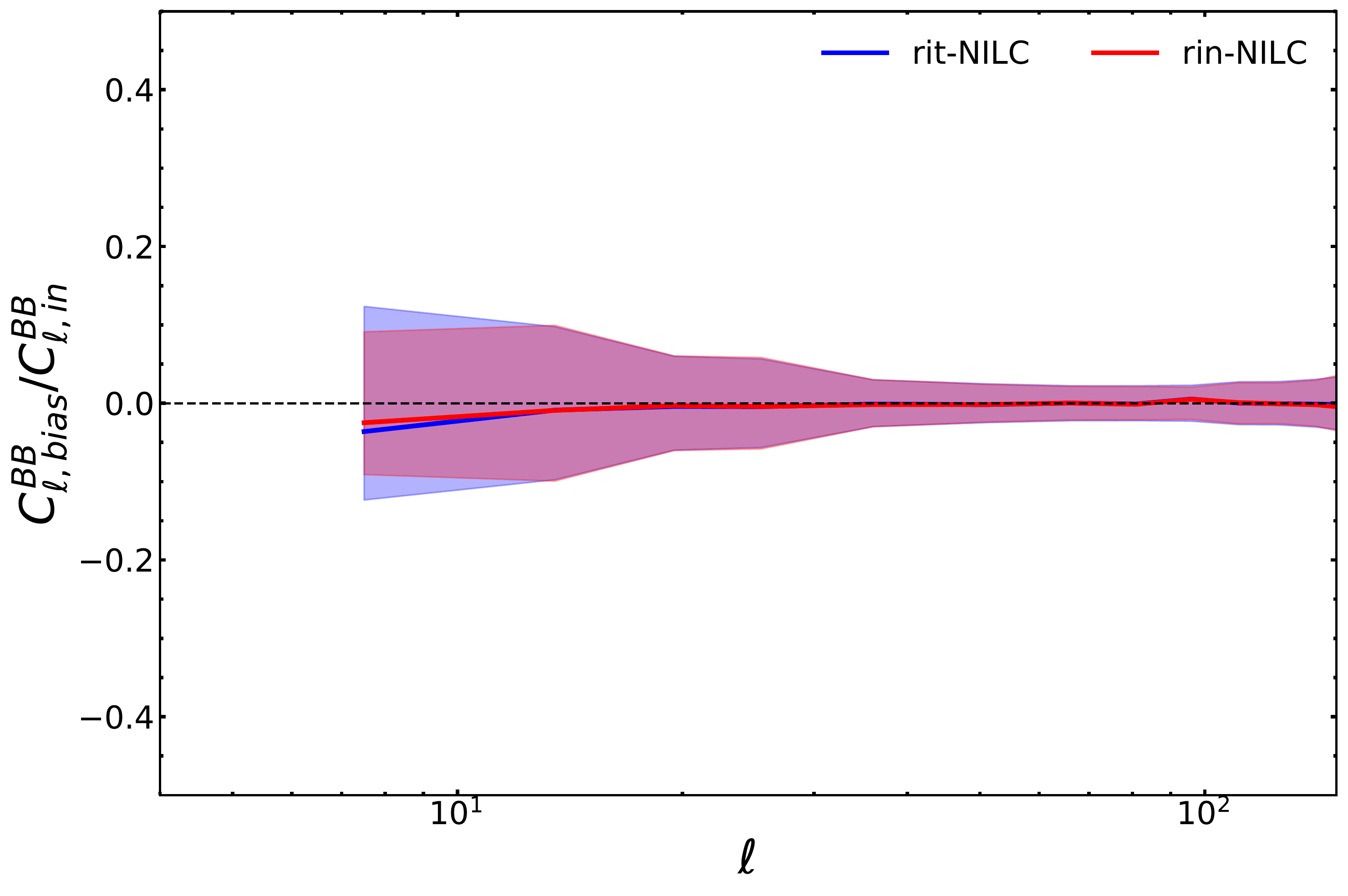}
	\includegraphics[width=0.45\textwidth]{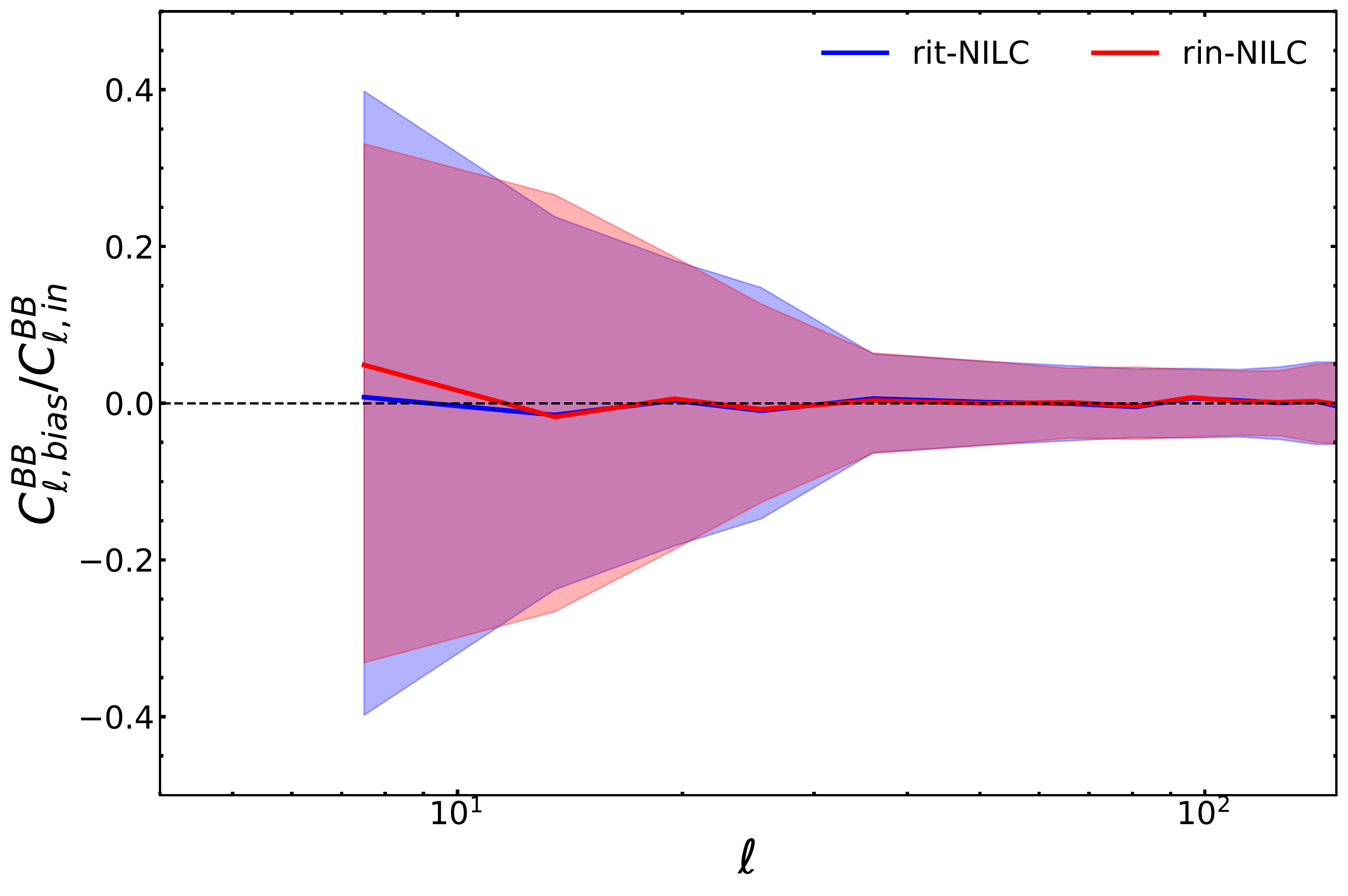}
    \includegraphics[width=0.45\textwidth]{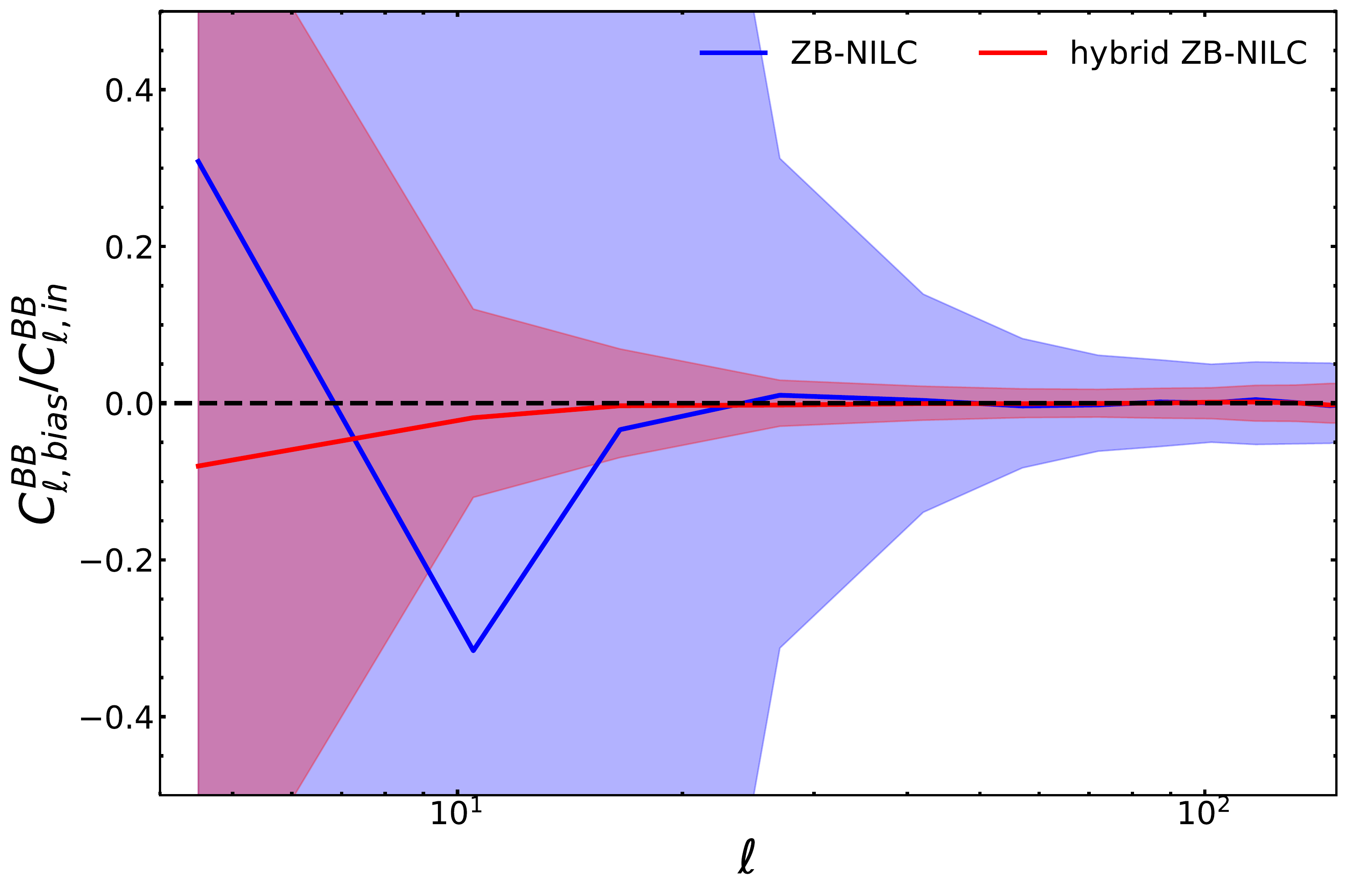}
    \hfill
    \includegraphics[width=0.45\textwidth]{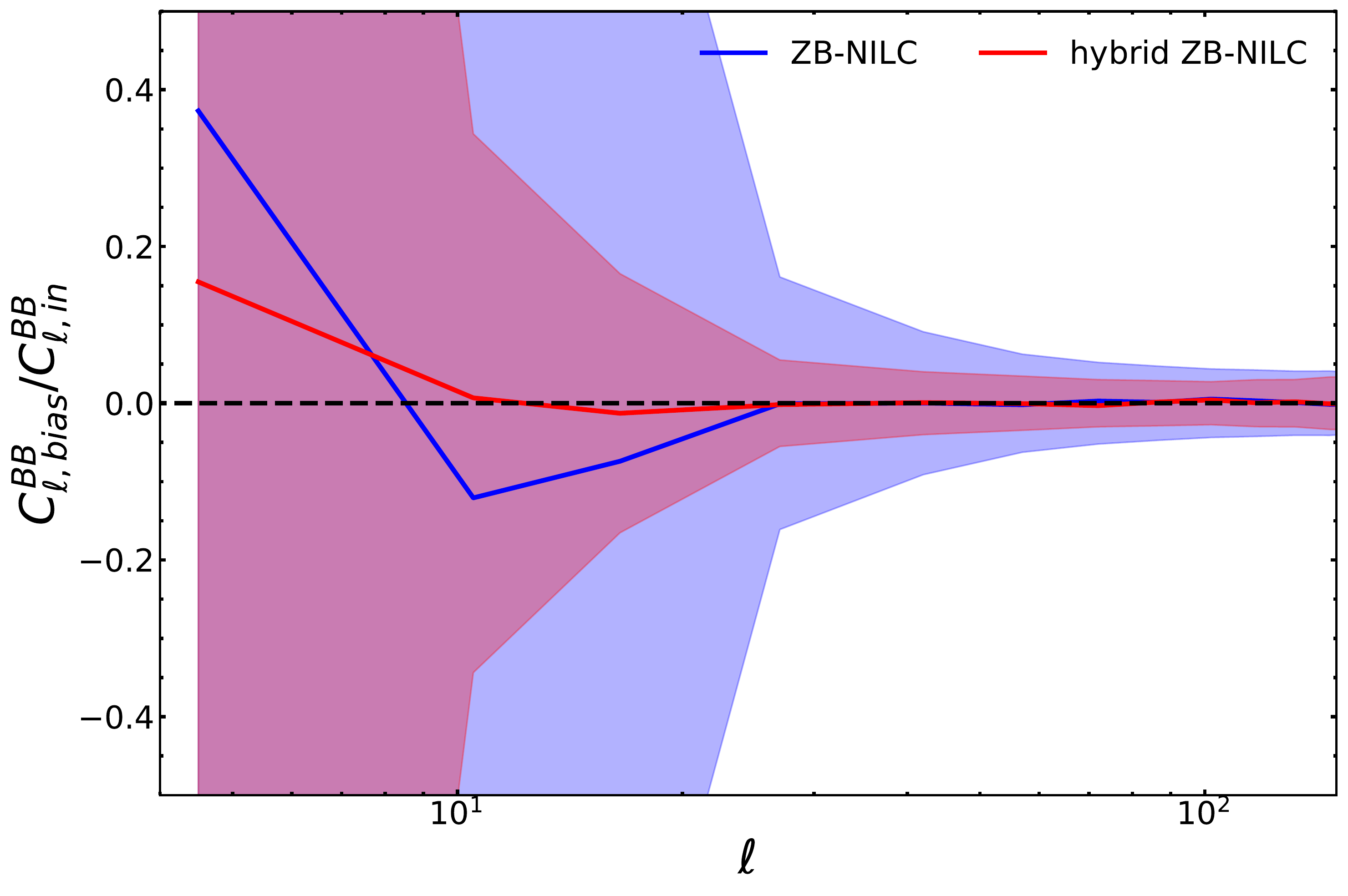}	
    \caption{SWIPE+Planck data-set. The ratio between the bias $C_{\ell}^{bias}$ (see Eq. \ref{eq:bias_app}) and the input CMB power spectrum $C_{\ell}^{cmb}$. All angular power spectra are computed as the average among $200$ different simulated NILC CMB reconstructions and adopting the masks employed in Sect. \ref{sec:results}. The shaded regions display the relative uncertainty computed as $\sigma(C_{\ell}^{bias})/(C_{\ell}^{cmb}\cdot \sqrt{N_{sims}})$, where $N_{sims}$ is the number of simulations. Top: results of the application of rit-NILC (blue) and rin-NILC (red). Bottom: bias from ZB-(blue) and hybrid ZB-NILC (red). The Galactic emission is simulated with the \texttt{d0s0} (left) and the \texttt{d1s1} (right) sky models.}
\label{fig:bias_swipe}
\end{figure*}
The NILC weights are estimated employing the empirical covariance matrix of the input multi-frequency channels (see Eq. \ref{eq:NILC_weights}). \\
If the covariance matrix is not correctly estimated due to the finite size of the domain over which it is computed, empirical correlations between the CMB modes and the residual contaminants are generated, leading to the de-projection of some of the CMB signal together with foregrounds and noise. \\
This phenomenon is known as \emph{NILC bias} and can cause a negative bias in the reconstructed CMB angular power spectrum, especially on the largest scales.  \\
\begin{figure*}[t]
	\centering
	\includegraphics[width=0.45\textwidth]{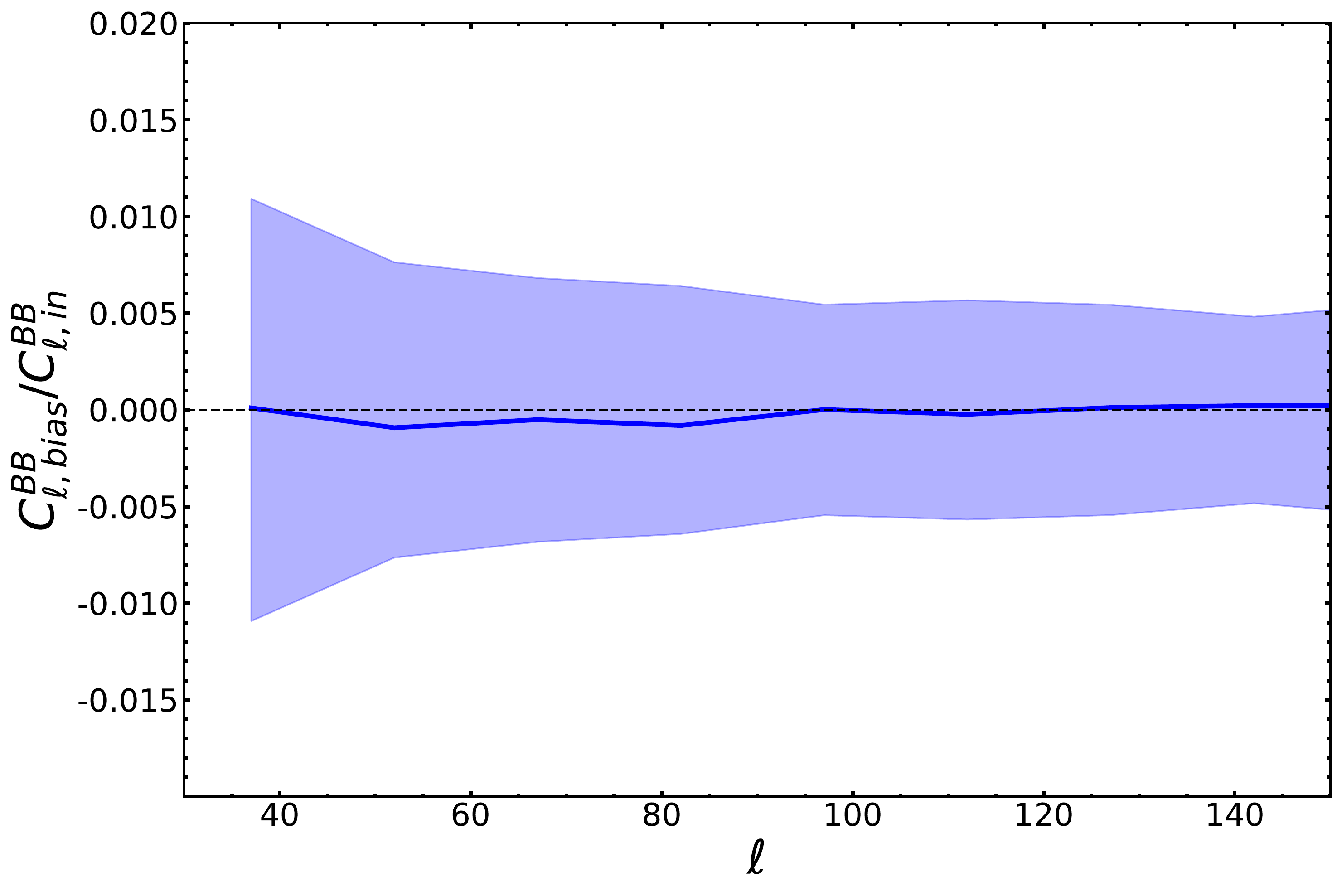}
	\hspace{1.4 cm}
	\includegraphics[width=0.45\textwidth]{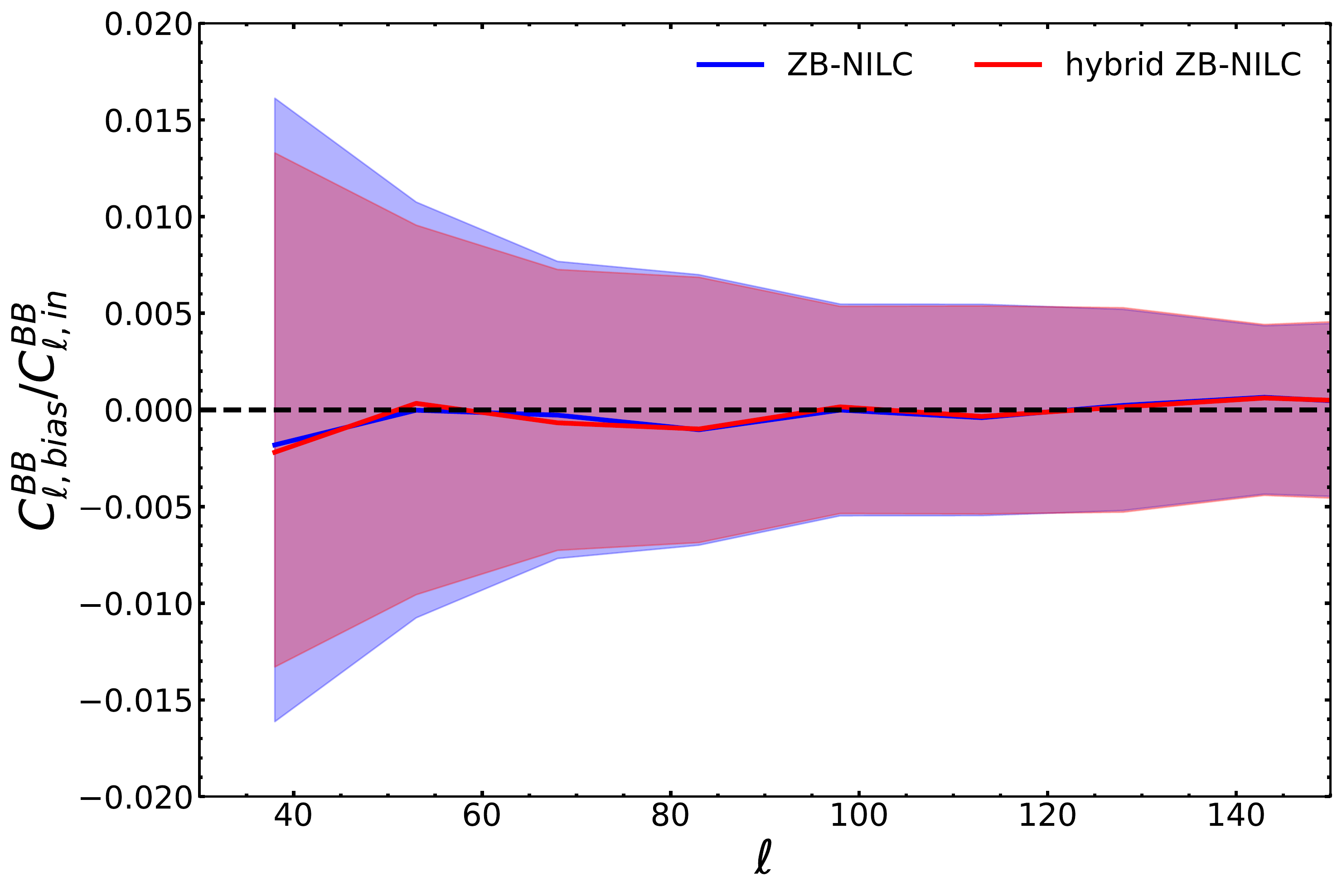}
	\caption{SO SAT data-set. The ratio between the bias $C_{\ell}^{bias}$ (see Eq. \ref{eq:bias_app}) and the input CMB power spectrum $C_{\ell}^{cmb}$. All angular power spectra are computed as the average among $200$ different simulated NILC CMB reconstructions and adopting the masks employed in Sect. \ref{sec:results}. The shaded regions display the relative uncertainty computed as $\sigma(C_{\ell})/(C_{\ell}^{cmb}\cdot \sqrt{N_{sims}})$. Left: results of the application of r-NILC; right: bias from the ZB-(blue) and hybrid ZB-NILC (red) methods. In all cases input foregrounds emission is simulated assuming the \texttt{d1s1} sky model.}
\label{fig:bias_nilc_so}
\end{figure*}
To avoid such an effect, enough modes have to be sampled in the computation of the covariance matrix to preserve the input CMB signal. In general, this problem is tackled by identifying a proper domain to compute the sample average of the product of needlet coefficients. Furthermore, usually the first needlet band is built from the sum in quadrature of several harmonic filters at low multipoles (see Eq. \ref{eq:b_merge}). \\
In practise, after the application of a component separation method on the $B$-mode data, one would like to extract cosmological information from the angular power spectrum ($C_{\ell}^{out}$) of the BB cleaned map. This quantity is composed of several terms: 
\begin{equation}
C_{\ell}^{out} = C_{\ell}^{cmb} + C_{\ell}^{fgds} + C_{\ell}^{noi} + 2\cdot C_{\ell}^{c-f} + 2\cdot C_{\ell}^{c-n} + 2\cdot C_{\ell}^{n-f},
\end{equation}
where $C_{\ell}^{cmb},\ C_{\ell}^{fgds}$ and $C_{\ell}^{noi}$ are the angular power spectra of CMB, residuals of foregrounds and noise, while the other terms represent the corresponding correlations among these components. \\
With de-noising techniques (cross-spectra of half-mission solutions or Monte Carlo simulations) one can eliminate $C_{\ell}^{noi}$, while the foregrounds contamination ($C_{\ell}^{fgds}$) can be marginalised at the likelihood level. Therefore, our final estimate of the angular power spectrum of the NILC CMB solution reads:
\begin{equation}
    \hat{C}_{\ell}^{out} = C_{\ell}^{cmb} + 2\cdot C_{\ell}^{c-f} + 2\cdot C_{\ell}^{c-n} + 2\cdot C_{\ell}^{n-f}.
\label{eq:bias}
\end{equation}
If NILC is properly implemented, the correlation terms should be very low and the CMB well reconstructed. \\
In this paper, we analyse simulated data. Thus, it is possible to assess the goodness of the CMB angular power spectrum reconstruction by computing:
\begin{equation}
    C_{\ell}^{bias} = C_{\ell}^{out} - C_{\ell}^{fgds} - C_{\ell}^{noi}-C_{\ell}^{cmb},
\label{eq:bias_app}
\end{equation} 
which should be compatible with zero at all angular scales. \\
The relative bias $C_{\ell}^{bias}/C_{\ell}^{cmb}$ is shown in Figs. \ref{fig:bias_swipe} and \ref{fig:bias_nilc_so} for the different cases considered in Sect. \ref{sec:results}: the application of r-NILC to the LSPE+Planck data-set assuming the \texttt{d0s0} or \texttt{d1s1} sky model (see the top panels of Figs. \ref{fig:results_liu_nilc_swi_d0s0} and \ref{fig:results_liu_nilc_swi_d1s1}) and to SO SAT data-set with the \texttt{d1s1} Galactic model (see the top panels of Fig. \ref{fig:results_liu_nilc_so_d1s1}); the application of ZB- and Hybrid ZB-NILC in the same cases (see the bottom panels of Figs. \ref{fig:results_liu_nilc_swi_d0s0}, \ref{fig:results_liu_nilc_swi_d1s1} and
\ref{fig:results_liu_nilc_so_d1s1}). \\
We can observe that the bias is always fully compatible with zero given the uncertainty on the reconstructed mean $C_{\ell}^{bias}$, which is estimated by dividing the dispersion of the angular power spectra of the bias among the different CMB reconstructions by the square root of the number of simulations: $\sigma(C_{\ell}^{bias})/ \sqrt{N_{sims}}$.

\end{appendix}

\end{document}